\documentclass[fleqn,usenatbib]{mnras}

\usepackage{newtxtext,newtxmath}
\usepackage[T1]{fontenc}

\DeclareRobustCommand{\VAN}[3]{#2}
\let\VANthebibliography\thebibliography
\def\thebibliography{\DeclareRobustCommand{\VAN}[3]{##3}\VANthebibliography}

\usepackage{graphicx}
\usepackage{amsmath}
\usepackage{longtable}
\usepackage{subfig}

\title[Reconciling MOSDEF \& KBSS-MOSFIRE]{Reconciling the Results of the $z\sim2$ MOSDEF and KBSS-MOSFIRE Surveys$^{1}$}

\author[J. N. Runco et al.]{Jordan N. Runco,$^{2}$\thanks{E-mail: jrunco@astro.ucla.edu}
Naveen A. Reddy, $^{3}$
Alice E. Shapley,$^{2}$
Charles C. Steidel,$^{4}$\newauthor
Ryan L. Sanders,$^{5,6}$
Allison L. Strom,$^{7}$
Alison L. Coil,$^{8}$
Mariska Kriek,$^{9,10}$\newauthor
Bahram Mobasher,$^{3}$
Max Pettini$^{11}$
Gwen C. Rudie,$^{12}$
Brian Siana,$^{3}$\newauthor
Michael W. Topping,$^{2,13}$
Ryan F. Trainor,$^{14}$
William R. Freeman,$^{3}$
Irene Shivaei,$^{13}$\newauthor
Mojegan Azadi,$^{15}$
Sedona H. Price,$^{16}$
Gene C. K. Leung,$^{17}$
Tara Fetherolf,$^{3}$\newauthor
Laura de Groot,$^{18}$
Tom Zick,$^{9}$
Francesca M. Fornasini,$^{19}$
Guillermo Barro$^{20}$
\\
$^{1}$Based on data obtained at the W.M. Keck Observatory, which is operated as a scientific partnership among the California Institute of \\ Technology, the University of California,  and the National Aeronautics and Space Administration, and was made possible by the generous  \\ financial support  of the W.M. Keck Foundation.\\
$^{2}$Physics \& Astronomy Department, University of California: Los Angeles, 430 Portola Plaza, Los Angeles, CA 90095, USA\\
$^{3}$Department of Physics \& Astronomy, University of California, Riverside, 900 University Avenue, Riverside, CA 92521, USA\\
$^{4}$Cahill Center for Astronomy and Astrophysics, California Institute of Technology, 1200 E California Blvd, MC249-17, Pasadena, CA 91125, USA\\
$^{5}$Department of Physics, University of California, Davis, One Shields Ave, Davis, CA 95616, USA\\
$^{6}$Hubble Fellow\\
$^{7}$Department of Astrophysical Sciences, 4 Ivy Lane, Princeton University, Princeton, NJ 08544, USA\\
$^{8}$Center for Astrophysics and Space Sciences, University of California, San Diego, 9500 Gilman Dr., La Jolla, CA 92093-0424, USA\\
$^{9}$Astronomy Department, University of California, Berkeley, CA 94720, USA\\
$^{10}$Leiden Observatory, Leiden University, PO Box 9513, NL-2300 RA Leiden, The Netherlands \\
$^{11}$Institute of Astronomy, University of Cambridge, Madingley Road, Cambridge, CB3 0HA, UK\\
$^{12}$The Observatories of the Carnegie Institution for Science, 813 Santa Barbara Street, Pasadena, CA 91101, USA\\
$^{13}$Department of Astronomy/Steward Observatory, 933 North Cherry Ave, Rm N204, Tucson, AZ 85721-0065, USA\\
$^{14}$Department of Physics and Astronomy, Franklin \& Marshall College, 637 College Ave., Lancaster, PA 17603, USA\\
$^{15}$Harvard-Smithsonian Center for Astrophysics, 60 Garden Street, Cambridge, MA, 02138, USA \\
$^{16}$Max-Planck-Institut f\"ur Extraterrestrische Physik, Postfach 1312, Garching, 85741, Germany \\
$^{17}$Department of Astronomy, The University of Texas at Austin, 2515 Speedway Blvd Stop C1400, Austin, TX 78712, USA\\
$^{18}$Department of Physics, The College of Wooster, 1189 Beall Avenue, Wooster, OH 44691, USA\\
$^{19}$Department of Physics and Astronomy, Stonehill College, 320 Washington Street, Easton, MA 02357, USA \\
$^{20}$Department of Physics, University of the Pacific, 3601 Pacific Ave, Stockton, CA 95211, USA
}

\date{Accepted XXX. Received YYY; in original form ZZZ}

\pubyear{2021}

\begin{document}
\label{firstpage}
\pagerange{\pageref{firstpage}--\pageref{lastpage}}
\maketitle

\begin{abstract}
The combination of the MOSDEF and KBSS-MOSFIRE surveys represents the largest joint investment of Keck/MOSFIRE time to date, with $\sim$3000 galaxies at $1.4\lesssim z\lesssim3.8$, roughly half of which are at $z\sim2$. MOSDEF is photometric- and spectroscopic-redshift selected with a rest-optical magnitude limit, while KBSS-MOSFIRE is primarily selected based on rest-UV colors and a rest-UV magnitude limit. Analyzing both surveys in a uniform manner with consistent spectral-energy-distribution (SED) models, we find that the MOSDEF $z\sim2$ targeted sample has higher median $M_{\ast}$ and redder rest U$-$V color than the KBSS-MOSFIRE $z\sim2$ targeted sample, and smaller median SED-based SFR and sSFR (SFR(SED) and sSFR(SED)). Specifically, MOSDEF targeted a larger population of red galaxies with U$-$V and V$-$J $\geq$1.25, while KBSS-MOSFIRE contains more young galaxies with intense star formation. Despite these differences in the $z\sim2$ targeted samples, the subsets of the surveys with multiple emission lines detected and analyzed in previous work are much more similar. All median host-galaxy properties with the exception of stellar population age --- i.e., $M_{\ast}$, SFR(SED), sSFR(SED), $A_{\rm{V}}$, and UVJ colors --- agree within the uncertainties. Additionally, when uniform emission-line fitting and stellar Balmer absorption correction techniques are applied, there is no significant offset between both samples in the [O~\textsc{III}]$\lambda$5008/H$\beta$ vs. [N~\textsc{II}]$\lambda$6585/H$\alpha$ diagnostic diagram, in contrast to previously-reported discrepancies. We can now combine the MOSDEF and KBSS-MOSFIRE surveys to form the largest $z\sim2$ sample with moderate-resolution rest-optical spectra and construct the fundamental scaling relations of star-forming galaxies during this important epoch.
\end{abstract}

\begin{keywords}
galaxies: evolution --- galaxies: high-redshift --- galaxies: ISM
\end{keywords}

\section{Introduction} \label{sec:intro}

Rest-frame optical emission-line spectroscopy provides a wealth of information about a galaxy, including its dust extinction, star-formation rate (SFR), virial and non-virial dynamics (e.g., large-scale outflows), active galactic nucleus (AGN) activity, and properties of the ionized interstellar medium (ISM) such as the metallicity, electron density ($n_{\rm{e}}$), ionizing spectrum, and ionization parameter ($U$; i.e., the ratio of ionizing photon density to hydrogen, and therefore electron, density). Understanding the emission-line properties of star-forming galaxies throughout cosmic history is thus essential for characterizing the formation and evolution of the stellar and gaseous contents of galaxies. One of the most important time periods for studying rest-optical emission-line spectra of galaxies is at $z\sim2$, which hosts the peak level star formation in the universe \citep{mad14}. During this epoch, the modern Hubble sequence was not yet fully established and many key features of current galaxy patterns were being set, including the bimodal distribution of galaxy colors and correlations between structural parameters and stellar properties.

In the last decade, the commissioning of multi-object near-IR spectrographs on large ground-based telescopes has given us the ability to characterize the rest-optical emission-line properties of these high-redshift galaxies in large statistical samples. 
Two surveys taking advantage of the MultiObject Spectrometer For Infra-Red Exploration (MOSFIRE; \citealt{mcl12}) instrument on the 10 m Keck I telescope are the MOSFIRE Deep Evolution Field (MOSDEF; \citealt{kri15}) survey and Keck Baryonic Structure Survey (KBSS: \citealt{ste14}). 
Each of the two surveys contains $\sim$1500 galaxies with deep MOSFIRE observations at $1.4 \lesssim z \lesssim 3.8$, with roughly half the galaxies in each survey at $z\sim2$.

At $z\sim2$, Keck/MOSFIRE can observe the full rest-optical wavelength range from $3700-7000$ \AA, including all of the rest-optical emission-lines needed to track the location of $z\sim2$  galaxies on the ``BPT'' diagrams. The [O~\textsc{III}]$\lambda$5008/H$\beta$ vs. [N~\textsc{II}]$\lambda$6585/H$\alpha$ diagram (first introduced by \citealt{bal81} and which we refer to hereafter as the ``[N~\textsc{II}] BPT diagram''), and the [O~\textsc{III}]$\lambda$5008/H$\beta$ vs. [S~\textsc{II}]$\lambda\lambda$6718,6733/H$\alpha$ diagram (first introduced in \citealt{vei87} and which we refer to hereafter as the ``[S~\textsc{II}] BPT diagram'') can both be used to distinguish between AGN and star formation activity as the main ionizing source for the ISM of a galaxy. The ionizing spectrum shape as well as the typical ionization parameter are different in gas photoionized by hot stars as opposed to by an AGN. Therefore, star-forming galaxies and AGNs occupy distinct regions within these rest-optical emission-line diagrams.

In early studies with Keck/NIRSPEC, it was found that star-forming galaxies at $z>1$ show elevated [N~\textsc{II}]$\lambda$6585/H$\alpha$ at fixed [O~\textsc{III}]$\lambda$5008/H$\beta$ (or vice versa; e.g. \citealt{sha05, erb06b, liu08}) on average compared to local galaxies in the Sloan Digital Sky Survey (SDSS; \citealt{yor00}), indicating a redshift dependence in the [N~\textsc{II}] BPT diagram.  Subsequent work based on larger samples collected with MOSFIRE (e.g., \citealt{ste14, sha15}), has since confirmed the systematic [N~\textsc{II}] BPT offset. There have been numerous explanations that have been proposed for this systematic offset, including variations in the physical properties of galaxies such as H~\textsc{II} region ionizing spectra at fixed metallicities, H~\textsc{II} region ionization parameter, gas-phase N/O abundance ratio differences, H~\textsc{II} region electron densities and density structure, unresolved AGN activity, shocks, and galaxy selection effects (e.g., \citealt{liu08, bri08, wri10, kew13, yeh13, jun14, mas14, ste14, ste16, coi15, sha15, sha19, san16, str17, str18, fre19, kas19, top20a, run21}).  
Results based on both KBSS and recent MOSDEF observations favor the idea that the main driver for the observed offset is a harder ionizing spectrum at fixed nebular oxygen abundance, which arises due to $\alpha$-enhancement in the massive stars in star-forming galaxies at $z\sim2$ \citep{ste14, ste16, str17,sha19, top20a, red21, run21}.

Understanding this offset is vital because rest-optical emission-lines are often used to infer many physical properties of the ISM. Locally, many calibrations exist between rest-optical emission-lines and ISM properties (e.g., gas-phase oxygen abundance; \citealt{pet04}). Without a complete understanding of the [N~\textsc{II}] BPT offset, it will remain unclear if these local calibrations can be used in the $z > 1$ universe \citep{bia18,san20b}.

While the MOSDEF and KBSS studies have converged on the factors physically driving the [N~\textsc{II}] BPT offset, there are still differences in key results. Most notable is that the two surveys have found quantitatively different offsets between the emission-line sequences for $z\sim2$ and local galaxies on the [N~\textsc{II}] BPT diagram, with the KBSS sample showing higher emission-line ratios on average \citep{ste14, sha15, str17, run21}. 
This discrepancy suggests that there might be fundamental differences between galaxies contained in the two surveys. 

To fully understand the driver of the [N~\textsc{II}] BPT offset and the correlation between $z\sim2$ rest-optical emission-lines and galaxy properties (e.g., gas-phase oxygen abundance), a data set of unprecedented size is needed. 
To achieve such a sample, this study represents the beginning of a collaboration between the MOSDEF and KBSS teams. 
With over 100 nights of Keck time and $\sim$3000 galaxies observed between $1.4 \lesssim z \lesssim 3.8$ (about half at $z\sim2$), the combination of these two surveys represents the largest total investment of Keck/MOSFIRE time. Combining the two data sets would yield a sample with a statistical size capable of determining robust trends in galaxy properties and relationships in the $z\sim2$ universe analogous to what exists locally.

However, because of differences in survey selection methods and results from previous studies (primarily concerning the [N~\textsc{II}] BPT diagram), a thorough comparison of the two samples to search for possible biases is necessary before they can be combined. 
Here we perform a uniform comparison of the two samples to understand if any systematic differences between them exist, and if so, how to merge the samples in a way that mitigates those differences for future work. 
In this study, we analyze the MOSDEF and KBSS samples using consistent stellar population synthesis modeling assumptions applied to emission-line-corrected photometry, as well as the same methodology for emission-line fitting. This controlled approach eliminates any systematic biases stemming from different analysis methods, and isolates differences between the samples themselves. 
From the consistent SED fitting, we compare galaxy properties between the two samples. 
From the consistent emission-line measurements, we compare the locations of MOSDEF and KBSS samples on the [N~\textsc{II}] and [S~\textsc{II}] BPT diagrams and quantify the differences in their offsets from the local sequence. 
The combination of these two analyses enables the discovery of any systematic differences between the two samples of galaxies.

In Section \ref{sec:sample_selection}, we present and compare the selection methods for the MOSDEF and KBSS surveys and discuss the $z\sim2$ samples from each survey that are analyzed in this study.
Section \ref{sec:methodology} describes the methodology for the SED and emission-line fitting analyses, and 
Section \ref{sec:spec_sample_analysis} presents the results from these fitting methods for the $z\sim2$ spectroscopic samples with multiple emission-line detections that have been analyzed in previous studies.
Section \ref{sec:discussion} discusses the results for the $z\sim2$ spectroscopic samples with multiple emission-line detections and compares their SED properties with those of the larger sample of $z\sim2$ targeted galaxies from which they are drawn. 
Section \ref{sec:summary} presents a summary of key results and looks ahead to future work. 
Appendix \ref{sec:csf_models} provides SED fitting results for the MOSDEF and KBSS $z\sim2$ samples assuming constant star formation (CSF) histories instead of the delayed-$\tau$ star formation histories described in Section \ref{subsec:sed_fitting}. 
Finally, Appendix \ref{sec:emline_fitting_example} provides examples of spectra to highlight differences between the results of the two emission-line fitting methods discussed in this study. 
We adopt the following abbreviations for emission-line ratios used frequently throughout the paper.
\begin{equation}
\label{eqn:N2_abbreviation}
    \rm{N2 = [N~\textsc{II}]\lambda6585/H\alpha}
\end{equation}
\begin{equation}
\label{eqn:S2_abbreviation}
    \rm{S2 = [S~\textsc{II}]\lambda\lambda6718,6733/H\alpha}
\end{equation}
\begin{equation}
\label{eqn:O3_abbreviation}
    \rm{O3 = [O~\textsc{III}]\lambda5008/H\beta}
\end{equation}
\begin{equation}
\label{eqn:O3N2_abbreviation}
    \rm{O3N2 = O3/N2}
\end{equation}
\begin{equation}
\label{eqn:O3S2_abbreviation}
    \rm{O3S2 = O3/S2}
\end{equation}
All emission-line wavelengths are in vacuum.  Throughout this paper, we adopt a $\Lambda$-CDM cosmology with $H_0$ = 70 km s$^{-1}$ Mpc$^{-1}$, $\Omega_{\rm{m}}$ = 0.3, and $\Omega_\Lambda$ = 0.7.  Also, we assume the solar abundance pattern from \citet{asp09}.

\section{Observations \& Sample Selection} \label{sec:sample_selection}

Here we provide overviews of the MOSDEF (Section \ref{subsec:mosdef_survey}) and KBSS (Section \ref{subsec:kbss_survey}) surveys. 
We describe the sample selection methods of the surveys and the $z\sim2$ spectroscopic samples that we adopt from previous studies (KBSS: \citealt{str17}; MOSDEF: \citealt{san18}). 
In addition, we highlight some of the key differences between the surveys in Section \ref{subsec:mosdef_kbss_sample}. 

Some notable similarities between the two surveys are that they both utilized Keck/MOSFIRE and therefore both obtained moderate spectral resolution ($R =$ 3000-3650) data, both surveys were conducted over roughly the same period (2012-2016), and both contain approximately the same number of galaxies (N$\sim$1500).

\subsection{The MOSDEF Survey} \label{subsec:mosdef_survey}

The MOSDEF survey was a 48.5-night Keck/MOSFIRE observing program spanning multiple years (2012-2016). Approximately 1500 galaxies were targeted across three redshift ranges: $1.37 \leq z \leq 1.70$, $2.09 \leq z \leq 2.61$, and $2.95 \leq z \leq 3.80$. 
These ranges were selected to optimise the detection of strong rest-optical emission-lines (e.g., [O~\textsc{II}]$\lambda\lambda$3727,3730, H$\beta$, [O~\textsc{III}]$\lambda\lambda$4960,5008, H$\alpha$, [N~\textsc{II}]$\lambda$6585, and [S~\textsc{II}]$\lambda\lambda$6718,6733) within windows of atmospheric transmission. 
Roughly half of the galaxies in the survey are in the middle redshift range at $z\sim2.3$. In this redshift range, $\sim$1/3 of the galaxies were targeted based on existing spectroscopic redshifts, while the remaining $\sim$2/3 were targeted based on photometric redshifts drawn from the 3D-HST survey \citep{mom16}. 
The galaxies were selected from five well-studied extragalactic legacy fields covered by the CANDELS and 3D-HST surveys: AEGIS, COSMOS, GOODS-N, GOODS-S, and UDS \citep{gro11, koe11, mom16}.
The survey is $H$-band (F160W) magnitude limited, with brightness limits of $H_{\rm{AB}} =$ 24.0, 24.5, and 25.0, respectively, for the lowest, middle and highest redshift ranges.
Galaxies were observed for 1.5 hours in each filter ($Y$, $J$, and $H$ for a subset) in the lowest redshift bin $z\sim1.5$. The observing time was increased to 2 hours per filter for the middle ($z\sim2.3$; $J$, $H$, and $K_{\rm{s}}$ bands) and highest ($z\sim3.4$; $H$ and $K_{\rm{s}}$ bands) redshift bins.
For a full description of MOSDEF observing details and sample selection, see \citet{kri15}. 

In this study, we use the MOSDEF subsample from \citet{san18}, which contains 260 star-forming galaxies at $1.9 \leq z \leq 2.7$ ($z_{\rm{med}}$ = 2.29) with S/N of both H$\alpha$ and H$\beta$ $\geq$ 3. 
Galaxies with sky lines significantly contaminating H$\alpha$ or H$\beta$ were not included, nor those showing evidence of AGN activity based on their X-ray luminosity, IR colors, or if N2 $\geq$ 0.5 \citep{coi15, aza17, leu17}. 
Galaxies with $\log_{10}(M_{\ast}/M_{\odot})<9.0$ are rejected due to lack of completeness below that mass cutoff (see Section \ref{subsec:sample_incompleteness} for more details). 

The current study, which utilizes updated SED and emission-line fitting methods, finds small differences in stellar mass (i.e., $M_{\ast}$) and emission-line S/N from the values used in \citet{san18}. Nine of the 260 galaxies from \citet{san18} have updated $\log_{10}$($M_{\ast}/M_{\odot}$) $<$ 9.0 values, and are removed from this study. Additionally, one galaxy is not in the 3D-HST v4.1 catalog, which is needed for broadband SED fitting, and is also removed from the sample. Hereafter in this work, we define the resulting sample of 250 galaxies as the ``MOSDEF $z\sim2$ spectroscopic sample.'' 

Figure \ref{fig:mosdef_kbss_survey_properties} (top left panel) features a redshift histogram of all MOSDEF galaxies at $1.9 \leq z \leq 2.7$, which we refer to hereafter as the ``MOSDEF $z\sim2$ targeted sample''. These include galaxies targeted to lie at $2.09 \leq z \leq 2.61$ \citep{kri15} based on existing photometric or spectroscopic redshifts, and galaxies serendipitously detected on MOSDEF slits with redshifts measured in the range $1.9 \leq z \leq 2.7$. Blue indicates the distribution of the 250 galaxies in the $z\sim2$ spectroscopic sample that meet the selection criteria from \citet{san18}. Green indicates the 422 galaxies that have a confirmed MOSFIRE redshift, but do not meet all of the selection criteria outlined above (i.e., S/N, AGN, and/or $M_{\ast}$ cuts). Purple indicates the 114 galaxies that do not have a confirmed MOSFIRE redshift. The photometric redshift from the 3D-HST catalogs is shown for these galaxies. The median redshift for all MOSDEF galaxies at $1.9 \leq z \leq 2.7$ is the same as that for the 250 galaxy spectroscopic sample (i.e., $z_{\rm{med}}$ = 2.29). 

There are 536 galaxies in the MOSDEF $z\sim2$ targeted sample that were not included in the spectroscopic sample described here. Therefore, there are 786 galaxies in the MOSDEF $z\sim2$ targeted sample. Galaxies were removed due to several different criteria, as listed above. There were 114 galaxies lacking a robust MOSFIRE redshift. Of the remaining 422 galaxies, 97 galaxies were flagged as AGNs and removed. Of the 325 non-AGNs, 61 galaxies met all of the criteria to make it into the \citet{san18} sample, but were removed due to sky-line contamination of H$\alpha$ and/or H$\beta$. This removal was performed based on visual inspection of the spectra. In addition, 26 galaxies were removed because they have $\log_{10}$($M_{\ast}/M_{\odot}$) $<$ 9.0; 98 have S/N$_{\rm{H\beta}}$ and S/N$_{\rm{H\alpha}}$ $<$ 3, and are not retained; finally, 140 galaxies were removed because they have S/N$_{\rm{H\beta}}$ $<$ 3 and S/N$_{\rm{H\alpha}}$ $\geq$ 3.
Of the 325 galaxies that have a robust MOSFIRE redshift and are not AGN, 73\% are removed due to S/N$_{\rm{H\beta}}$ $\leq3$.

\subsection{The KBSS Survey} \label{subsec:kbss_survey}

The Keck Baryonic Structure Survey (KBSS) began as a redshift survey of star-forming galaxies in 15 independent fields, subtending a total area of 0.25 square degrees. Each KBSS survey region
is centered on a bright ($V \sim 16-17$) QSO with $z = 2.7\pm0.1$. Galaxies were selected for spectroscopy using photometric pre-selection based on rest-frame far-UV color criteria that had been described and tested in earlier work \citep{ste03, ste04, ade04}. The rest-UV spectroscopy was conducted using LRIS-B \citep{ste04} over the years 2002-2009, deliberately focusing on the redshift range $2 \lesssim z \lesssim 2.7$ to maximize the sensitivity of the background QSO sightline to the IGM and CGM within the survey volume (e.g., \citealt{rud12, rud19, tur14}). Most of the spectroscopically-observed galaxies satisfied the ``BX'' or ``MD'' color selection criteria, yielding galaxy redshifts $\langle z \rangle = 2.17\pm 0.33$ and $\langle z \rangle = 2.63\pm 0.28$, respectively. In addition to the UV color criteria, the KBSS UV spectroscopic sample was limited to galaxies with apparent magnitude ${\cal R_{\rm AB}}$[6830 \AA]$\le 25.5$ to ensure that the resulting spectra would yield redshifts for galaxies both with and without observed Ly$\alpha$ emission. The full KBSS UV spectroscopic sample consists of 2341 objects (2285 star-forming galaxies and 66 AGN) with $\langle z \rangle = 2.354\pm0.432$. Of these, 1435 (61.3\%) have $1.9 \le z \le 2.7$, and 950 (40.6\%) have $2.1 \le z \le 2.6$. The latter sub-sample comprised the 
highest priority targets for near-IR spectroscopy, for the reasons outlined above. 

The KBSS-MOSFIRE survey \citep{ste14, str17} was initiated soon after MOSFIRE was commissioned in 2012 April. The survey had multiple goals, including obtaining redshifts for galaxies that had failed to yield redshifts from previous LRIS spectroscopy, especially those within a few arc minutes in projection from the central QSO and whose rest-UV colors suggested redshifts in the broader range $1.8 \lesssim z \lesssim 3.5$. Additional photometric selection criteria were imposed to ensure that the targeted sample at $z\sim2$ would roughly uniformly span the full range of star-forming galaxy properties: first, we used the galaxy ${\cal R} - K_{\rm s}$ color ($\sim 2000$ \AA\ $-$ 6500 \AA\ in the rest-frame) for galaxies with $z\sim2.3$, with preference given to those with ${\cal R}-K_{\rm s} > 2$. Also added as potential MOSFIRE targets were galaxies satisfying ${\cal R}-K_{\rm s} > 2$ that were drawn from a region in $U_{\rm n}G{\cal R}$ color space adjacent to the BX and MD region, in an effort to include relatively massive galaxies likely to be more heavily reddened by dust; these  were given the name ``RK'' galaxies (see \citealt{str17}). None of the RK objects had UV-based redshifts prior to their observation with MOSFIRE. 

Unlike MOSDEF, KBSS-MOSFIRE attempted to obtain the complete set of strong nebular lines only for galaxies in the range $z=2-2.7$. However, MOSFIRE slitmasks also included galaxies without previous spectroscopic redshifts, either because they had never been observed with LRIS, or when previous LRIS observations did not result in a secure redshift. For objects without prior redshifts measurements, once a redshift was measured from initial MOSFIRE spectroscopy, the object was retained as a high priority for subsequent MOSFIRE observations only if it fell in the range $2 \le z \le 2.7$. 

MOSFIRE mask design was done iteratively as the survey progressed. Initially, the highest priority targets were objects with known redshifts $2.1 \le z \le 2.6$ drawn from the KBSS rest-UV spectroscopic survey. Any remaining space on each mask was assigned to galaxies in the following order of priority:
\begin{enumerate}
\item Galaxies without previous redshifts, drawn from the BX, MD, and RK color selection; BX and MD candidates without known redshifts were favored if they also had ${\cal R}-K_{\rm s} > 2$.  
\item Galaxies with UV-based redshifts falling outside the primary $2 \le z \le 2.7$ range. 
\item Galaxies without UV-based redshifts drawn from color selection windows least likely to fall in the primary redshift range of interest (i.e., ``BM'', ``M'', ``C'' candidates).   
\end{enumerate}

\begin{figure*}
    \includegraphics[width=0.49\linewidth]{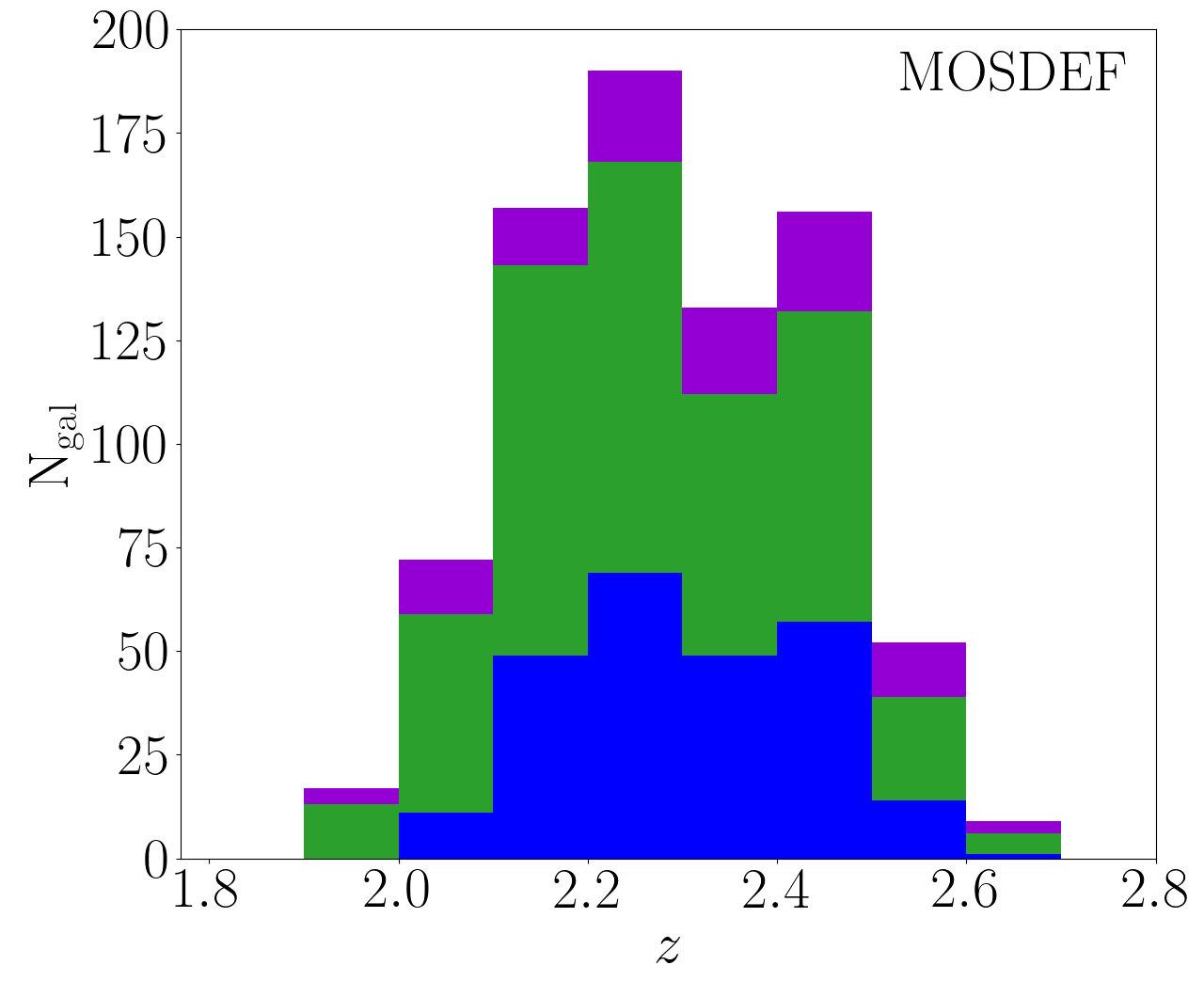}
    \includegraphics[width=0.49\linewidth]{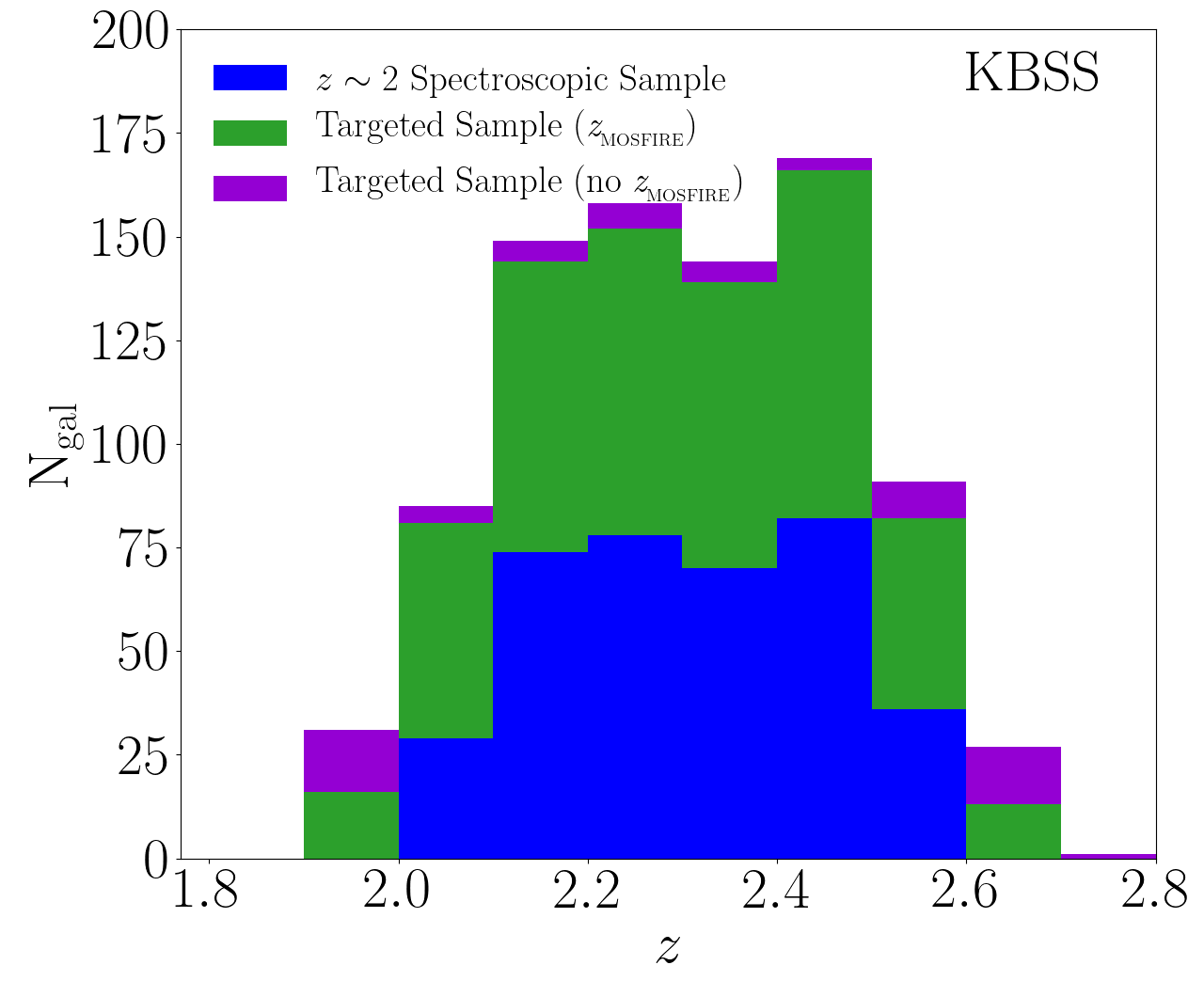}
    \includegraphics[width=0.49\linewidth]{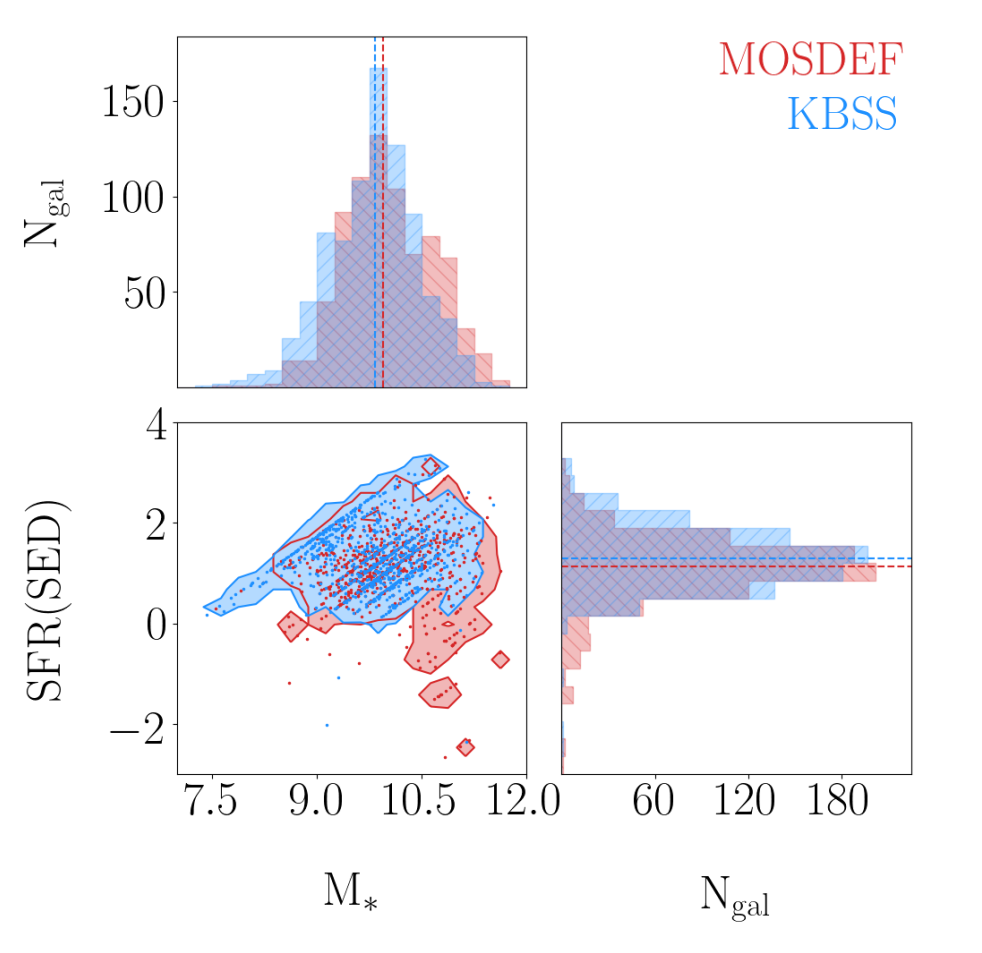}
    \includegraphics[width=0.49\linewidth]{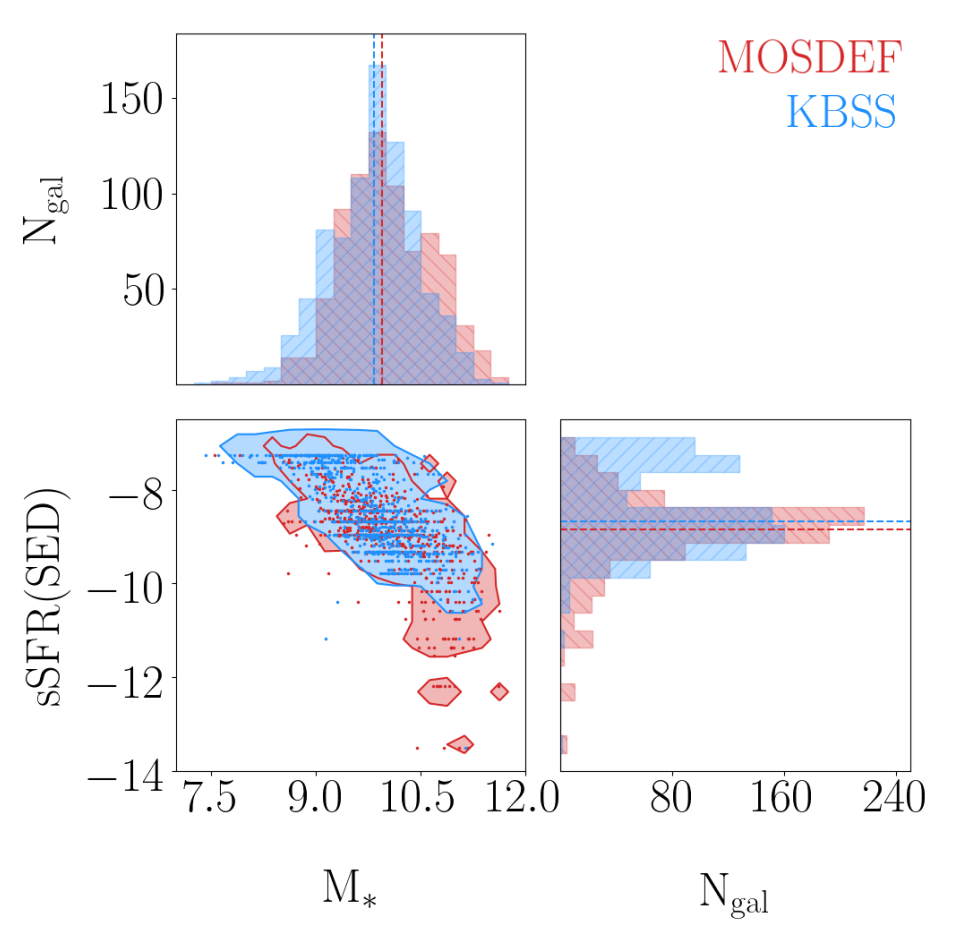}
    \caption{Top: Redshift histograms. Shown in these panels are distributions of galaxies at $1.9 \leq z \leq 2.7$ for the MOSDEF (left) and KBSS (right) surveys. Blue indicates the $z\sim2$ spectrosopic samples discussed in Sections \ref{subsec:mosdef_survey} (MOSDEF) and \ref{subsec:kbss_survey} (KBSS). Green indicates galaxies that were observed and have a MOSFIRE redshift, but were not included in the $z\sim2$ spectrosopic samples due to the various sample selection criteria discussed in Sections \ref{subsec:mosdef_survey} and \ref{subsec:kbss_survey}. Purple indicates galaxies that were targeted, but a robust MOSFIRE redshift was not obtained. For MOSDEF, the purple redshifts displayed are photometric redshifts. For KBSS, the purple redshifts displayed were obtained by previous rest-UV observations with Keck/LRIS. The LRIS redshifts were estimated from either Ly$\alpha$ or low-ionization interstellar absorption lines. 
    Bottom: Corner plots comparing SFR(SED) vs. $M_{\ast}$ (left) and sSFR(SED) vs. $M_{\ast}$ (right) for the MOSDEF (red) and KBSS (blue) $z\sim2$ targeted samples. The dashed lines on the 1D histograms mark the median values for the distributions, and the contours on the diagonal trace the 3$\sigma$ regions in 2D space. Individual data points are also shown in the 2D distribution. The range of datapoints is limited by the fact that $M_{\ast}$ and SFR(SED) are correlated and are both determined by the normalization of the SPS model to the photometry. In addition, for the 2D panels the data crowd in the upper-left portion of the distribution (left panel) and at high sSFR(SED) (right panel) because of the lower age limit for the stellar populations. In total, there are 786 MOSDEF and 850 KBSS galaxies. Note that two MOSDEF galaxies with log$_{10}$(SFR(SED)/M$_{\odot}$/yr$^{-1}$) $<$ $-$3.0 are excluded from the left panel and one MOSDEF galaxy with log$_{10}$(sSFR(SED)/yr$^{-1}$) $<$ $-$14.0 is excluded from the right panel to better show the data.}
    \label{fig:mosdef_kbss_survey_properties}
\end{figure*}

The status of all objects in a given survey field was evaluated after each MOSFIRE observing run, and objects were re-prioritized based on the S/N achieved for each emission line of interest; for $H$- and $K_{\rm{s}}$-band observations these included H$\beta$, [O~\textsc{III}]$\lambda\lambda$4960,5008 in $H$~band, H$\alpha$ and [N~\textsc{II}]$\lambda\lambda$6550,6585 in $K_{\rm{s}}$~band. High priority targets (based on redshift) were retained for inclusion on new masks until all relevant lines were successfully detected.  

For the purposes of comparison with MOSDEF, we consider the KBSS-MOSFIRE $z\sim 2$ targeted sample to include all MOSFIRE-observed galaxies whose redshifts were known from prior rest-UV spectroscopy to be in the range $1.9 \le z \le 2.7$ (621 objects in total, of which 558 yielded MOSIFRE redshifts and 63 did not), plus the sample of objects that had not been identified prior to MOSFIRE observation that fall in the primary $1.9 \le z \le 2.7$ redshift range (237 objects). The $z\sim2$ targeted sample thus defined has 858 objects, of which 795 yielded MOSFIRE redshifts\footnote{The $z\sim2$ targeted KBSS-MOSFIRE sample does not include objects without UV redshifts for which no significant emission lines were detected in the MOSFIRE spectra. Many are targets observed in only 1 MOSFIRE atmospheric band as filler, and would be unlikely to be assigned to subsequent masks if initial spectroscopy showed no sign of emission in the desired range.}.
We do not have the broadband photometry required for SED fitting for 8/858 galaxies (seven have a MOSFIRE redshift while one does not). Therefore, the final KBSS $z\sim 2$ targeted sample to be used in this analysis consists of 850 galaxies, 788 of which have a MOSFIRE redshift. In comparison, the MOSDEF $z\sim 2$ targeted sample contains 786 galaxies, of which 672 have a MOSFIRE redshift.

KBSS did not dedicate a specific amount of time to observing each galaxy, instead choosing to stay in a given field until all rest-optical emission-lines were detected in the majority of galaxies in that field. 
The median observing time for a galaxy in the $H$-band (covering H$\beta$ and [O~\textsc{III}]$\lambda\lambda$4960,5008) was 2.44 hours. The maximum time spent on a galaxy was 10.64 hours. For the $K_{\rm s}$-band (covering H$\alpha$, [N~\textsc{II}]$\lambda$6585, and [S~\textsc{II}]$\lambda\lambda$6718,6733), the median and maximum observing times were, respectively, 2.78 and 12.13 hours.  
For full KBSS-MOSFIRE observing details, see \citet{ste14} and \citet{str17}.

In this study, we use the KBSS-MOSFIRE sample from \citet{str17}, which contains 377 galaxies at $1.9 < z < 2.7$ ($z_{\rm{med}}$ = 2.30) with spectral coverage of the H$\beta$, [O~\textsc{III}]$\lambda$5008, H$\alpha$, and [N~\textsc{II}]$\lambda$6585 emission lines (i.e., the four required for the [N~\textsc{II}] BPT diagram). To be included in this sample, the S/N of H$\alpha$ must be $\geq$ 5, the S/N of H$\beta$ and [O~\textsc{III}]$\lambda$5008 must be $\geq$ 3, and [N~\textsc{II}]$\lambda$6585 must simply fall within the wavelength coverage.

Galaxies are identified as having an AGN based on X-ray luminosity, mid-IR photometry, strong UV emission in high-ionization lines (e.g., C~\textsc{IV}$\lambda$1549, C~\textsc{III}]$\lambda$1908, and N~\textsc{V}$\lambda$1240; \citealt{red08, hai11}), broad emission-line features in the rest-optical spectra, and/or N2 $\geq$ 0.5. 

To be consistent with the MOSDEF sample, we remove eight KBSS galaxies from the \citet{str17} sample flagged in the sample as AGN. 
Hereafter in this work, we define the resulting sample of 369 galaxies as the ``KBSS $z\sim2$ spectroscopic sample.''

Figure \ref{fig:mosdef_kbss_survey_properties} (top right panel) includes a redshift histogram of all KBSS galaxies at $1.9 \leq z \leq 2.7$, which we refer to hereafter as the ``KBSS $z\sim2$ targeted sample''. Blue indicates the distribution of the 369 galaxies in the $z\sim2$ spectroscopic sample that meet the selection criteria from \citet{str17}. 
Green indicates galaxies that have a confirmed MOSFIRE redshift, but do not meet all of the selection criteria (i.e., emission-line S/N and broadband photometry criteria) outlined above. 
Purple indicates galaxies that do not have a confirmed MOSFIRE redshift. For these galaxies, the redshifts shown are based on Keck/LRIS measurements of either the Ly$\alpha$ or low-ionization interstellar absorption features. The median redshift for all KBSS galaxies at $1.9 \leq z \leq 2.7$ (i.e., $z_{\rm{med}}$ = 2.29) is slightly lower than for the 369 galaxy spectroscopic sample (2.31). Out of these 369 galaxies, 357 (96.7\%) are  $U_{\rm{n}}G\mathcal{R}$ color selected and 12 (3.3\%) are ``RK'' galaxies.

\subsection{MOSDEF \& KBSS Sample Comparison} \label{subsec:mosdef_kbss_sample}

The selection methods for the MOSDEF and KBSS surveys are distinct. MOSDEF is selected based on photometric redshifts (or spectroscopic redshifts when available), with a magnitude limit in the rest-optical ($H$~band), while KBSS-MOSFIRE is a UV-color-selected sample with a magnitude limit in the $\mathcal{R}$~band. As discussed in \citet{shi15}, the MOSDEF magnitude-limited selection method is incomplete in the low-mass regime. At the same time, the KBSS rest-UV sample can be biased against high-mass or dusty galaxies whose $U_{\rm{n}}G\mathcal{R}$ colors place them outside of the nominal rest-UV selection windows (e.g., \citealt{red05, str17}). This difference in the $M_{\ast}$ regimes probed by the two samples can be seen in the bottom panels of Figure \ref{fig:mosdef_kbss_survey_properties}. This figure contains corner plots showing the 1D histograms and 2D distributions of SFR(SED) vs. $M_{\ast}$ (bottom left) and sSFR(SED) vs. $M_{\ast}$ (bottom right) for both samples. The sample medians are marked on the 1D histograms and the 3$\sigma$ regions are traced by contours in the 2D space. It is clear that KBSS more heavily targets the low mass regime while MOSDEF targets span a wider range of stellar populations in the high mass regime. These features of the two samples suggest that they are complementary to one another in the low- and high-mass extremes, and that the combination of the MOSDEF and KBSS $z\sim2$ surveys would be complete across a wide $M_{\ast}$ range of $\sim10^{8.5}-10^{11.5}$ M$_{\odot}$. 
Additionally, the MOSDEF and KBSS samples are complementary in sSFR(SED) space, as KBSS is better sampled at $\log_{10}$(sSFR(SED)/yr$^{-1}$) $\geq$ $-8.0$ (indicating more intense star formation), especially at the low-mass end, while MOSDEF has more coverage at $\log_{10}$(sSFR(SED)/yr$^{-1}$) $\leq$ $-9.0$ (indicating less intense star formation), especially at the high-mass end. 
Between the low- and high-mass regimes (i.e., $\sim10^{9.5}-10^{10.5} M_{\odot}$) the contour maps overlap for the two samples.
Our methods for measuring these galaxy stellar population properties are presented in Section \ref{subsec:sed_fitting} and a detailed comparisons of these properties for the two samples are discussed in Sections \ref{subsec:spec_sample_galaxy_properties} and \ref{subsec:targeted_sample_sed_properties}.

Despite the differences in selection of the MOSDEF and KBSS $z\sim2$ targeted catalogs, there are notable similarities between the two surveys. The surveys have comparable $z\sim2$ catalog sizes (786 galaxies for MOSDEF and 850 for KBSS) and median redshifts (2.29 for both surveys). Additionally, the majority of galaxies in each survey would meet the selection criteria of the other survey. 
Most of the KBSS $z\sim2$ targeted sample (817/850 galaxies or 96.1\%) has $H_{\rm{AB}} <$ 24.5 (i.e., the main MOSDEF selection criterion). 
Table \ref{tab:ugr_space} shows the percent of the MOSDEF sample that falls into each region of the $U_{\rm{n}}G\mathcal{R}$ diagram as defined in \citet{ste03, ste04, str17}. 
For MOSDEF galaxies, we passed best-fit stellar population model SEDs (see Section \ref{subsec:sed_fitting}) through the $U_{\rm{n}}G\mathcal{R}$ filters to estimate synthetic magnitudes using the IRAF \citep{tod86, tod93} routine, {\it sbands}.
The statistics for the MOSDEF sample are compared to those of the KBSS sample. 
The distribution of MOSDEF and KBSS galaxies in the various regions on the $U_{\rm{n}}G\mathcal{R}$ diagram is very similar. Most notable is that MOSDEF has a similar percentage of BX galaxies (82.8\%) as KBSS (80.5\%). The BX region on the $U_{\rm{n}}G\mathcal{R}$ diagram is statistically where $z\sim2$ galaxies are most likely to be \citep{ste04}. Also, 773/786 galaxies (98.3\%) in the MOSDEF $z\sim2$ targeted sample have $\mathcal{R}<25.5$ and fall into one of the defined regions of $U_{\rm{n}}G\mathcal{R}$ color space. Of the remaining 13 galaxies in the ``Other'' category, 12 have $\mathcal{R}>25.5$ while the other one has $\mathcal{R}<25.5$ but did not fall into any of the defined regions on the $U_{\rm{n}}G\mathcal{R}$ diagram. Finally, as we detail in the next section, the spectroscopic incompleteness of the MOSDEF survey leads to an even greater similarity between the distributions of galaxy properties for the MOSDEF and KBSS $z\sim2$ spectroscopic samples that have been analyzed in, e.g., \citet{san18} and \citet{str17}.

\begin{table}
    \centering
    \begin{tabular}{rrr}
        \multicolumn{3}{c}{$U_{\rm{n}}G\mathcal{R}$ Diagram Statistics} \\
        \hline\hline
        $U_{\rm{n}}G\mathcal{R}$ Region & N/850 of KBSS & N/786 of MOSDEF \\
        (1) & (2) & (3) \\
        \hline
 BX & 684 (80.5\%) & 651 (82.8\%) \\
 BM & 10 (1.2\%) & 4 (0.5\%) \\
 M/MD & 97 (11.4\%) & 75 (9.5\%) \\
 C/D & 23 (2.7\%) & 0 \\
 RK & 28 (3.3\%) & 43 (5.5\%) \\
 Other & 8 (0.9\%) & 13 (1.7\%) \\
        \hline
    \end{tabular}
    \caption{
  Col. (1): Region of $U_{\rm{n}}G\mathcal{R}$ space, as defined in \citet{ste03, ste04, str17}.
  Col. (2): Percent of the KBSS sample in that region of $U_{\rm{n}}G\mathcal{R}$ space.
  Col. (3): Percent of the MOSDEF sample in that region of $U_{\rm{n}}G\mathcal{R}$ space.
  Note that to fall into one of the regions on the $U_{\rm{n}}G\mathcal{R}$ diagram, the galaxy must have $\mathcal{R}<25.5$. Some of the galaxies in the ``Other'' category fall into one of the defined $U_{\rm{n}}G\mathcal{R}$ regions, but have $\mathcal{R}>25.5$.}
    \label{tab:ugr_space}
\end{table}

\subsection{MOSDEF Sample Incompleteness} \label{subsec:sample_incompleteness}

For MOSDEF, it has been known since the design of the survey that both the number of targets and the success rate drop below $\log_{10}(M_{\ast}/M_{\odot})$ = 9.0 due to the $H$-band magnitude selection limits \citep{kri15}. Accordingly, \citet{san18}, who analyze a galaxy sample that our current MOSDEF sample is intended to represent, removed any galaxies below a mass limit of
$\log_{10}(M_{\ast}/M_{\odot})$ = 9.0.
Our new SED fitting methodology utilizing emission-line subtracted photometry resulted in slightly different stellar masses from what was used in \citet{san18}, with a small fraction (3.5\%) now below that mass limit. As stated above, we remove these galaxies to be consistent with the well documented MOSDEF incompleteness in this mass regime.  
KBSS does not have this incompleteness at low mass. Since the goal of this study is to compare MOSDEF and KBSS samples that have been analyzed in previous work \citep{str17, san18}, we do not remove the galaxies with $\log_{10}(M_{\ast}/M_{\odot})$ $<$ 9.0 from the KBSS $z\sim2$ spectroscopic sample.

We note that the MOSDEF survey is also incomplete in the red, massive regime. The success rate of obtaining a $z_{\rm{MOSFIRE}}$ declines for galaxies with red rest-frame optical/near-IR colors, where rest-frame U$-$V and V$-$J are both greater than $\sim$1.25 (we discuss the UVJ diagram and how U$-$V and V$-$J colors are estimated in Section \ref{subsec:sed_fitting}). Due to the rest-UV sample selection methods (including the $\mathcal{R}\leq 25.5$ cut off), the KBSS survey does not target galaxies in this regime. Most of the RK galaxies are not in this regime as well. 
The spectroscopic incompleteness of the MOSDEF survey makes the MOSDEF and KBSS $z\sim2$ spectroscopic samples more similar than the MOSDEF and KBSS $z\sim2$ targeted samples are. We explore this topic more in Section~\ref{subsec:targeted_sample_sed_properties}.

\subsection{SDSS Comparison Sample} \label{subsec:sdss_sample}

When discussing the [N~\textsc{II}] and [S~\textsc{II}] BPT diagrams, we compare the emission lines from our high-redshift MOSDEF and KBSS samples with similar measurements from galaxies in the local universe. We pull archival emission-line measurements from the SDSS Data Release 7 (DR7; \citealt{aba09}), specifically from the MPA-JHU DR7 release of spectrum measurements\footnote{https://wwwmpa.mpa-garching.mpg.de/SDSS/DR7/}. 
The SDSS sample is restricted to a redshift range of $0.04 \leq z \leq 0.10$, and AGN are removed using equation 1 from \citet{kau03} or if $\mbox{N2} > 0.5$. On the [N~\textsc{II}] ([S~\textsc{II}]) BPT diagrams, a S/N $\geq$ 3 is required for the H$\beta$, [O~\textsc{III}]$\lambda$5008, H$\alpha$, and [N~\textsc{II}]$\lambda$6585 ([S~\textsc{II}]$\lambda\lambda$6718,6733) features. These criteria result in comparison samples of 103,422 and 102,543 SDSS galaxies, respectively, on the [N~\textsc{II}] and [S~\textsc{II}] BPT diagrams.

\section{Methodology} \label{sec:methodology}

For this study, it is essential to analyze the MOSDEF and KBSS samples using the same methodology. Doing so eliminates any bias introduced from the analysis process, and can therefore isolate any systematic intrinsic differences between the two samples. Section \ref{subsec:sed_fitting} introduces how the host galaxy properties are obtained using SED fitting, and Section \ref{subsec:emline_fitting} describes the emission-line measurements.

\subsection{SED Fitting} \label{subsec:sed_fitting}

Both MOSDEF and KBSS samples are covered by multi-wavelength photometry enabling robust SED fitting; however, the specific sources of the photometry vary between surveys. 
For MOSDEF, the specific set of ground-based and {\it HST} optical and near-IR photometric bands varies from field to field (AEGIS, COSMOS, GOODS-N, GOODS-S, and UDS), though all have {\it Spitzer}/IRAC coverage \citep{ske14}. 
All KBSS galaxies have $U_{\rm{n}}$, $G$, $\mathcal{R}$, $J$, and $K_{\rm{s}}$ imaging. Additionally, all 15 fields have \textit{Spitzer}/IRAC coverage (typically two channels per field; \citealt{red12}), 10 fields have at least one pointing from \textit{Hubble}/WFC3-IR F160W \citep{law12, mos15}, and 8 fields have $J$1, $J$2, $J$3, $H$1, $H$2 imaging using \textit{Magellan}-FourStar \citep{per13}.
While both MOSDEF and KBSS have broadband photometry spanning a similar wavelength range (rest-UV to rest-IR), we note that MOSDEF has a denser coverage with 15 photometry points on average compared to 7 for KBSS.

In this study, we apply the same SED fitting method to both the MOSDEF and KBSS samples. We use the SED fitting code FAST \citep{kri09} to obtain many of the key galaxy properties compared between the two galaxy samples. For this modeling, we adopt the Flexible Stellar Population Synthesis (FSPS) library from \citet{con10}, assume a \citet{cha03} stellar initial mass function (IMF), a \citet{cal00} dust attenuation curve, and delayed-$\tau$ star-formation histories where SFR(SED) $\propto$ $t \times e^{-t/\tau}$. Here, $t$ is the time since the onset of star formation, and $\tau$ is the characteristic star-formation timescale. Both $t$ and $\tau$ are allowed to vary in the FAST modeling. In addition to $t$ and $\tau$, we allow the amount of dust attenuation ($A_{\rm{V}}$) and stellar mass of the galaxy ($M_{\ast}$) to vary and set the metallicity to 0.019, which is defined to be solar metallicity in the \citet{con10} library.

We use emission-line subtracted photometry, where the flux of the strong rest-optical emission lines measured directly from the MOSFIRE spectra (e.g. H$\beta$, [O~\textsc{III}]$\lambda\lambda$4960,5008, and H$\alpha$) has been subtracted from the photometric measurements. These strong lines could bias the shape of the SED fit on the red side of the Balmer break, resulting in the fitting code favoring older ages than it would based solely on the stellar continuum.

From the SED fitting, we compare estimates of $M_{\ast}$, SFR(SED), sSFR(SED), $A_{\rm{V}}$, $t/\tau$ for the MOSDEF and KBSS samples. All SFR values are those inferred from the SED fits. These values are listed in Section \ref{subsec:spec_sample_galaxy_properties} for the $z\sim2$ spectroscopic samples and Section \ref{subsec:targeted_sample_sed_properties} for the $z\sim2$ targeted samples.

Using the best-fit SEDs obtained from FAST, we also estimate the rest-frame UVJ colors of the MOSDEF and KBSS samples using the IRAF \citep{tod86, tod93} function {\it sbands}. 
The UVJ diagram is a useful tool because the combination of the U$-$V and V$-$J colors breaks the degeneracy for age and reddening (i.e., distinguishing between red quiescent galaxies and dusty star-forming galaxies). Additionally, at red (i.e., high) U$-$V values, increasing V$-$J traces increasing sSFR \citep[][]{wil09}. 
In the star-forming track of the UVJ diagram extending from low U$-$V and V$-$J up towards higher U$-$V and V$-$J, low mass galaxies populate the bottom left region (i.e., blue U$-$V and V$-$J colors) while high mass star-forming galaxies occupy the upper right region (i.e., red U$-$V and V$-$J colors). Quiescent galaxies lie in the upper left portion of the diagram \citep[][]{wil09}.

We note that we experimented with CSF models similar to those utilized in past KBSS studies (e.g., \citealt{ste14,ste16,str17}) and some MOSDEF studies (e.g., \citealt{du18, red18}). Appendix \ref{sec:csf_models} provides a description of both the CSF models and the results for the $z\sim2$ spectroscopic and targeted samples. 
The results from the FAST delayed-$\tau$ and CSF models do not perfectly agree with each other; however, the differences are minor (particularly for the spectroscopic samples) and the key takeaways of this study are the same regardless of which SED fitting method we use.

We also experiment with models where star formation increases with time (hereafter referred to as $\tau$-rising models), as this star formation history can provide a better fit to $z\sim2-3$ galaxies assuming constant star formation \citep{red12}. 
Similar to the CSF models, the differences in results between the delayed-$\tau$ and $\tau$-rising models are minor, and the key takeaways of this study would remain unchanged if we used the results from the latter.

\subsection{Emission-line Measurements} \label{subsec:emline_fitting}

We performed measurements for the MOSDEF and KBSS spectra using a custom emission-line fitting IDL code developed by the MOSDEF team that fits a single Gaussian to each emission-line in the spectra \citep{red15}. The FWHM estimates are allowed to vary independently for each emission-line, but with the following restrictions. The lower limit for the FWHM is the instrumental resolution, which is determined by sky lines. The upper limit is 0.5 \AA\space larger than the FWHM measured for the line with the highest S/N (usually H$\alpha$). 
The best-fit FAST models obtained through SED fitting were used to correct H$\alpha$ and H$\beta$ for the underlying contribution of stellar Balmer absorption. Specifically, the continuum under each line was fixed to the best-fit model from FAST. If the best-fit Gaussian model deviates significantly from the data, the emission-line fitting code will pick the integrated bandpass flux as the preferred estimate of the emission-line flux. 
This custom emission-line fitting code has been used for past MOSDEF studies (e.g., \citealt{sha15, sha19, san16, san18}); however, it has been since notably updated for more accurate estimates of the contribution of stellar Balmer absorption. 
For the MOSDEF $z\sim2$ spectroscopic sample, the median decrease in Balmer absorption correction is 36.4\%, when compared with the previous iteration of the MOSDEF code, which results in a 4.1\% median decrease in H$\beta$ flux.

The current emission-line fitting code used in this study is different from that used in earlier published work for both MOSDEF and KBSS, which results in small variations in the emission-line S/N compared to past studies. 
For MOSDEF, updates to the estimates of the stellar Balmer absorption correction result in a H$\beta$ S/N $<$ 3 for three galaxies that previously had a higher S/N and were thus included in the \citet{san18} sample. For KBSS, the utilization of a different code compared to the one used in \citet{str17} results in 24 galaxies with H$\beta$ S/N $<$ 3, nine galaxies with [O~\textsc{III}]$\lambda$5008 S/N $<$ 3, and two additional galaxies with H$\beta$ and [O~\textsc{III}]$\lambda$5008 S/N $<$ 3. Although their S/N falls below the original sample selection criteria, we still include these galaxies in our study and investigate their SED properties. However, we include only the subset of galaxies with S/N $\geq$ 3 for all relevant emission-lines in our analysis of the [N~\textsc{II}] and [S~\textsc{II}] BPT diagrams.

We note that one difference between the MOSDEF and KBSS data reduction approaches lies in the 2D to 1D extraction methods. MOSDEF uses an ``optimal'' \citep{fre19} extraction while the \citet{str17} catalogs uses a boxcar extraction. We tested how this difference affects the measured emission-line flux by extracting MOSDEF spectra using the boxcar method and refitting the emission lines. For the four emission lines on the [N~\textsc{II}] BPT diagram, we find that the boxcar method has a higher median flux of 1-3\% compared to the optimal extraction method. For the line ratios (i.e., [O~\textsc{III}]$\lambda$5008/H$\beta$ and [N~\textsc{II}]$\lambda$6585/H$\alpha$), the difference is less than 1\%. Since the difference is negligible, we continue to use the optimal extraction for MOSDEF and boxcar extraction for KBSS, as these are the methods used for the previously published survey results.

\begin{figure*}
    \centering
     \subfloat[]{
       \includegraphics[width=0.32\linewidth]{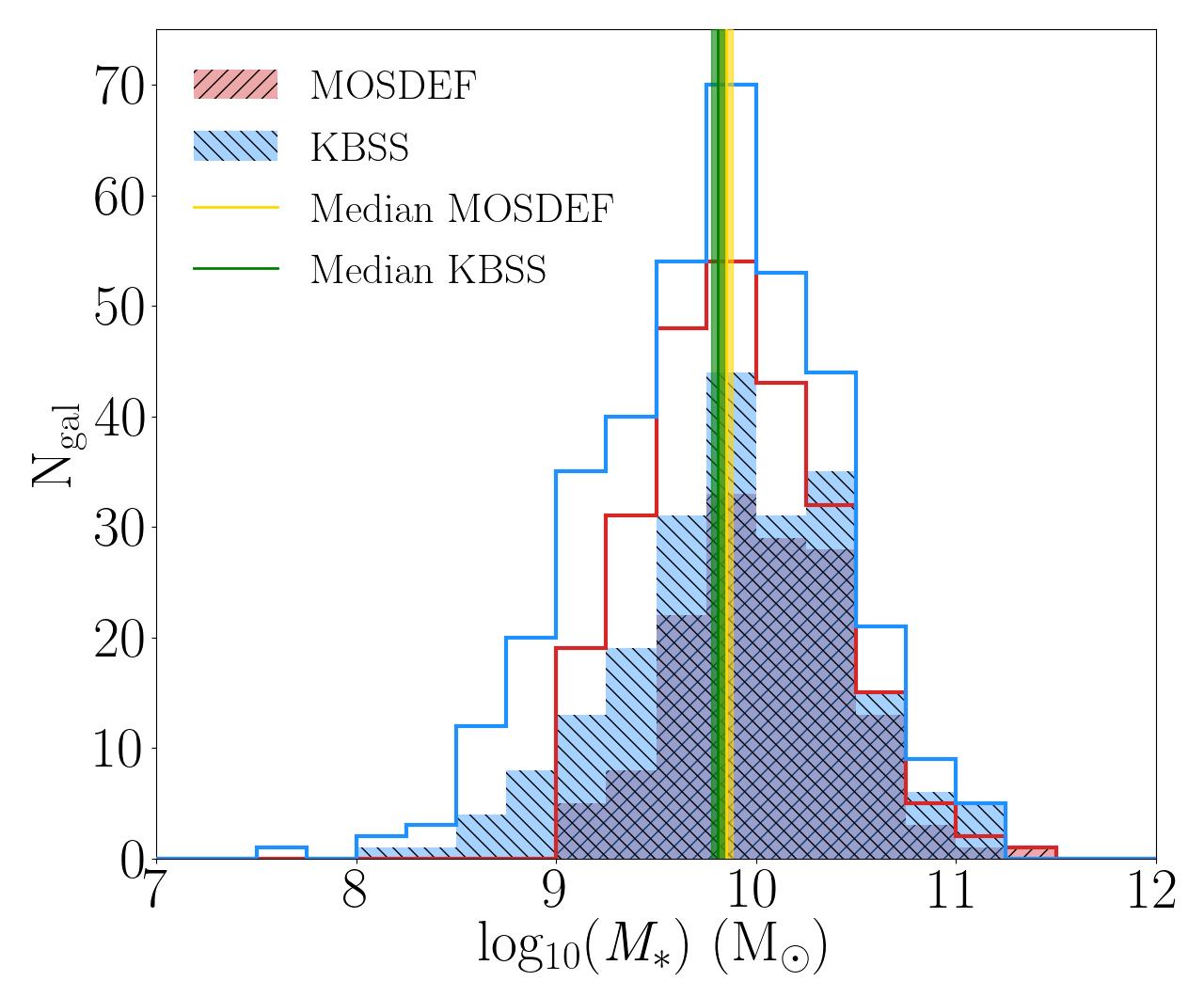}
     }
     \hfill
     \subfloat[]{
       \includegraphics[width=0.32\linewidth]{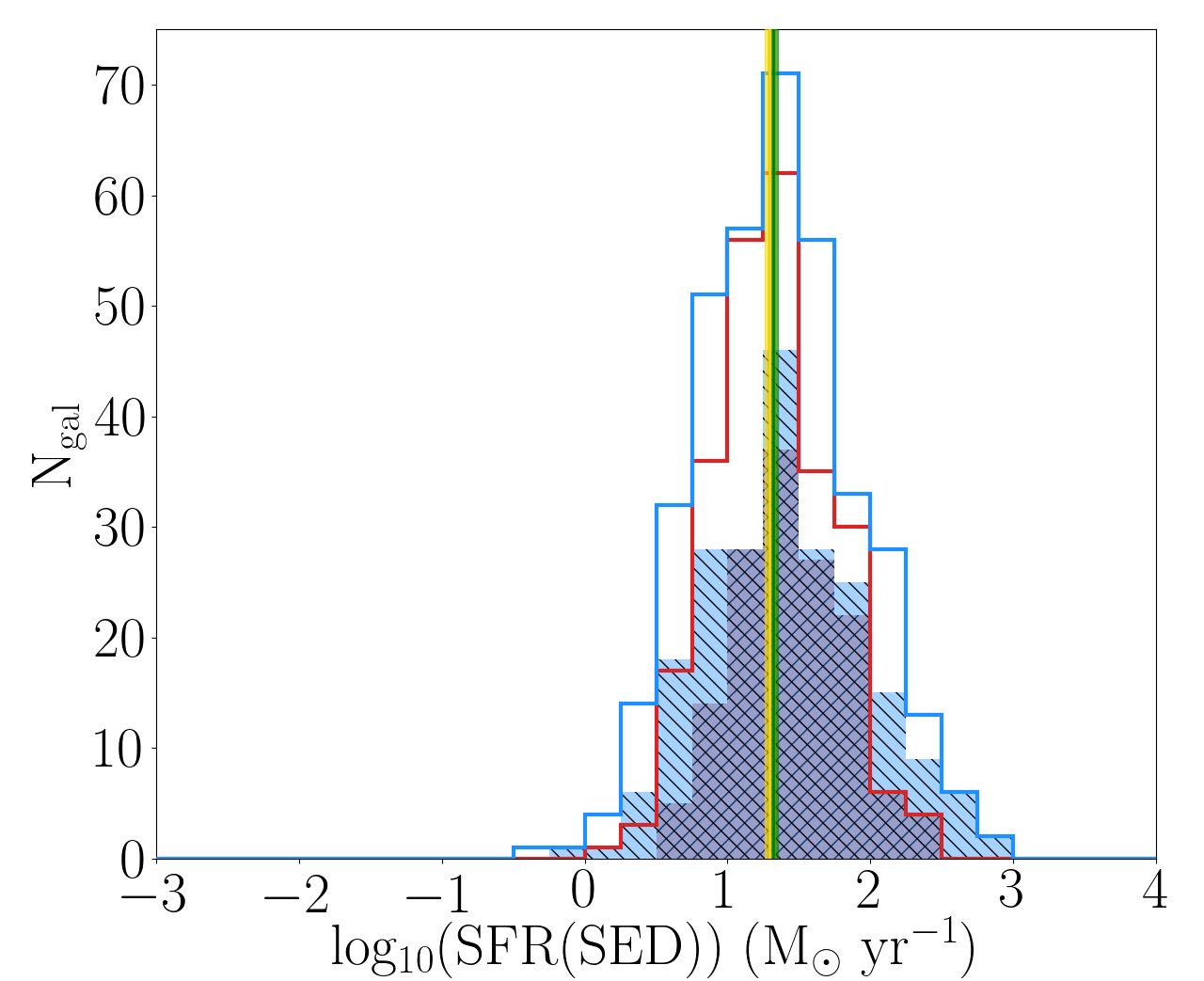}
     }
     \hfill
     \subfloat[]{
       \includegraphics[width=0.32\linewidth]{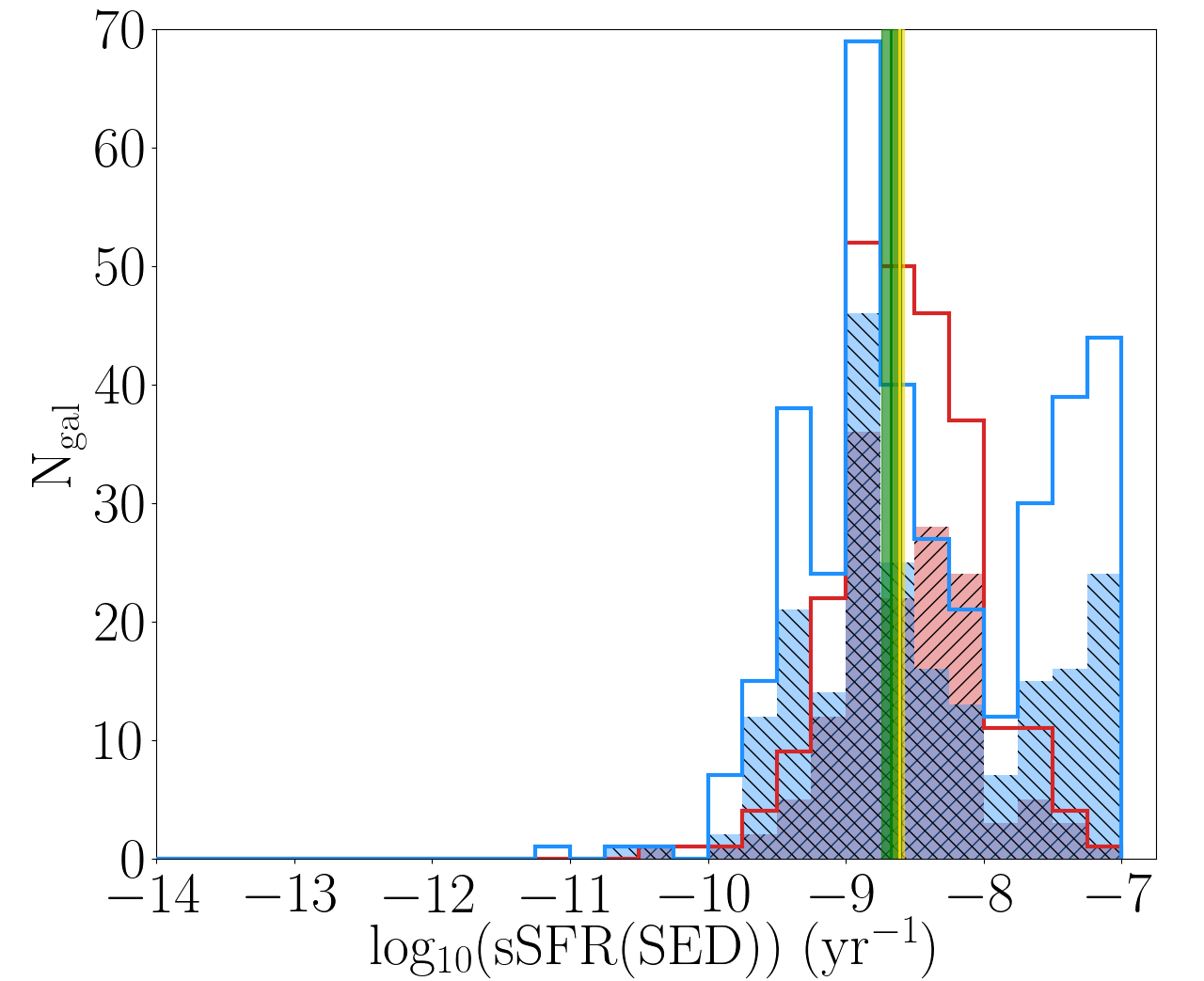}
     }
     \hfill
     \subfloat[]{
       \includegraphics[width=0.32\linewidth]{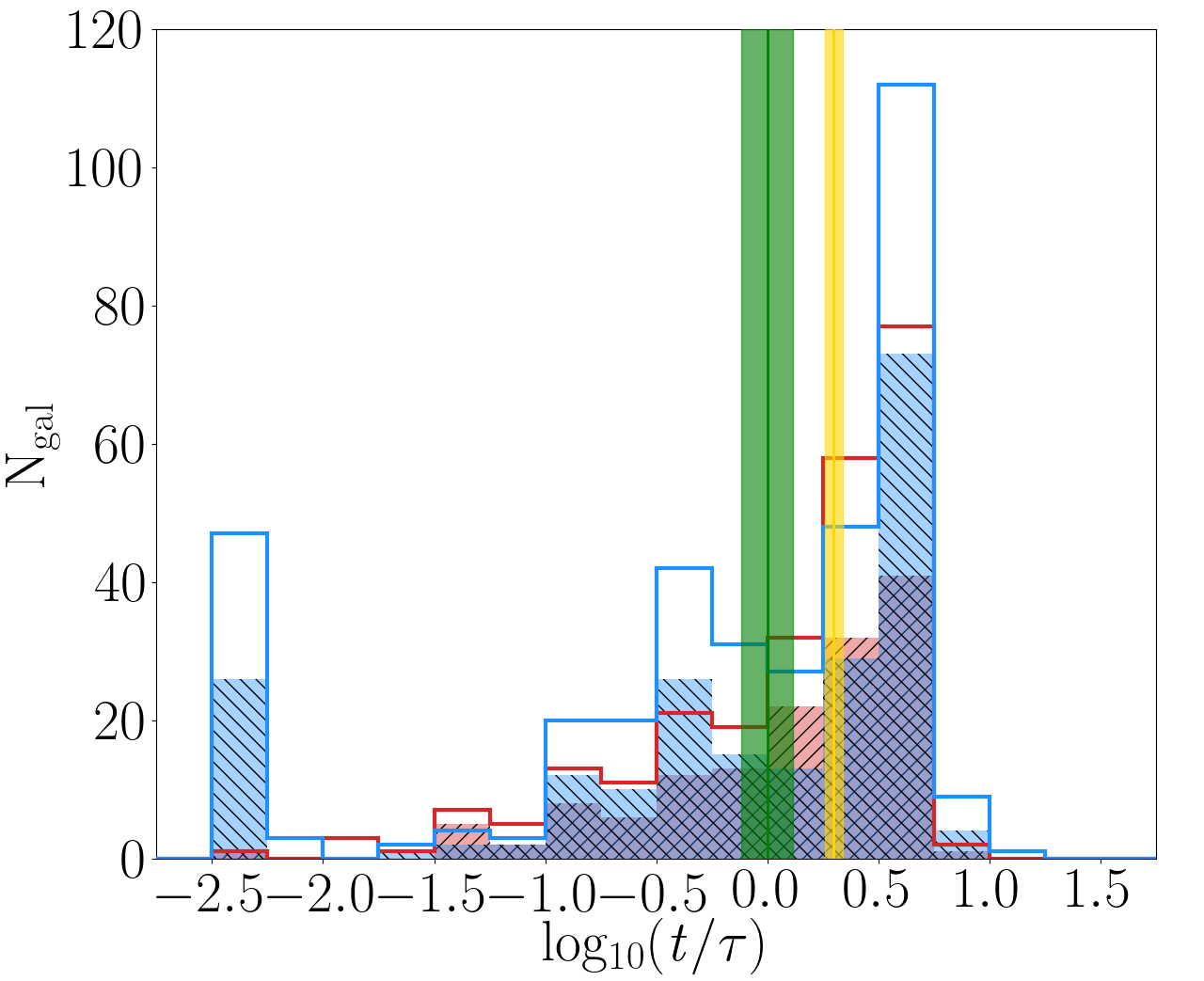}
     }
     \hfill
     \subfloat[]{
       \includegraphics[width=0.32\linewidth]{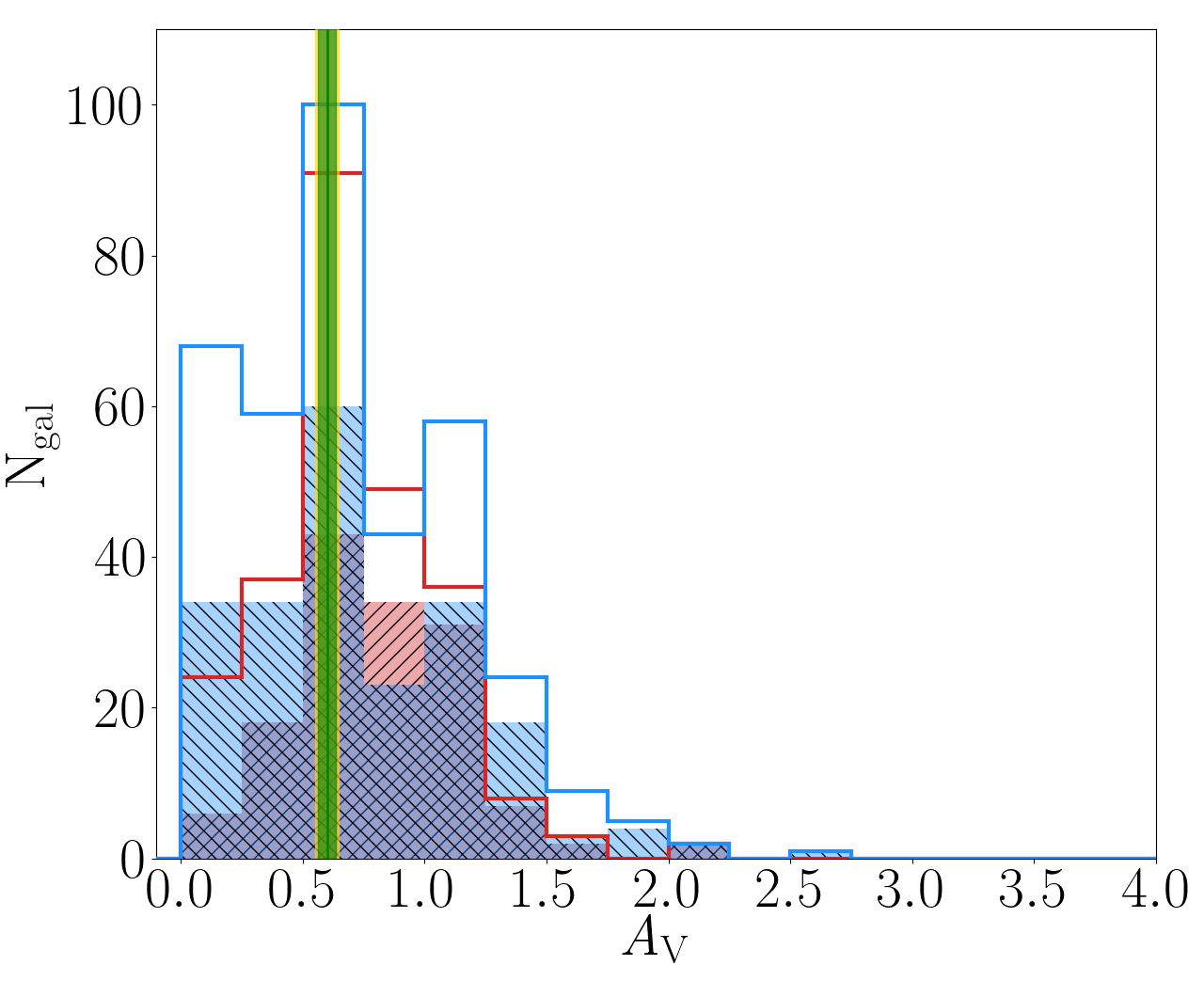}
     }
     \hfill
     \subfloat[]{
       \includegraphics[width=0.32\linewidth]{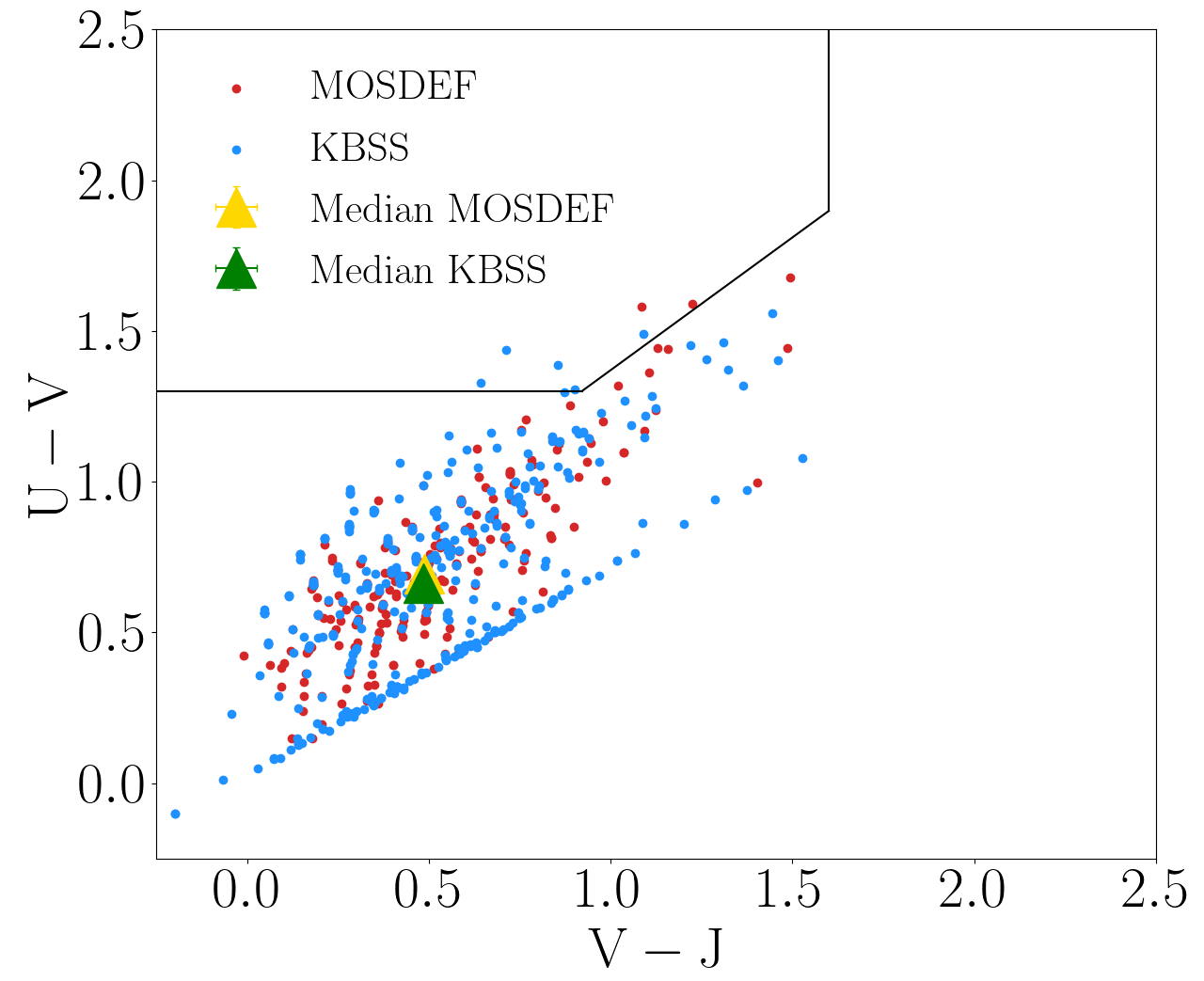}
     }
    \caption{Distribution of physical properties for the MOSDEF (red) and KBSS (blue) $z\sim2$ spectroscopic samples. The red and blue skeletal histograms represent the full spectroscopic samples (i.e., 250 galaxies for MOSDEF and 369 galaxies for KBSS), while the filled regions identify the subset of the spectroscopic samples that are on the [N~\textsc{II}] BPT diagram (i.e., 143 galaxies for MOSDEF and 213 galaxies for KBSS). The sample medians, with uncertainties for are shown in yellow and green for the MOSDEF (KBSS) full spectroscopic samples (i.e., the larger empty histograms), and are given in Table \ref{tab:z2_spec_sample_properties}. The following galaxy properties are shown: (a) $M_{\ast}$, (b) SFR(SED), (c) sSFR(SED), (d) log$_{10}$($t/\tau$) of the stellar population using a delayed-$\tau$ star formation model, (e) $A_{\rm{V}}$, and (f) the UVJ diagram. The quiescent galaxy region in the upper left is enclosed in a solid curve, while star-forming galaxies occupy the bottom half and upper right-hand corner of the diagram. We find that all KBSS and MOSDEF sample median properties agree within the 1$\sigma$ uncertainty, except for $t/\tau$. The x-axes for panels a-e and both axes for panel f are the same as in Figure \ref{fig:z2_targeted_sample_properties} to show how the distributions of the $z\sim2$ spectroscopic sample compared to those for the $z\sim2$ targeted samples.}
    \label{fig:z2_spec_sample_properties}
\end{figure*}

\begin{table*}
    \centering
    \begin{tabular}{rrrrr}
        \multicolumn{5}{c}{Median Values for Physical Properties of the MOSDEF and KBSS $z\sim2$ Spectroscopic Samples} \\
        \hline\hline
        Physical Property & MOSDEF Median & KBSS Median & p-value & Statistical significance \\
        (1) & (2) & (3) & (4) & (5) \\
        \hline
 log$_{10}(M_{\ast}/M_{\odot}$) & 9.85 $\pm$ 0.03 & 9.81 $\pm$ 0.03 & 0.0061 & 2.74$\sigma$ \\
 log$_{10}$($t/\tau$) & 0.30 $\pm$ 0.04 & 0.00 $\pm$ 0.12 & 8.9e$-05$ & 3.75$\sigma$ \\
 log$_{10}$(SFR(SED)/$M_{\odot}$/yr$^{-1}$) & 1.29 $\pm$ 0.04 & 1.32 $\pm$ 0.03 & 0.064 & 1.85$\sigma$ \\
 log$_{10}$(sSFR(SED)/yr$^{-1}$) & $-$8.61 $\pm$ 0.04 & $-$8.67 $\pm$ 0.07 & 8.7e$-10$ & 6.13$\sigma$ \\
 $A_{\rm{V}}$ & 0.60 $\pm$ 0.05 & 0.60 $\pm$ 0.04 & 5.5e$-05$ & 4.03$\sigma$ \\
 U$-$V & 0.69 $\pm$ 0.02 & 0.66 $\pm$ 0.02 & 0.059 & 1.89$\sigma$ \\
 V$-$J & 0.49 $\pm$ 0.02 & 0.48 $\pm$ 0.03 & 0.18 & 1.34$\sigma$ \\
        \hline
    \end{tabular}
    \caption{
  Col. (1): Physical property of the MOSDEF and KBSS $z\sim2$ spectroscopic samples (i.e., the skeletal histograms) shown in Figure \ref{fig:z2_spec_sample_properties}.
  Col. (2): Median value with uncertainty of the MOSDEF $z\sim2$ spectroscopic sample.
  Col. (3): Median value with uncertainty of the KBSS $z\sim2$ spectroscopic sample.
  Col. (4): Two-tailed p-value, based on the K-S test, estimating the probability that the null hypothesis can be rejected.
  Col. (5): Statistical significance (i.e., the $\sigma$ value) corresponding to the p-value. }
    \label{tab:z2_spec_sample_properties}
\end{table*}

\begin{figure}
    \centering
    \includegraphics[width=0.98\linewidth]{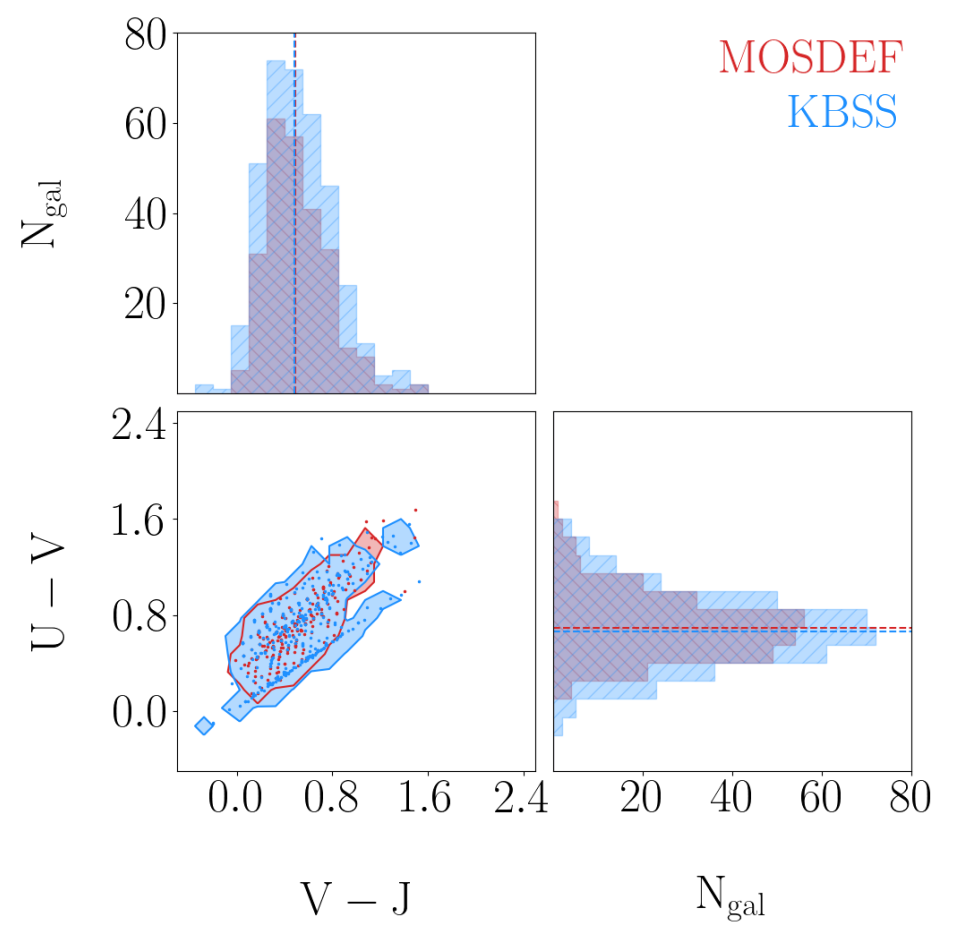}
    \caption{Corner plot comparing the distributions of U$-$V and V$-$J colors for the MOSDEF (red) and KBSS (blue) $z\sim2$ spectroscopic samples. The dashed lines on the 1D histograms mark the median values for the distributions of U$-$V and V$-$J colors given in Table \ref{tab:z2_targeted_sample_properties}, and the contours in the 2D panel trace the 3$\sigma$ regions in UVJ color-color space. The individual data points are also shown in the 2D UVJ distribution. It is shown here that the MOSDEF and KBSS $z\sim2$ spectroscopic samples occupy a very similar region of UVJ space, in addition to having sample medians that agree within the uncertainties.} \label{fig:z2_zspec_sample_uvj_corner}
\end{figure}

\begin{figure*}
    \includegraphics[width=0.49\linewidth]{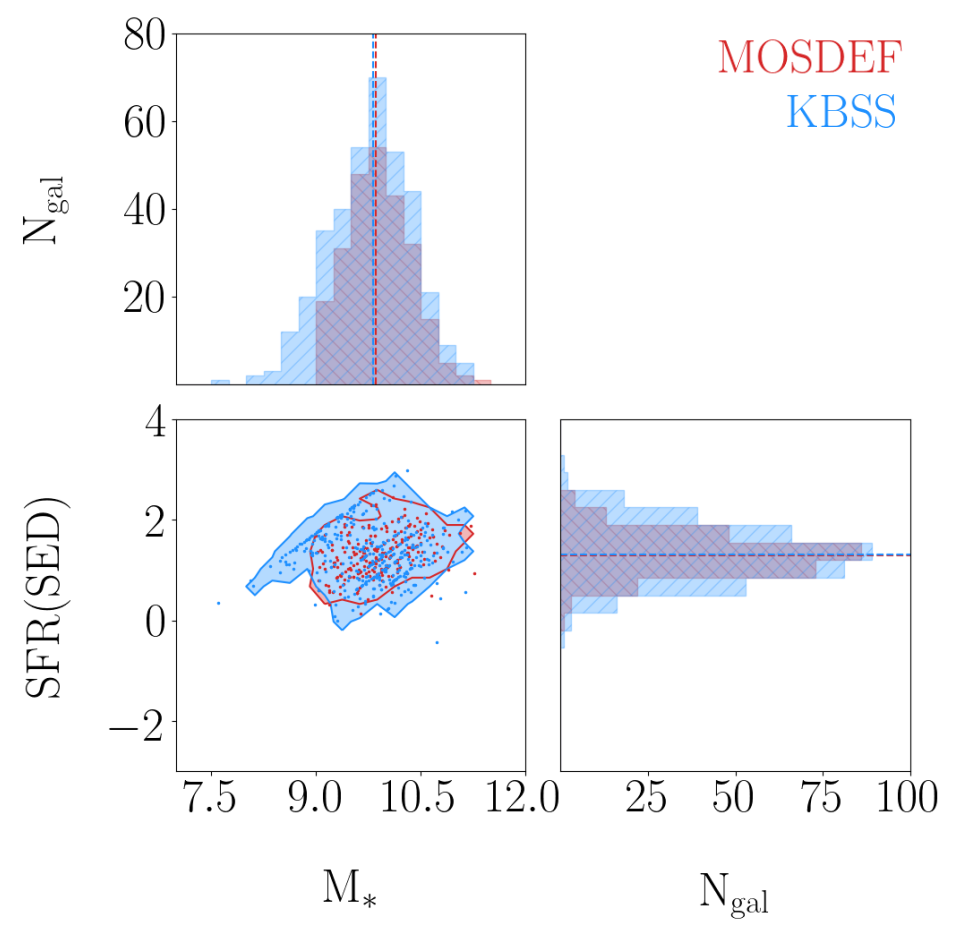}
    \includegraphics[width=0.49\linewidth]{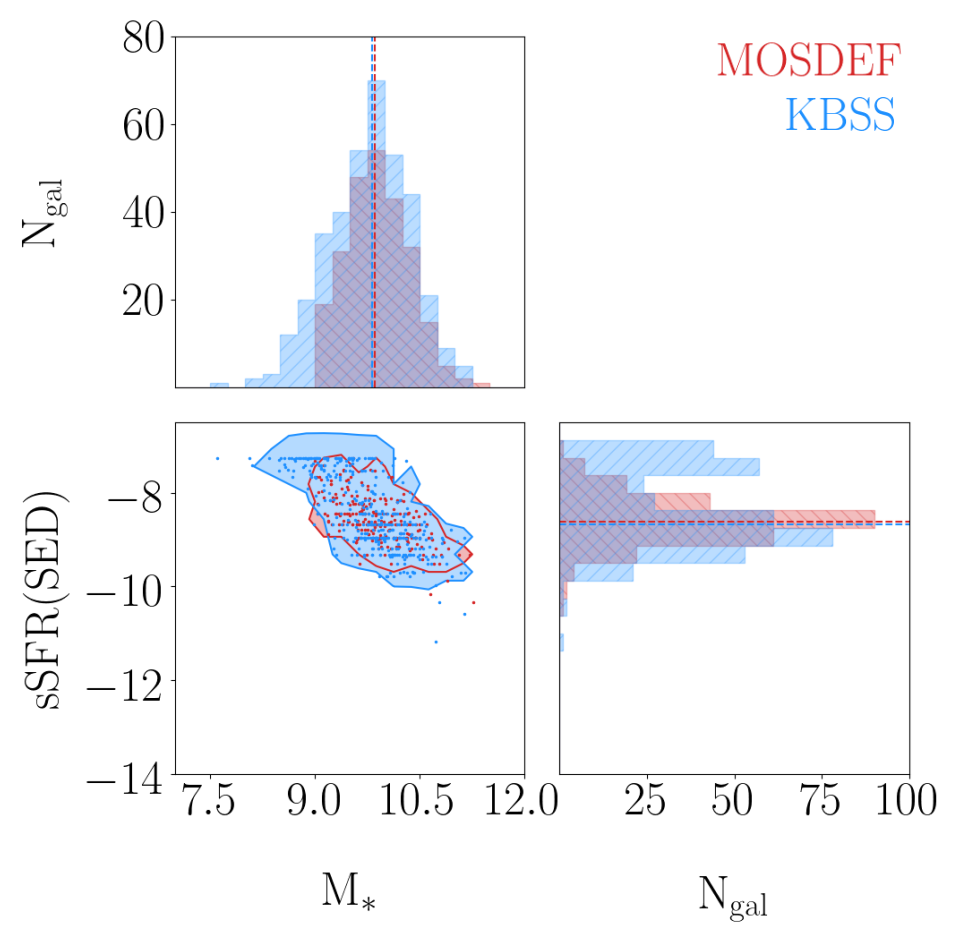}
    \caption{Corner plots comparing SFR(SED) vs. $M_{\ast}$ (left) and sSFR(SED) vs. $M_{\ast}$ (right) for the MOSDEF (red) and KBSS (blue) $z\sim2$ spectroscopic samples. The dashed lines on the 1D histograms mark the median values for the distributions, and the contours on the diagonal trace the 3$\sigma$ regions in 2D space. The individual data points are also shown in the 2D distribution.
    As stated for Figure \ref{fig:mosdef_kbss_survey_properties}, the distribution of datapoints is limited by the fact that both $M_{\ast}$ and SFR(SED) are determined by the normalization of the SPS model to the photometry. Additionally, in 2D space the datapoints crowd in the upper-left part of the distribution (left panel) and at high sSFR(SED) (right panel) because of the lower age limit for the stellar populations. We have the same axes limits as the bottom two panels in Figure \ref{fig:mosdef_kbss_survey_properties} so the distribution of galaxies in the $z\sim2$ spectroscopic samples can easily be compared with the $z\sim2$ targeted samples.}
    \label{fig:mosdef_kbss_spectroscopic_properties}
\end{figure*}

\begin{figure}
    \centering
    \includegraphics[scale=0.25]{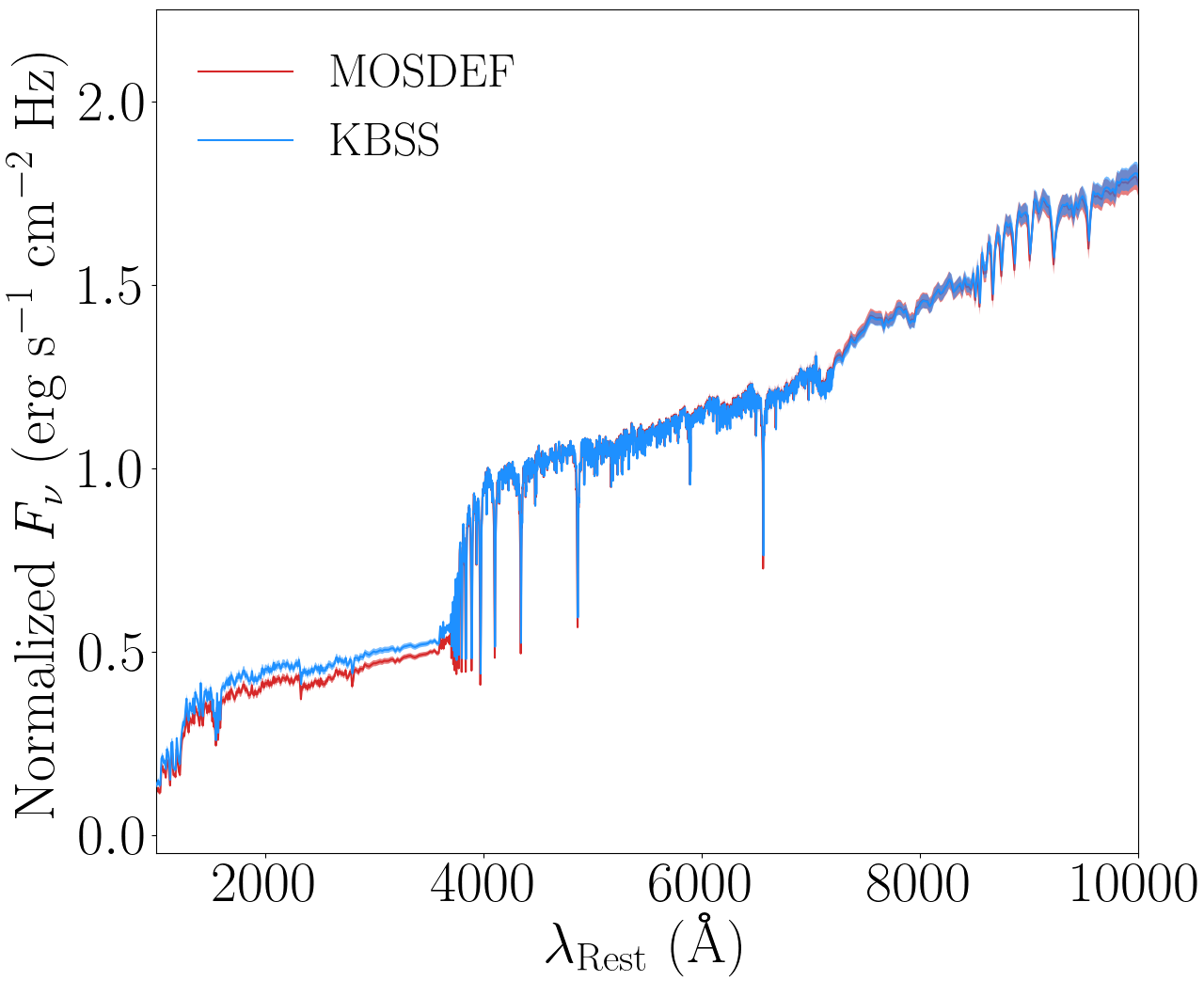}
    \caption{Stacked SEDs for the MOSDEF (red) and KBSS (blue) $z\sim2$ spectroscopic samples. The SEDs are normalized at 4550 \AA\space to show any differences in the global spectral shape. The $1\sigma$ scatter of the average SEDs is shown with shaded regions for both samples.} \label{fig:z2_spec_sample_stacked_seds}
\end{figure}

\begin{figure*}
    \includegraphics[width=0.49\linewidth]{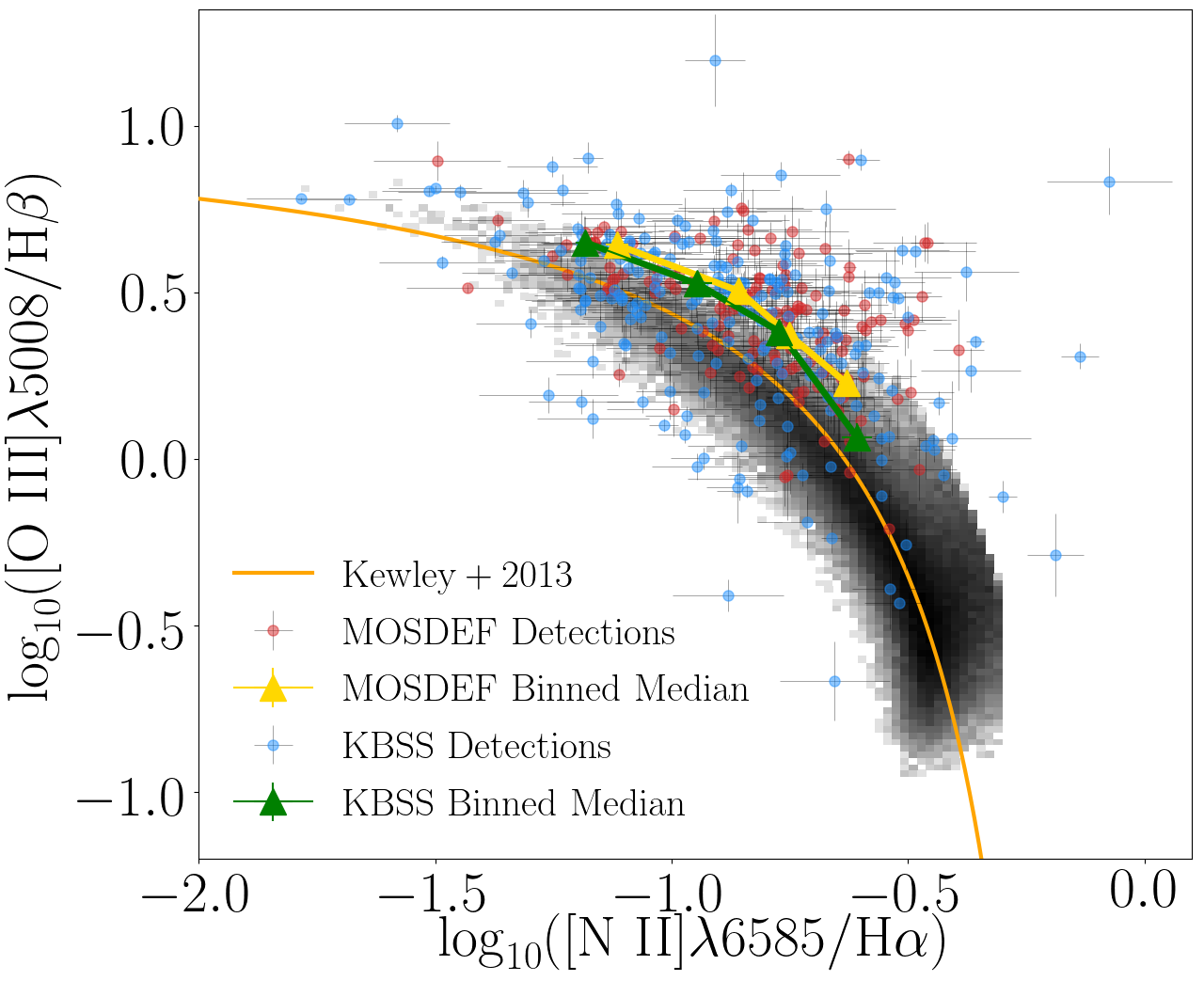}
    \includegraphics[width=0.49\linewidth]{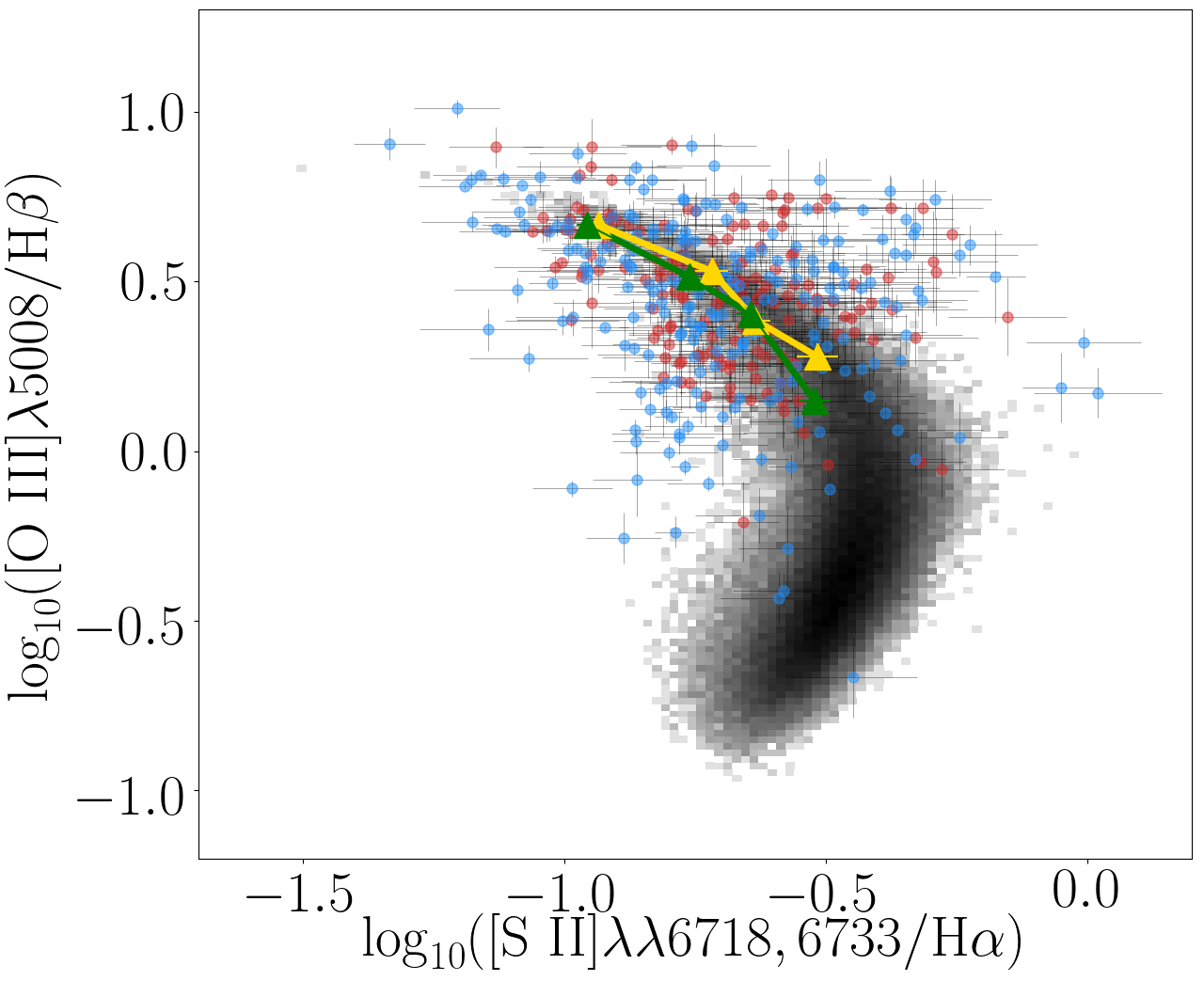}
    \caption{The MOSDEF (red) and KBSS (blue) $z\sim2$ spectroscopic samples on the [N~\textsc{II}] BPT (left) and [S~\textsc{II}] BPT (right) diagrams. The grayscale 2D histogram indicates local SDSS galaxies. Galaxies must have S/N $\geq$ 3 for the H$\beta$, [O~\textsc{III}]$\lambda$5008, H$\alpha$, and [N~\textsc{II}]$\lambda$6585 ([S~\textsc{II}]$\lambda\lambda$6718,6733) emission-lines to be included in the [N~\textsc{II}] BPT ([S~\textsc{II}] BPT) diagram. The yellow and green lines represent binned medians where the sample is binned in log$_{10}$(O3N2) (left) and log$_{10}$(O3S2) (right) for the MOSDEF and KBSS samples. The orange curve on the [N~\textsc{II}] BPT diagram is a fit to the $z\sim0$ star-forming locus \citep{kew13}.} \label{fig:z2_spec_sample_bpt_diagrams}
\end{figure*}

\section{The Physical Properties of the $\lowercase{z}\sim2$ Spectroscopic Samples} \label{sec:spec_sample_analysis}

In this section, we compare the MOSDEF and KBSS $z\sim2$ spectroscopic samples introduced in Sections \ref{subsec:mosdef_survey} (MOSDEF) and \ref{subsec:kbss_survey} (KBSS). Specifically, in Section \ref{subsec:spec_sample_galaxy_properties} we analyze the host galaxy properties of the samples derived from SED fitting. In Sections \ref{subsec:niibpt} and \ref{subsec:siibpt}, we investigate the locations of the $z\sim2$ spectroscopic samples on, respectively, the [N~\textsc{II}] and [S~\textsc{II}] BPT diagrams.

\subsection{SED Fitting Properties} \label{subsec:spec_sample_galaxy_properties}

Figure \ref{fig:z2_spec_sample_properties} (panels a-e) displays histograms of the galaxy properties for the MOSDEF and KBSS $z\sim2$ spectroscopic samples obtained through the FAST SED fitting. 
The red and blue lines give the distributions for the full MOSDEF (250 galaxies) and KBSS (369 galaxies) $z\sim2$ spectroscopic samples. The red and blue filled portions of the distributions indicate the subset of galaxies from each sample (143 MOSDEF and 213 KBSS galaxies) that have S/N $\geq$ 3 for H$\beta$, [O~\textsc{III}]$\lambda$5008, H$\alpha$, and [N~\textsc{II}]$\lambda$6585 and are included on the [N~\textsc{II}] BPT diagram. 
Sample medians are displayed as vertical lines for every galaxy property, and we derive 1$\sigma$ uncertainties on the medians through bootstrap resampling. These medians are estimated using all galaxies in the $z\sim2$ spectroscopic samples, not only the subsets of galaxies on the [N~\textsc{II}] BPT diagram, however the [N~\textsc{II}] BPT subsamples yield sample medians that do not significantly change the key results of this study. 
These values are given in Table \ref{tab:z2_spec_sample_properties}. We also list the probabilities (i.e., p-values), based on the Kolmogorov-Smirnov (K-S) test, of the null hypothesis that the two samples are drawn from the same distribution. The samples are also shown on the UVJ diagram (panel f). We include the \citet{wil09} boundary between quiescent and star forming galaxies on the UVJ diagram. 
Figure \ref{fig:z2_zspec_sample_uvj_corner} provides a corner plot of the U$-$V and V$-$J colors showing both 1D histograms of each parameter and the distribution of data points in 2D space overlaid with 3$\sigma$ contours.

Overall, we find that the MOSDEF and KBSS $z\sim2$ spectroscopic samples have very similar galaxy property distributions and median values. 
The KBSS sample has a smaller (i.e., younger) median $t/\tau$ compared to the MOSDEF sample. The p-value corresponds to an offset of greater than 3$\sigma$ indicating that this difference is significant.  The median values for all other host galaxy properties agree within the uncertainties. 
We note that despite the sample medians for sSFR(SED) and $A_{\rm{V}}$ agreeing within the uncertainties, the p-values correspond to a significance greater than 3$\sigma$ indicating that the two samples do not appear to occupy the same parameter space for these properties. 
Of the KBSS sample, 125 galaxies (33.9\%) have $\log_{10}$(sSFR(SED)/yr$^{-1}$) $\geq$ $-8.0$ compared to only 27 galaxies (10.8\%) of the MOSDEF sample. 
Also, we find that 50 galaxies (13.6\%) in the KBSS sample have stellar population age estimates of $\log_{10}$($t/\tau$) $\leq-2.0$. 
There is only one MOSDEF galaxy (0.4\% of the sample) in this low $t/\tau$ regime. The KBSS galaxies with $\log_{10}$($t/\tau$) $\leq$ $-$2.0 are the same galaxies that have $\log_{10}$(sSFR(SED)/yr$^{-1}$) $\geq -$8.0. This correspondence indicates that the KBSS sample has a subset of galaxies that is very young with intense star formation, for which there is no analogous subset of galaxies in the MOSDEF sample. 
For $A_{\rm{V}}$, we find that 158 MOSDEF galaxies (63.2\% of the sample) have $0.5 \leq A_{\rm{V}} \leq 1.0$ compared to 161 KBSS galaxies (only 43.6\% of the sample).
Otherwise, the distributions of galaxy properties between the two samples are very similar, indicated by p-values at less than 3$\sigma$ significance.

Figure \ref{fig:mosdef_kbss_spectroscopic_properties} shows corner plots of SFR(SED) vs. $M_{\ast}$ (left panel) and sSFR(SED) vs. $M_{\ast}$ (right panel) for the MOSDEF and KBSS $z\sim2$ spectroscopic samples. The sample medians from Table \ref{tab:z2_spec_sample_properties} are shown on the 1D histogram, and the 3$\sigma$ contour for each sample is outlined in the 2D distribution.
These differences echo those observed for the $z\sim2$ targeted samples (see Figure \ref{fig:mosdef_kbss_survey_properties}). The difference at low mass is accentuated by the fact that we implemented a mass cut at 10$^{9}$ M$_{\odot}$ for MOSDEF and did not for KBSS. However, the MOSDEF sample cut was implemented because of the known incompleteness at low mass. We emphasize that our goal here is to compare the properties of samples that have been analyzed in previously published work \citep[i.e.,][]{san18,str17}, and that these different selections at the low-mass end simply reflect what was adopted in these earlier papers. As for the difference in sSFR(SED), the fact that KBSS contains more galaxies at high sSFR(SED) than MOSDEF matches our observation when comparing the distributions in panel (c) of Figure \ref{fig:z2_spec_sample_properties}. 

The high $M_{\ast}$, low sSFR(SED) end of the 3$\sigma$ contours is approximately the same for both samples. This is different from what we find for the $z\sim2$ targeted samples, where MOSDEF extends to higher mass and lower sSFR(SED) compared to KBSS in Figure \ref{fig:mosdef_kbss_survey_properties}. The difference is due to the MOSDEF sample, where the 3$\sigma$ contour for the $z\sim2$ spectroscopic sample does not extend to as high mass and as low sSFR(SED) as the $z\sim2$ targeted sample. This again points out the MOSDEF incompleteness with respect to the massive, low sSFR(SED) galaxies targeted in the survey (see Section \ref{subsec:targeted_sample_sed_properties} further discussion on this topic). 

Figure \ref{fig:z2_spec_sample_stacked_seds} shows the stacked SEDs of the MOSDEF and KBSS $z\sim2$ spectroscopic samples. Similar to Figure 13 in \citet{str17}, we normalize each spectrum by the flux at 4550 \AA\space, sum the normalized spectra in each sample, then divide by the sample size to account for the uneven sizes of the MOSDEF and KBSS samples. We find that the stacked SEDs of the two samples are nearly identical. It is reasonable that the differences in rest-frame flux of the UV, optical, and near-IR regimes are negligible given that the sample medians for $\log_{10}$($t/\tau$) and UVJ colors agree within the uncertainties. 
The MOSDEF Balmer break is larger compared to KBSS; however, given the separation in median $t/\tau$ between the two samples, it is unclear why the difference in the Balmer breaks is not more pronounced.

\subsection{The [N~\textsc{II}] BPT Diagram} \label{subsec:niibpt}

The left panel of Figure \ref{fig:z2_spec_sample_bpt_diagrams} shows the [N~\textsc{II}] BPT diagram for the MOSDEF and KBSS samples. We include median lines binned by $\log_{10}$(O3N2) for both samples. There are equal numbers of galaxies in each bin. The binning scheme was adopted because it divides the samples into subgroups segregated roughly along the local star-forming sequence. Only galaxies with S/N $\geq$ 3 for the H$\beta$, [O~\textsc{III}]$\lambda$5008, H$\alpha$, [N~\textsc{II}]$\lambda$6585 emission-lines are included on the figure and in the binned medians. 
There are 143 MOSDEF and 213 KBSS galaxies that meet the S/N requirements to be included in this plot. 

We find that the two binned medians for the MOSDEF and KBSS samples have an almost identical offset from the local sequence. When discussing the [N~\textsc{II}] BPT diagram, we define the term ``offset'' as the orthogonal distance from the \citet{kew13} fit of the local star-forming locus.
The MOSDEF and KBSS bins have respectively,a median offset of 0.12 $\pm$ 0.02 and 0.10 $\pm$ 0.02 dex (0.02 $\pm$ 0.02 dex separation) perpendicular to the \citet{kew13} fit. 
The overlap of the offsets within the uncertainties is very different from those presented in past MOSDEF \citep{sha15} and KBSS \citep{ste14, str17} work. These earlier studies have cited KBSS as having a larger offset from the local sequence than MOSDEF by $\sim$0.1 dex. We credit the convergence of the [N~\textsc{II}] BPT offsets to consistent emission-line fitting between the two samples. Both samples underwent changes to line ratios compared to past measurements. 

Here we find that median O3 ratio for the MOSDEF sample is 0.04 dex higher compared to past MOSDEF studies \citep[e.g.,][]{sha15,san16}. When restricting the $z\sim2$ spectroscopic sample to galaxies included on the [N~\textsc{II}] BPT diagram, the updated (and more robust) methodology for estimating stellar Balmer absorption results in a median decrease of 6.7\% in the H$\beta$ fluxes. 
This updated methodology has a negligible effect on the H$\alpha$ flux and therefore results in a negligible change to the N2 ratio.  The KBSS sample is characterized by a 0.07 dex lower N2 compared to results presented in past KBSS studies. For galaxies with [N~\textsc{II}]$\lambda$6585 S/N $\geq3.0$ in both this study and past KBSS studies \citep{ste14, str17}, we find a median decrease of 12.8\% in [N~\textsc{II}]$\lambda$6585 flux estimates. In summary, when using a uniform emission-line fitting and Balmer absorption correction methodology we find that both higher O3 for the MOSDEF sample and lower N2 for the KBSS sample combine to erase the apparent differences between the two samples in the [N~\textsc{II}] BPT diagram that serve to make the median excitation sequences from each survey indistinguishable within the uncertainties.

\subsection{The [S~\textsc{II}] BPT Diagram} \label{subsec:siibpt}

The right panel of Figure \ref{fig:z2_spec_sample_bpt_diagrams} shows the [S~\textsc{II}] BPT diagram for the MOSDEF and KBSS samples. We include median lines binned in $\log_{10}$(O3S2) for both samples. Each bin has an equal number of galaxies. Similar to the case for the [N~\textsc{II}] BPT diagram, the binning scheme was adopted because it divides the samples into subgroups segregated roughly along the local star-forming sequence. 
Also similar to the [N~\textsc{II}] BPT diagram, only galaxies with S/N $\geq$ 3 for the emission-lines relevant to the [S~\textsc{II}] BPT diagram (i.e., H$\beta$, [O~\textsc{III}]$\lambda$5008, H$\alpha$, and [S~\textsc{II}]$\lambda\lambda$6718,6733) are included in the figure and the binned medians. 
There are 156 MOSDEF and 228 KBSS galaxies that meet the S/N requirements to be included in this plot.

Aside from in the bin with the highest S2, the binned medians for the two samples are very similar. Furthermore, the MOSDEF and KBSS median line ratios are more similar than in previous studies \citep{ste14, sha15, str17, san18}. The MOSDEF sample shows a 0.04 dex increase in O3 (similar to the [N~\textsc{II}] BPT diagram), a result of the updated method for estimating stellar Balmer absorption corrections. 
For KBSS, there is an 0.04 dex decrease in the S2 ratio due to smaller flux estimates by 8.6\% on average for the [S~\textsc{II}]$\lambda\lambda$6718,6733 doublet than in the methodology from \citet{str17}. This change is analogous to the effect we observed with the [N~\textsc{II}]$\lambda$6585 flux estimates, which will be discussed in Section~\ref{subsec:discussion_niibpt_offset}. Overall, the changes observed in the [S~\textsc{II}] BPT diagram reflect similar shifts in MOSDEF and KBSS to those we found in the [N~\textsc{II}] BPT diagram.

\section{Discussion} \label{sec:discussion}

This section begins with a discussion of the [N~\textsc{II}] BPT offset. In Section \ref{subsec:niibpt}, we have shown that the median offset from the local sequence for the $z\sim2$ MOSDEF and KBSS spectroscopic samples agree within the uncertainties when using consistent analysis methods. Section \ref{subsec:discussion_niibpt_offset} discusses the significance of this result, compares the results here to previous MOSDEF and KBSS studies, and investigates the bias introduced from utilizing different emission-line fitting methodologies.

We then turn our attention to the MOSDEF and KBSS $z\sim2$ targeted samples, from which the $z\sim2$ spectroscopic samples were drawn.
Section \ref{subsec:targeted_sample_sed_properties} compares the host galaxy properties of the $z\sim2$ targeted samples. Additionally, we discuss differences found between the $z\sim2$ targeted samples (i.e., the full set of galaxies the MOSDEF and KBSS surveys intended to observe) and the smaller $z\sim2$ spectroscopic samples (i.e., the subsets of the $z\sim2$ targeted samples with high S/N spectra). This comparison is significant because it reveals if the galaxies targeted by the MOSDEF and KBSS surveys are fundamentally different from the subsets of galaxies with high S/N  analyzed in past publications (e.g., \citealt{ste14, sha15, str17, san18}).

\subsection{The [N~\textsc{II}] BPT Offset} \label{subsec:discussion_niibpt_offset}

We have shown that, when subject to identical analysis methods, the MOSDEF and KBSS samples are much more similar than reported in previous work.
As discussed in Section \ref{subsec:emline_fitting}, we used an updated version of the MOSDEF emission-line fitting code for this study. 
Using the same code for both samples yielded offsets of 0.12 $\pm$ 0.02 and 0.10 $\pm$ 0.02 dex from the \citet{kew13} fit (0.02 $\pm$ 0.02 dex separation) for, respectively, the MOSDEF and KBSS samples.

As a consistency check, we fit the emission lines of both samples using the same Gaussian fitting assumptions as the {\tt MOSPEC} code that KBSS previously used \citep{ste14, str17}. {\tt MOSPEC} is a 1D spectral analysis code written in IDL and developed to analyze the MOSFIRE spectra of faint galaxies. Because ``{\tt MOSPEC}'' and ``MOSDEF'' are very similar acronyms, we will hereafter refer to the {\tt MOSPEC} code used by the KBSS team as the ``KBSS {\tt MOSPEC} code'' for clarity. 
The key difference between the two codes is that the KBSS {\tt MOSPEC} code fixes the centroids of every emission-line in a given band to a single redshift. Also, the velocity widths of all emission-lines are fixed to be the same in each band (i.e., $H$ and $K_{\rm{s}}$).
This restriction assigns H$\beta$ and each line in the [O~\textsc{III}]$\lambda\lambda$4960,5008 doublet the same velocity widths in the H~band, and H$\alpha$, [N~\textsc{II}]$\lambda$6585, and each line in the [S~\textsc{II}]$\lambda\lambda$6718,6733 doublet the same velocity width in the $K_{\rm{s}}$-band. As discussed in Section \ref{subsec:emline_fitting}, the centroids and widths are allowed more freedom to vary within the MOSDEF code \citep{kri15, red15}. Appendix \ref{sec:emline_fitting_example} provides examples of the two codes fitting the same set of spectra to highlight the similarities and differences between them.

It is important to emphasize the motivations that led to the way these two codes were written.  
For the MOSDEF code, more freedom was allowed in the fitting (e.g., in the centroids and the FWHMs) to fully marginalize over errors in these parameters for the highest S/N line, given that in some cases these lines were relatively weak and/or did not have a Gaussian shape. 
For KBSS, which observed the same galaxies many times at different PAs and for a longer period, the line shape can be better quantified.  
Thus, it is more likely that all lines will provide an accurate probe of the FWHM and redshift. 

We find that using the KBSS {\tt MOSPEC} code results in a $\sim$13\% median increase in [N~\textsc{II}]$\lambda$6585 flux estimates for both the MOSDEF and KBSS samples compared to when using the MOSDEF code. This difference corresponds to higher N2 line ratio estimates for both samples as well. A similar effect is observed for the [S~\textsc{II}]$\lambda\lambda$6718,6733 doublet. 

When using the KBSS {\tt MOSPEC} code, the MOSDEF and KBSS samples, respectively, have a median offset of 0.15 $\pm$ 0.02 and 0.12 $\pm$ 0.02 dex (0.03 $\pm$ 0.02 dex separation) perpendicular to the \citet{kew13} fit. The separation of the two samples with the KBSS {\tt MOSPEC} code is within the uncertainties of that found with the MOSDEF code (i.e., 0.02 $\pm$ 0.02 dex). 
Note that due to the changes to the N2 and O3 flux ratios, both the MOSDEF and KBSS samples are more offset from the local SDSS sequence by 0.03 and 0.02 dex, respectively, when using the KBSS {\tt MOSPEC} code compared to when the MOSDEF code is used. 
As noted above, this shift arises primarily from the KBSS {\tt MOSPEC} code yielding larger [N~\textsc{II}]$\lambda$6585 flux estimates compared to the MOSDEF code. There is no significant shift in O3 regardless of which line fitting code is adopted.

When compared to the [N~\textsc{II}]$\lambda$6585 and [S~\textsc{II}]$\lambda\lambda$6718,6733 doublet bandpass integrated line flux, we find that the MOSDEF Gaussian fits generally estimates a lower flux on average while the KBSS Gaussian fits estimates a higher flux on average. In other words, the MOSDEF code underestimates the flux while the KBSS {\tt MOSPEC} code overestimates the flux for these weaker emission-lines compared to their integrated bandpass fluxes. 
The goal of this paper is not to determine which fitting method is more correct, but instead to show that different assumptions, both reasonable, when implemented into emission-line fitting codes, can lead to different flux measurements resulting in shifts of $\sim$0.02-0.03 dex on the [N~\textsc{II}] BPT diagram. These varying assumptions introduce inherent systematic uncertainty in the flux measurements that are not captured within the statistical uncertainties of the spectra.
However, it is important to note that we find nearly identical small relative offsets --- 0.02 $\pm$ 0.02 dex using the MOSDEF code and 0.03 $\pm$ 0.02 dex using the KBSS {\tt MOSPEC} code --- between the median MOSDEF and KBSS emission-line sequences on the [N~\textsc{II}] BPT diagram when the same line fitting code is used. 
Therefore, the same conclusions regarding the locus of $z\sim 2$ star-forming galaxies in the [N~\textsc{II}] BPT diagram would be reached regardless of which line fitting code is used for analysis. 

We note that the $\sim$0.1 dex offset cited in previous MOSDEF (e.g., \citealt{sha15}) and KBSS (e.g. \citealt{ste14, str17}) works is resolved based on updated (smaller) stellar Balmer absorption corrections for the MOSDEF sample as well as consistent emission-line fitting methods.

\begin{figure*}
    \centering
     \subfloat[]{
       \includegraphics[width=0.32\linewidth]{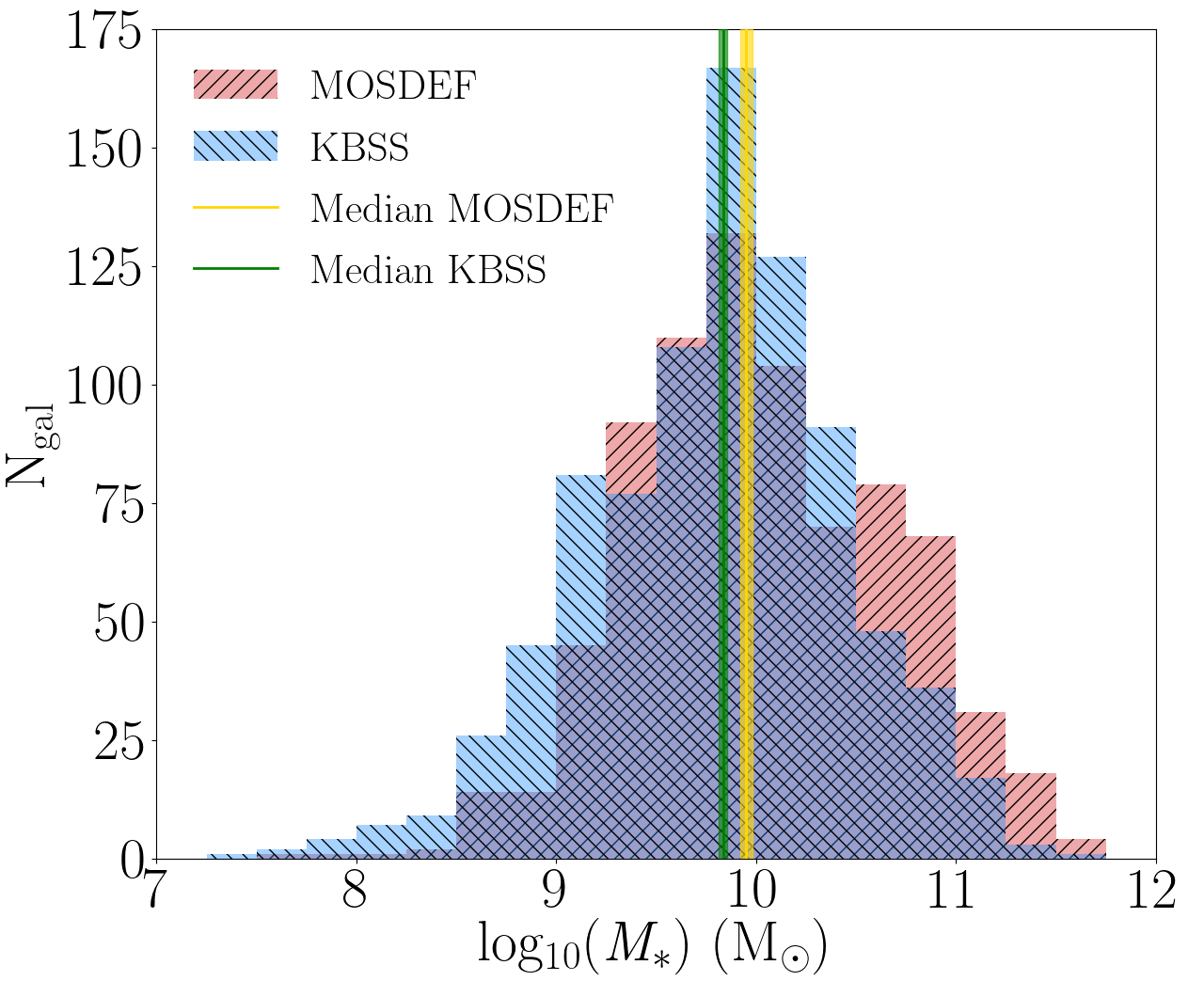}
     }
     \hfill
     \subfloat[]{
       \includegraphics[width=0.32\linewidth]{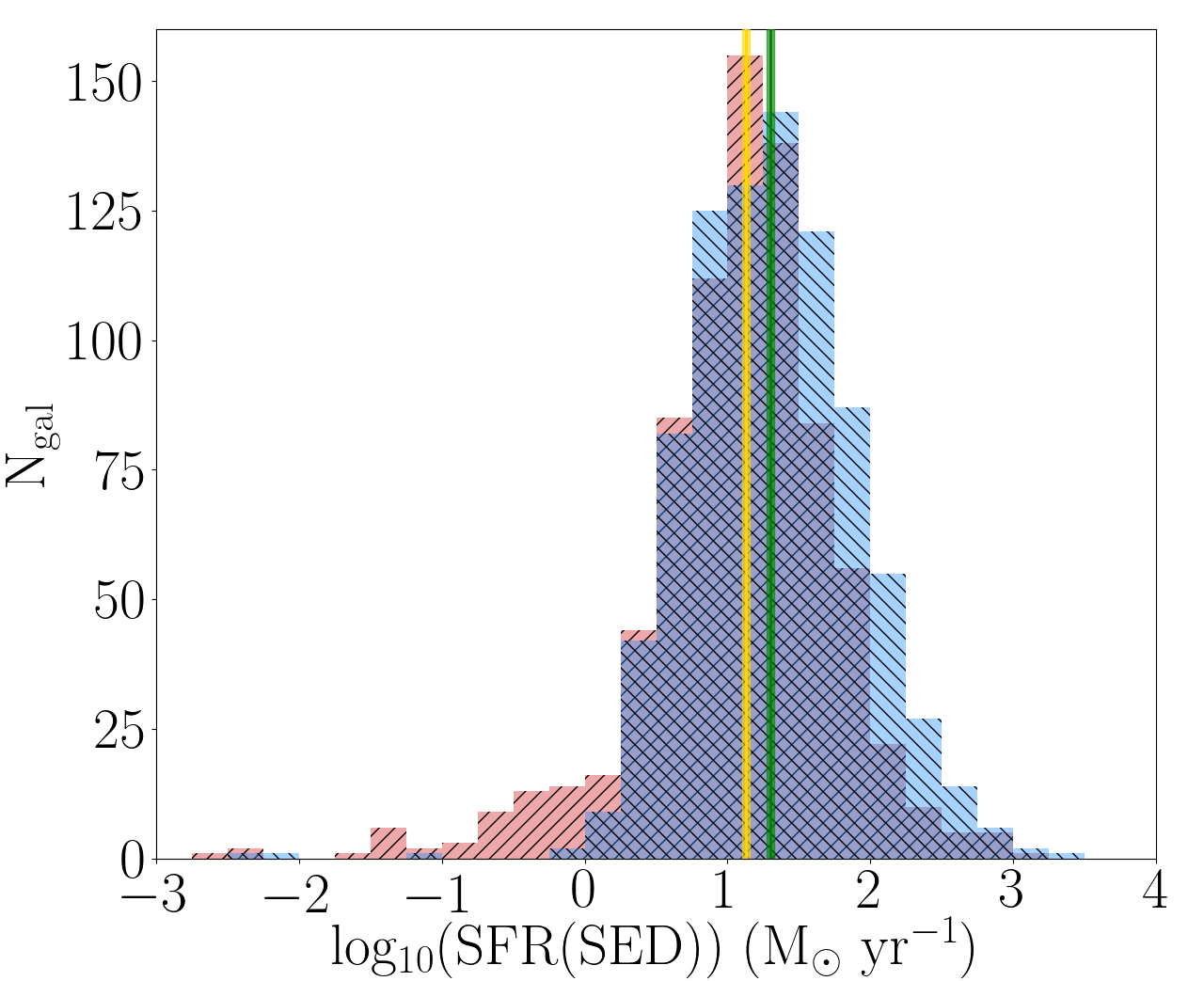}
     }
     \hfill
     \subfloat[]{
       \includegraphics[width=0.32\linewidth]{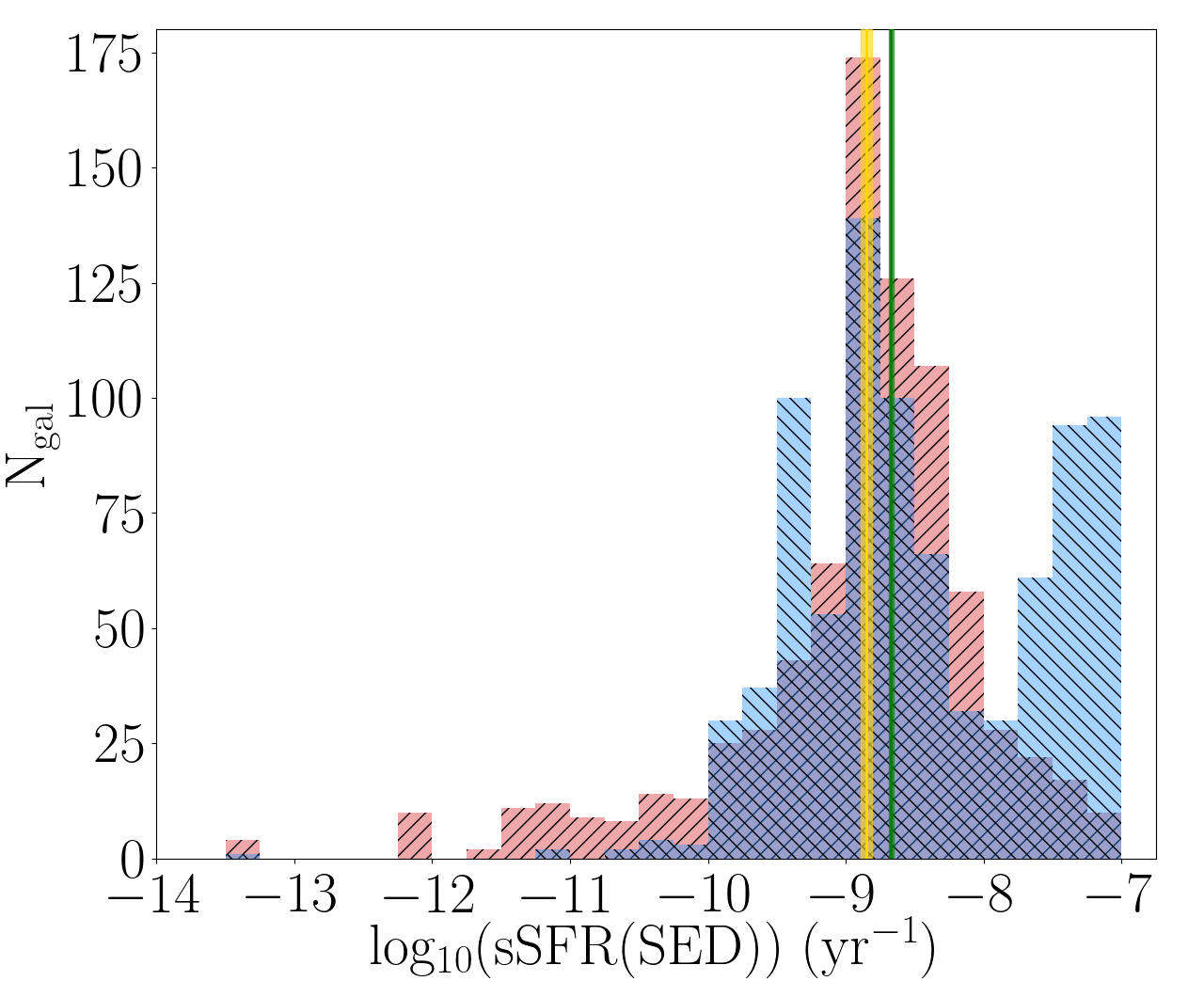}
     }
     \hfill
     \subfloat[]{
       \includegraphics[width=0.32\linewidth]{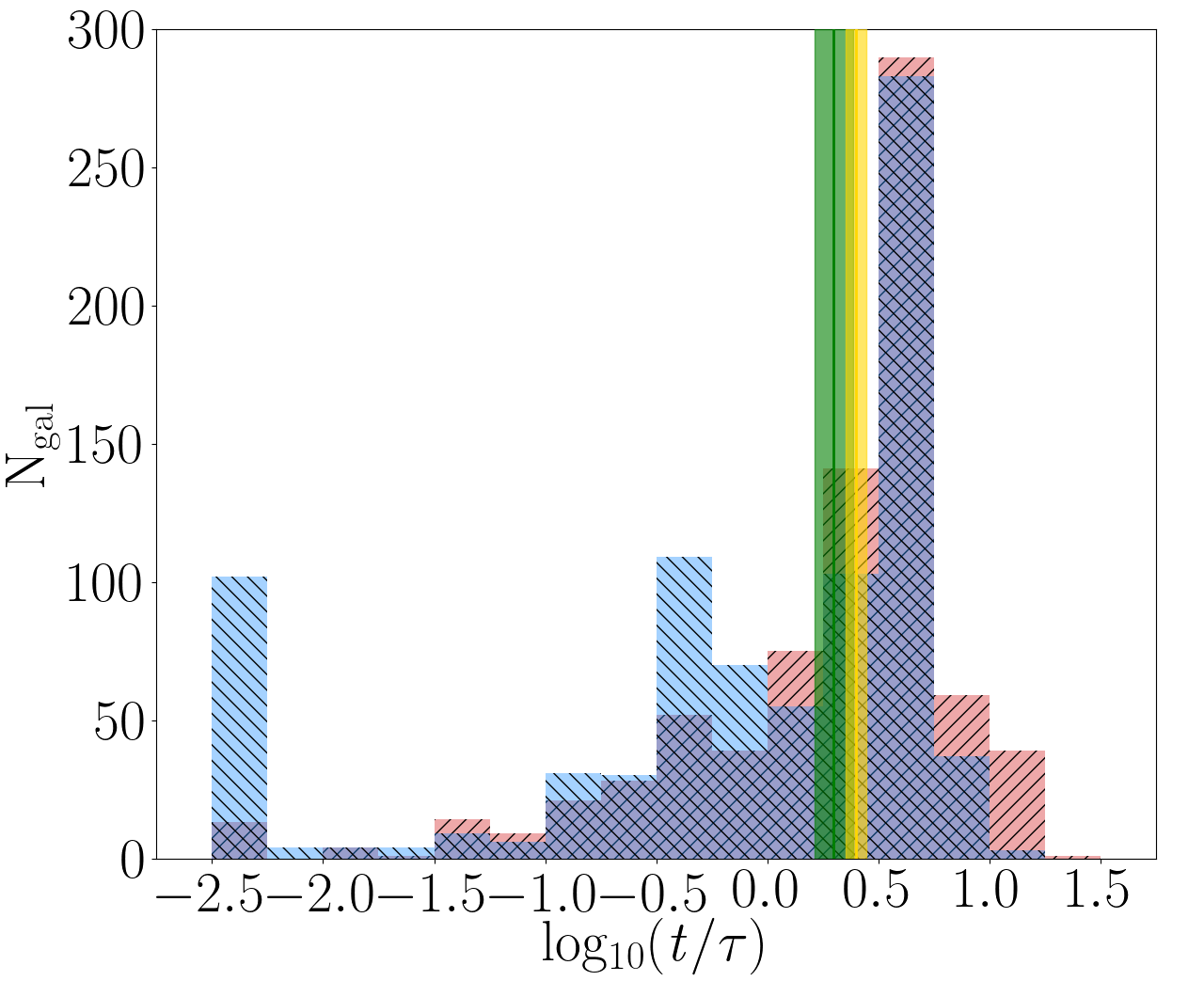}
     }
     \hfill
     \subfloat[]{
       \includegraphics[width=0.32\linewidth]{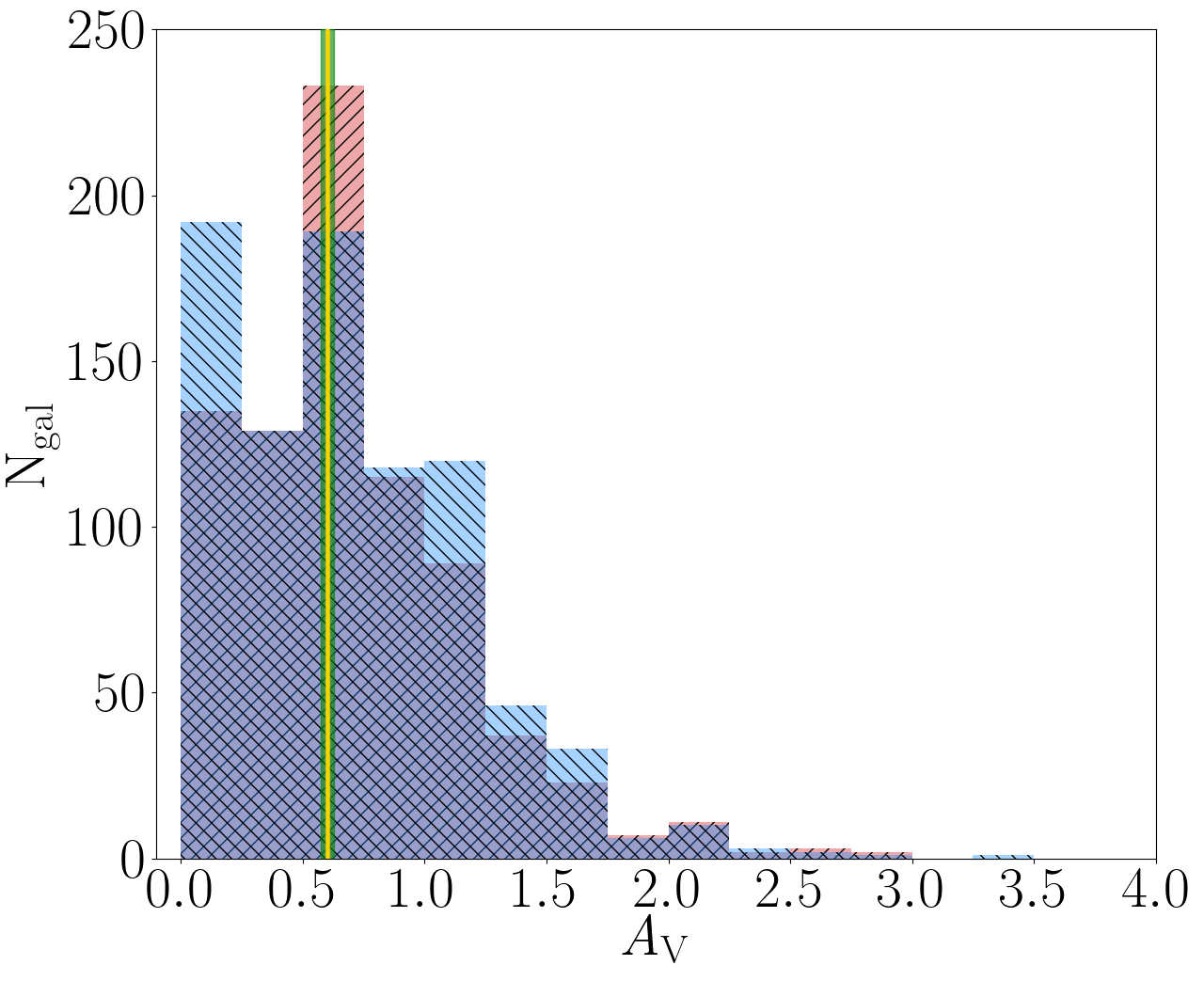}
     }
     \hfill
     \subfloat[]{
       \includegraphics[width=0.32\linewidth]{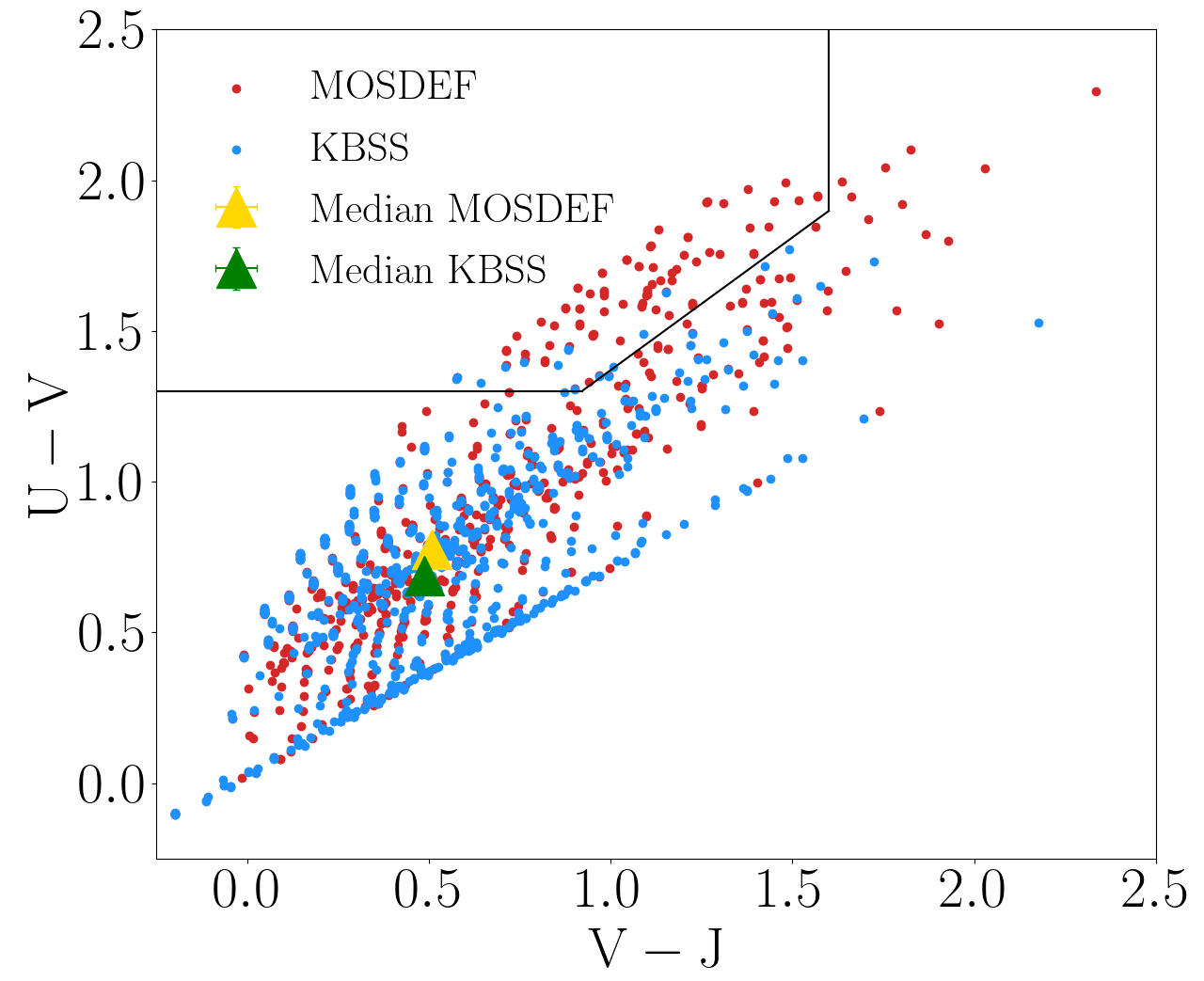}
     }
    \caption{Distribution of physical properties for the MOSDEF (red) and KBSS (blue) $z\sim 2$ targeted samples. The sample medians, with uncertainties are shown in yellow (green) for the MOSDEF (KBSS) samples, and are given in Table \ref{tab:z2_targeted_sample_properties}. The following galaxy properties are shown: (a) $M_{\ast}$, (b) SFR(SED), (c) sSFR(SED), (d) log$_{10}$($t/\tau$) of the stellar population using a delayed-$\tau$ star formation model, (e) $A_{\rm{V}}$, and (f) the UVJ diagram.  We find that the MOSDEF $z\sim2$ targeted sample has a higher $M_{\ast}$, lower SFR(SED) and sSFR(SED), and a redder U$-$V color compared to the KBSS $z\sim2$ targeted sample. Note that two MOSDEF galaxies with log$_{10}$(SFR(SED)/M$_{\odot}$/yr$^{-1}$) $<$ $-$3.0 are excluded from the panel (b), and one MOSDEF galaxy with log$_{10}$(sSFR(SED)/yr$^{-1}$) $<$ $-$14.0 is excluded from panel (c) to better show the data.}
    \label{fig:z2_targeted_sample_properties}
\end{figure*}

\begin{table*}
    \centering
    \begin{tabular}{rrrrr}
        \multicolumn{5}{c}{Median Values for Physical Properties of the MOSDEF and KBSS $z\sim2$ Targeted Samples} \\
        \hline\hline
        Physical Property & MOSDEF Median & KBSS Median & p-value & Statistical significance \\
        (1) & (2) & (3) & (4) & (5) \\
        \hline
 log$_{10}(M_{\ast}/M_{\odot}$) & 9.95 $\pm$ 0.03 & 9.84 $\pm$ 0.02 & 1.9e$-$7 & 5.21$\sigma$ \\
 log$_{10}$($t/\tau$) & 0.40 $\pm$ 0.05 & 0.30 $\pm$ 0.09 & 7.5e$-$18 & 8.61$\sigma$ \\
 log$_{10}$(SFR(SED)/M$_{\odot}$/yr$^{-1}$) & 1.13 $\pm$ 0.02 & 1.30 $\pm$ 0.02 & 4.2e$-$9 & 5.88$\sigma$ \\
 log$_{10}$(sSFR(SED)/yr$^{-1}$) & $-$8.85 $\pm$ 0.04 & $-$8.67 $\pm$ 0.01 & 1.1e$-$20 & 9.33$\sigma$ \\
 $A_{\rm{V}}$ & 0.60 $\pm$ 0.01 & 0.60 $\pm$ 0.03 & 6.2e$-$8 & 5.41$\sigma$ \\
 U$-$V & 0.78 $\pm$ 0.02 & 0.69 $\pm$ 0.01 & 3.6e$-$8 & 5.51$\sigma$ \\
 V$-$J & 0.51 $\pm$ 0.02 & 0.49 $\pm$ 0.01 & 3.9e$-$5 & 4.11$\sigma$ \\
        \hline
    \end{tabular}
    \caption{
  Col. (1): Physical property of the galaxies in the sample shown in Figure \ref{fig:z2_targeted_sample_properties}.
  Col. (2): Median value with uncertainty of the MOSDEF $z\sim2$ targeted sample.
  Col. (3): Median value with uncertainty of the KBSS $z\sim2$ targeted sample.
  Col. (4): Two-tailed p-value, based on the K-S test, estimating the probability that the null hypothesis can be rejected.
  Col. (5): Statistical significance (i.e. the $\sigma$ value) that the p-value corresponds to. }
    \label{tab:z2_targeted_sample_properties}
\end{table*}

\begin{figure}
    \centering
    \includegraphics[width=0.98\linewidth]{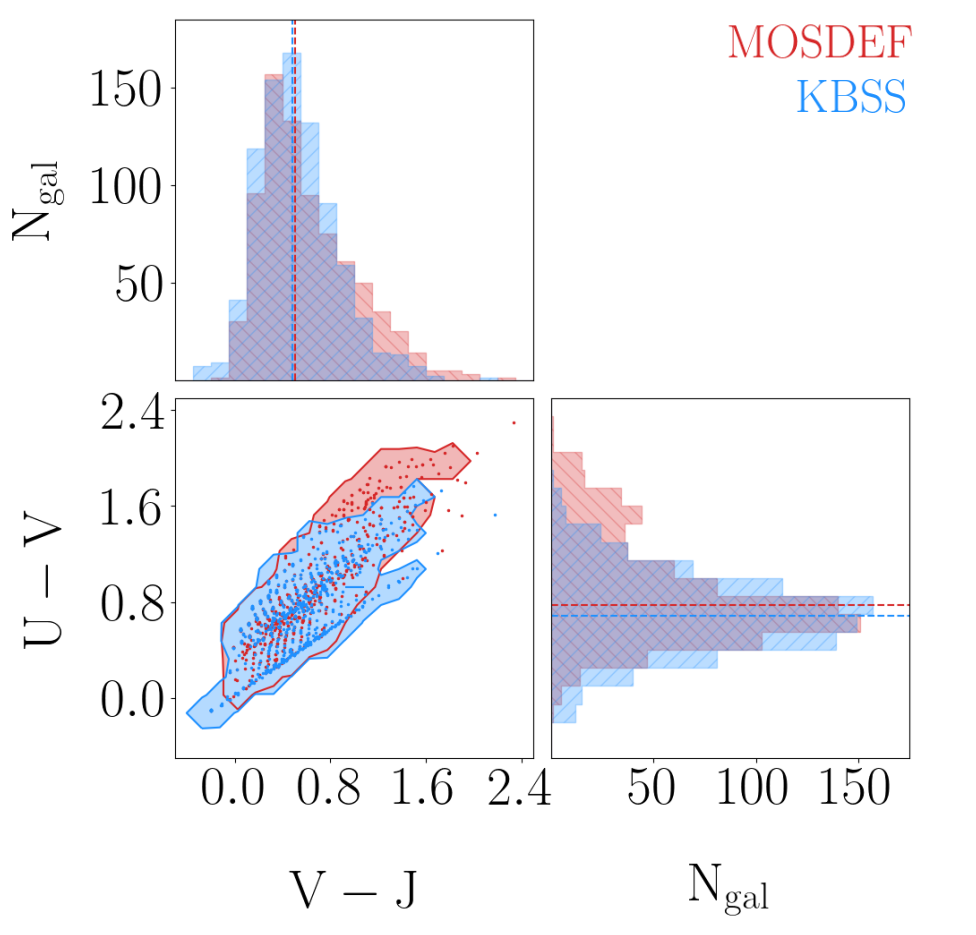}
    \caption{Corner plot comparing the distributions of U$-$V and V$-$J colors for the MOSDEF (red) and KBSS (blue) $z\sim2$ targeted samples. The dashed lines on the 1D histograms mark the median values for the distributions of U$-$V and V$-$J colors given in Table \ref{tab:z2_targeted_sample_properties}, and the contours in the 2D panel trace the 3$\sigma$ regions in UVJ color-color space. The individual data points are also shown in the 2D UVJ distribution. It is shown here that MOSDEF targeted a population of red galaxies with high U$-$V color that is not seen in the KBSS $z\sim2$ targeted sample. Additionally, MOSDEF (KBSS) extends to slightly redder (bluer) regions of V$-$J color space.} \label{fig:z2_targeted_sample_uvj_corner}
\end{figure}

\begin{figure}
    \centering
    \includegraphics[scale=0.25]{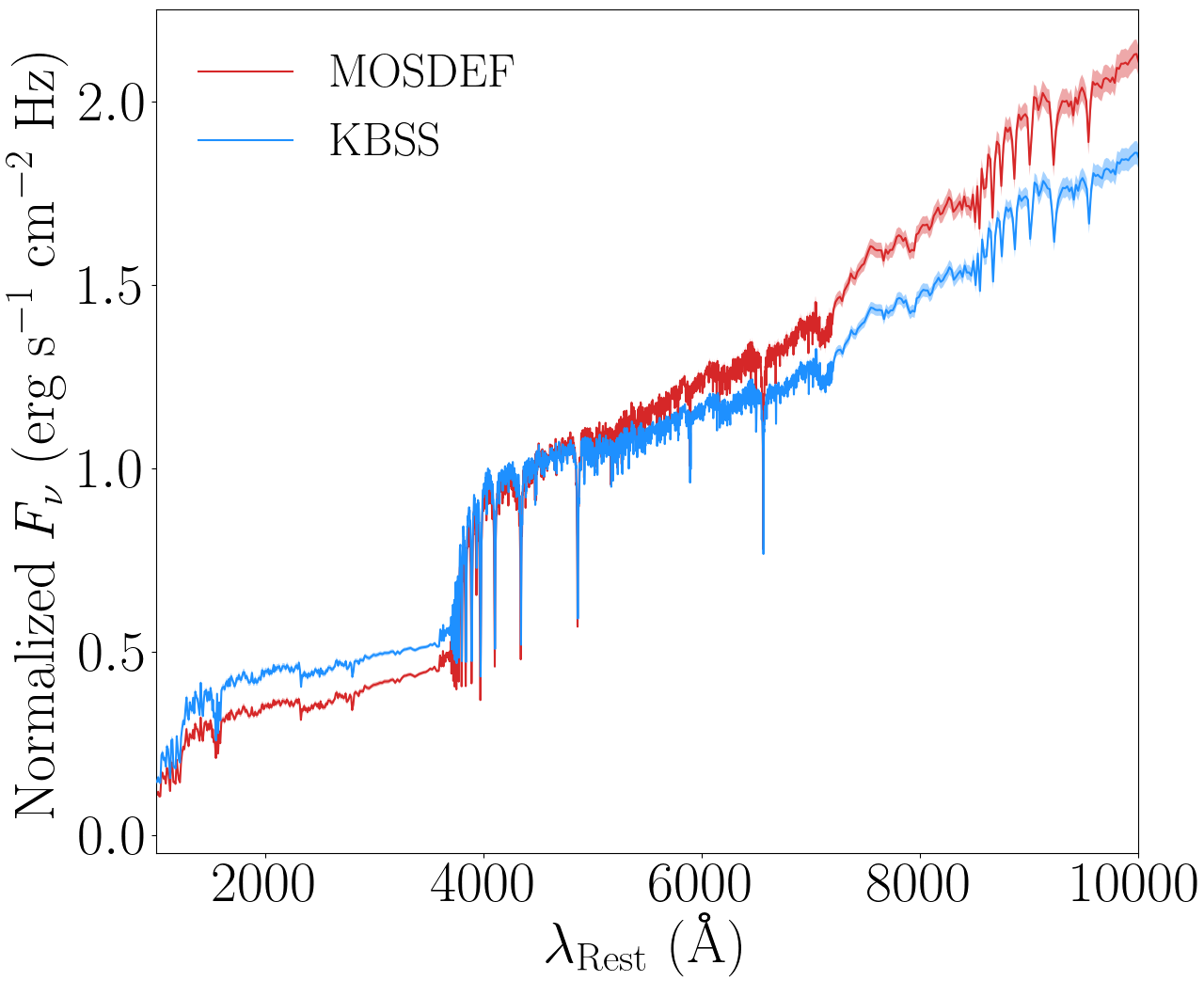}
    \caption{Stacked SEDs for the MOSDEF (red) and KBSS (blue) $z\sim2$ targeted samples. The SEDs are normalized at 4550 \AA\space to show any differences in the global spectral shape. The scatter of the SEDs is shown with shaded regions for both samples.} \label{fig:z2_targeted_sample_stacked_seds}
\end{figure}

\subsection{SED Properties of MOSDEF and KBSS $z\sim2$ Targeted Samples} \label{subsec:targeted_sample_sed_properties}

In Section \ref{subsec:spec_sample_galaxy_properties}, we compared the galaxy properties obtained through SED fitting for the $z\sim2$ spectroscopic MOSDEF and KBSS samples (i.e., the galaxies shown in blue in Figure \ref{fig:mosdef_kbss_survey_properties} and analyzed in previous studies \citealt{str17, san18}). Except for $t/\tau$ (i.e., age), the sample medians for every other galaxy property investigated agree within the uncertainties. 
Here, we expand our analysis and examine the SED properties of the full MOSDEF and KBSS $z\sim2$ targeted survey catalogs (i.e., all galaxies shown in Figure \ref{fig:mosdef_kbss_survey_properties}, including the galaxies where a $z_{\rm{MOSFIRE}}$ was not measurable). 
The goal of this analysis is to determine whether the different survey selection methods used by MOSDEF and KBSS resulted in galaxies with systematically different properties targeted in the surveys. As we discuss in Section \ref{subsec:mosdef_kbss_sample}, most of the MOSDEF $z\sim2$ galaxies would fall into the KBSS selection criteria (and vice versa). The majority (96.1\%) of the KBSS $z\sim2$ targeted sample satisfies the main selection criterion of the MOSDEF survey (i.e., $H_{\rm{AB}}<24.5$), while the distribution of the MOSDEF $z\sim2$ targeted sample on the $U_{\rm{n}}G\mathcal{R}$ diagram is very similar to KBSS with 98.5\% of MOSDEF galaxies having $\mathcal{R}<25.5$. 
This analysis will help to understand the biases of each survey with respect to the other, which is necessary for future studies using the combined sample.

We consider the same galaxy properties as in Section \ref{subsec:spec_sample_galaxy_properties}: $M_{\ast}$, SFR(SED), sSFR(SED), $\log_{10}$($t/\tau$) of the stellar population using a delayed-$\tau$ star formation model, $A_{\rm{V}}$, UVJ colors, and stacked SED shapes. The UVJ diagram along with histograms of these properties for MOSDEF and KBSS samples are shown in Figure \ref{fig:z2_targeted_sample_properties}. We derive 1$\sigma$ uncertainties on the medians through bootstrap resampling. Table \ref{tab:z2_targeted_sample_properties} lists the medians with uncertainties, two-tailed p-value based on the K-S test, and the level of significance (the $\sigma$ value) of the p-value. 
Additionally, Figure \ref{fig:z2_targeted_sample_uvj_corner} explores the UVJ colors in more detail by including both a 2D contour map in UVJ space and 1D histrograms of the U$-$V and V$-$J distributions. 
Finally, the stacked SEDs for the MOSDEF and KBSS $z\sim2$ targeted samples are shown in Figure \ref{fig:z2_targeted_sample_stacked_seds}. The stacked SEDs are made using the same methods described in Section \ref{subsec:spec_sample_galaxy_properties} (i.e., the SEDs are normalized at 4550 \AA \space to best show any variations in SED shape between the two samples). 

Based on visual inspection of the stacked SEDs, it is clear that on average the MOSDEF $z\sim2$ targeted sample has a larger Balmer break and higher flux redward of $\sim$5000 \AA. 
These observed differences in SED shapes for the $z\sim2$ targeted samples are not seen in the stacked SEDs for the $z\sim2$ spectroscopic samples, indicating that the MOSDEF and KBSS $z\sim2$ targeted samples are less similar than the $z\sim2$ spectroscopic samples. 
Additionally, we find that the MOSDEF $z\sim2$ targeted sample has a larger median $M_{\ast}$, smaller SFR(SED) and sSFR(SED) medians, and a redder U$-$V compared to the KBSS $z\sim2$ targeted sample. 
The medians for the remaining properties (i.e., $t/\tau$, $A_{\rm{V}}$, and V$-$J color) all agree within the uncertainties. 
The p-value from the K-S test for every galaxy property investigated in Figure \ref{fig:z2_targeted_sample_properties} has a significance of at least 4$\sigma$, indicating that 
the shape of the distribution for all galaxy properties considered --- even the ones with overlapping sample medians --- are significantly different. 

Examining the distribution of the galaxies in Figure \ref{fig:z2_targeted_sample_properties}, we find clear differences between the two surveys aside from the sample medians. In the KBSS $z\sim2$ targeted catalog, 281 galaxies (33.1\% of the sample) have $\log_{10}$(sSFR(SED)/yr$^{-1}$) $\geq-8.0$ and 106 galaxies (12.5\% of the sample) have $\log_{10}(t/\tau) \leq-2.0$. All 106 of the galaxies in this low $t/\tau$ regime are in the high sSFR(SED) regime. 
The MOSDEF $z\sim2$ targeted catalog is not well represented in these regions of parameter space, with only 77 galaxies (9.8\% of the sample) at $\log_{10}$(sSFR(SED)/yr$^{-1}$) $\geq-8.0$ and 14 galaxies (1.8\% of the sample) at $\log_{10}(t/\tau) \leq-2.0$. Similar to KBSS, all 14 of the MOSDEF $z\sim2$ targeted sample galaxies with low $t/\tau$ also have a high sSFR(SED).
It is apparent that the KBSS $z\sim2$ sample contains a population of very young galaxies with intense star formation that is mostly absent from the MOSDEF $z\sim2$ sample. 
At the same time, MOSDEF contains 59 galaxies (7.5\% of the sample) with both U$-$V and V$-$J $\geq$ 1.25, whereas KBSS comparatively only has 18 galaxies (2.1\% of the sample) in that area of the UVJ diagram. The 3$\sigma$ contours in UVJ space in Figure \ref{fig:z2_targeted_sample_uvj_corner} also show that the MOSDEF $z\sim2$ targeted sample probes a more reddened part of UVJ space than KBSS. 
Galaxies in this regime of the UVJ diagram tend to be more heavily dust obscured or have very low SFR(SED). 
The distributions of the $z\sim2$ targeted samples with $A_{\rm{V}} \geq 1$ are similar (22.1\% and 26.1\% of the MOSDEF and KBSS samples), indicating that the reddest galaxies in the MOSDEF $z\sim 2$ targeted sample --- i.e., those that are not well represented in the KBSS $z\sim 2$ targeted sample --- are not dominated by heavily dust obscured systems but rather those with low sSFR(SED). This difference can be attributed to the original selection method of the KBSS survey (i.e., rest-UV criteria based on $U_{\rm{n}}G\mathcal{R}$ colors), and is one of the reasons why the new subset of ``RK'' galaxies \citet{str17} was added to the survey. However, it is important to note that the ``RK'' galaxies are a small minority, only 28 galaxies corresponding to 3.3\%, of the 850 galaxy KBSS $z\sim2$ targeted sample.

Some of the results for the $z\sim2$ targeted samples in Figure \ref{fig:z2_targeted_sample_properties} are different from what we find for the $z\sim2$ spectroscopic samples in Section \ref{subsec:spec_sample_galaxy_properties}. 
The MOSDEF $z\sim2$ targeted sample has a higher median $M_{\ast}$ and $t/\tau$, a lower median SFR(SED) and sSFR(SED), and redder U$-$V color compared to the MOSDEF $z\sim2$ spectroscopic sample. The remaining host galaxy properties (i.e., $A_{\rm{V}}$ and V$-$J) agree within the median uncertainties. 
Additionally, the subset of the MOSDEF $z\sim2$ targeted sample with U$-$V and V$-$J colors $\geq$ 1.25 mentioned above are not in the $z\sim2$ spectroscopic sample (only 2 galaxies, 0.8\% of the sample, are in this regime). This indicates that the population of the MOSDEF survey that is red, old, massive with a relatively low SFR is mostly left out of the $z\sim2$ spectroscopic sample. For many of these galaxies, a $z_{\rm{MOSFIRE}}$ was not obtained, indicating that this is an incomplete regime within the MOSDEF survey that will need to be addressed with deeper spectroscopy. This region of UVJ space is mostly absent from the KBSS $z\sim2$ targeted sample as well. 

The KBSS $z\sim2$ targeted sample has a higher $t/\tau$ compared to the KBSS $z\sim2$ spectroscopic sample analyzed here. The remaining sample properties agree within the median uncertainties, including $M_{\ast}$, SFR(SED), sSFR(SED), $A_{\rm{V}}$, and UVJ colors. It does not appear that there is an analogous incompleteness within the KBSS catalog to what is observed for massive, low sSFR(SED) galaxies in MOSDEF. However, it is important to note that the KBSS $z\sim2$ targeted sample does not contain the red population that MOSDEF targeted but was ultimately unsuccessful with. This could be attributed to the differences in the MOSDEF and KBSS observing strategies. MOSDEF observed each object for the equivalent of 2 hours per filter in clear conditions in each band while KBSS retained objects for subsequent observation if all lines had not been detected. 

Despite the notable differences in the $z\sim2$ targeted samples described above, the MOSDEF and KBSS $z\sim2$ spectroscopic samples analyzed in the BPT diagrams have mostly similar host galaxy property distributions and median values. Because the targeting strategies of the surveys are so different, the similarity of the MOSDEF and KBSS $z\sim2$ spectroscopic samples reveals the challenges of detecting emission-lines for dusty and/or low sSFR galaxies at $z\sim2$. As the MOSDEF incompleteness is addressed with deeper spectroscopy for massive galaxies, and therefore a MOSDEF $z\sim2$ spectroscopic sample that better reflects its $z\sim2$ targeted sample is assembled,  it is unclear how the relative emission-line ratios of the MOSDEF and KBSS $z\sim2$ spectroscopic samples will change at the massive end.

\section{Summary} \label{sec:summary}

The combination of the MOSDEF and KBSS surveys represents the largest investment of Keck/MOSFIRE observing time, with $\sim$3000 galaxies (about half at $z\sim2$) observed over $\sim$100 nights. 
The statistical weight of the combined MOSDEF+KBSS sample is capable of determining robust trends in galaxy properties and physical relationships in the $z\sim2$ universe analogous to what currently exists locally. 
To understand both how the differences in the survey selection criteria affect the samples and how to account for any biases between the samples when combining them, this study presents a thorough comparison of the $1.9 \leq z \leq 2.7$ MOSDEF and KBSS samples. 
Using consistent broadband emission-line subtracted SED fitting and emission-line fitting, we compared the physical and emission-line properties of the previously published MOSDEF \citep{san18} and KBSS \citep{str17} samples. In addition, we investigate how well these samples represent the $z\sim2$ targeted samples from which they are drawn. Applying the same analysis methodology to both samples eliminates biases stemming from the analysis process, meaning any variations found between the samples are due to systematic differences between the samples due to the survey selection method. 

The main results are as follows:
\begin{enumerate}
    \item Despite the differences in galaxy selection methods for the MOSDEF and KBSS surveys, the $z\sim2$ spectroscopic samples have very similar galaxy properties. KBSS has a lower median $t/\tau$ of the stellar population assuming a delayed-$\tau$ star formation model compared to MOSDEF, but the medians for all other galaxy properties investigated ($M_{\ast}$, SFR(SED), sSFR(SED), $A_{\rm{V}}$, and UVJ colors) agree within the uncertainties. Additionally, the normalized stacked SEDs are very similar as well. The most notable sample difference is that there are 50 KBSS galaxies (13.6\% of the sample) that are very young ($\log_{10}(t/\tau) \leq-2.0$) with intense star formation ($\log_{10}$(sSFR(SED)/yr$^{-1}$) $\geq-8.0$) compared to only 1 such galaxy within the MOSDEF sample.
    \item The MOSDEF and KBSS $z\sim2$ spectroscopic samples have offsets, respectively, of 0.12 $\pm$ 0.02 and 0.10 $\pm$ 0.02 dex from the local SDSS sequence on the [N~\textsc{II}] BPT diagram using the emission-line fitting assumptions from the MOSDEF code. These offsets change to 0.15 $\pm$ 0.02 and 0.12 $\pm$ 0.02 dex, respectively, from the local SDSS sequence using the emission-line fitting assumptions from the KBSS {\tt MOSPEC} code. The binned medians are also more similar on the [S~\textsc{II}] BPT diagram compared to previous MOSDEF and KBSS studies. Because the two samples have similar patterns in their galaxy properties, this agreement on the [N~\textsc{II}] BPT diagram can be attributed to the utilization of consistent emission-line fitting. Updated methods for estimating stellar Balmer absorption resulted in smaller H$\beta$ flux estimates (therefore higher O3) for the MOSDEF sample compared to previous studies. Fitting the KBSS sample with the MOSDEF emission-line fitting code resulted in smaller [N~\textsc{II}]$\lambda$6585 and [S~\textsc{II}]$\lambda\lambda$6718,6733 flux estimates (therefore smaller N2 and S2) compared to past KBSS studies that used the KBSS {\tt MOSPEC} code. 
    \item The MOSDEF $z\sim2$ targeted sample has a larger median $M_{\ast}$, smaller SFR(SED) and sSFR(SED) medians, and a redder U$-$V compared to the KBSS $z\sim2$ targeted sample. These sample differences are highly robust ($>4\sigma$). Additionally, each survey occupies regions of parameter spaces that are complementary to those spanned by the other survey. In particular, 59 galaxies (7.5\%) in the MOSDEF $z\sim2$ targeted sample are red, displaying U$-$V and V$-$J colors $\geq$ 1.25 while only 18 galaxies (2.1\%) in the KBSS $z\sim2$ targeted sample are in this regime. On the other hand, 106 galaxies (12.5\%) in the KBSS $z\sim2$ targeted sample have $\log_{10}(t/\tau) \leq-2.0$ and $\log_{10}$(sSFR(SED)/yr$^{-1}$) $\geq -8.0$ compared to only 14 galaxies (1.8\%) in the MODSEF $z\sim2$ targeted sample. These differences indicate that the MOSDEF $z\sim2$ targeted sample contains more massive galaxies with low sSFR(SED) and KBSS includes more young galaxies with intense star formation. 
    \item While the KBSS $z\sim2$ spectroscopic sample reflects its $z\sim2$ targeted sample very well, there are some differences between the MOSDEF $z\sim2$ spectroscopic sample and its $z\sim2$ targeted sample. There is a population of the MOSDEF $z\sim2$ targeted sample that is red, old, massive (i.e., large U$-$V, V$-$J, $M_{\ast}$, and $t/\tau$) with a relatively low SFR that is not present in the $z\sim2$ spectroscopic sample. A $z_{\rm{MOSFIRE}}$ was not obtained for many of these galaxies, indicating that this is an incomplete regime in the MOSDEF survey. Deeper spectroscopy for these galaxies will be required to fill in this gap. 
\end{enumerate}

The results of this study also highlight the systematic uncertainties that arise based on how the measurements are performed on spectroscopic data of faint galaxies. The fact that the previously established difference in [N~\textsc{II}] BPT offset between the MOSDEF and KBSS surveys can in large part be attributed to how stellar Balmer absorption corrections and weak emission-line (e.g., [N~\textsc{II}]$\lambda$6585 and [S~\textsc{II}]$\lambda\lambda$6718,6733) fitting are performed, makes this abundantly clear. 
The different assumptions in the MOSDEF and KBSS fitting codes were both reasonable given the S/N of the spectra for each survey. 
However, the uncertainty that arises from minor differences in data analysis methods can be extremely difficult to estimate. 
Until standard methods to analyze basic quantities are adopted and commonly used (if that is even possible given that one's approach to fitting emission-lines depends on the spectral S/N), it is important to consider how minor differences in the data analysis process can affect results from different studies. 

With the MOSDEF and KBSS samples now better understood, it is possible to combine them to form a $z\sim2$ sample of unprecedented size. Future work will involve utilizing this joint sample to construct galaxy scaling relations for the largest rest-optical spectroscopic sample at $z\sim2$. 
Currently, there is no consensus on the redshift dependent offset of the mass-metallicity relationship (MZR), or if evolution exists between local and $z\sim1-3$ galaxies in the ``fundamental metallicity relationship'' (FMR). In particular, we will investigate the evolution of the mass-metallicity and fundamental metallicity relationships \citep[e.g.,][]{man10,san21,str21} using the joint MOSDEF-KBSS sample, enabling the most statistically robust characterization of these relationships at $z\sim2$.

In this study, we have shown that for large samples spanning multiple orders of magnitude in $M_{\ast}$ and SFR(SED), the median offset of $z\sim2$ galaxies perpendicular to the local sequence is approximately $0.10-0.15$ dex. A challenging systematic uncertainty to quantify and minimize moving forward is that stemming from analysis methods, as discussed above. However, the existence of the offset between $z\sim2$ and $z\sim0$ galaxies in the [N~\textsc{II}] BPT diagram persists regardless of the analysis method used, and therefore must be reckoned with as we seek to describe the evolution of rest-optical emission properties of galaxies at high redshift.

\section*{Acknowledgements}
We acknowledge support from NSF AAG grants AST1312780, 1312547, 1312764, 1313171, 1313472, 2009313, 2009085, and 2009278; grant AR-13907 from the Space Telescope Science Institute; and grant NNX16AF54G from the NASA ADAP program. We also acknowledge a NASA contract supporting the ``WFIRST Extragalactic Potential Observations (EXPO) Science Investigation Team'' (15-WFIRST15-0004), administered by GSFC. Support for this work was also provided through the NASA Hubble Fellowship grant \#HST-HF2-51469.001-A awarded by the Space Telescope Science Institute, which is operated by the Association of Universities for Research in Astronomy, Incorporated, under NASA contract NAS5-26555.
We thank the 3D-\textit{HST} collaboration, who provided spectroscopic and photometric catalogs used to select MOSDEF targets and to derive stellar population parameters.  We acknowledge the First Carnegie Symposium in Honor of Leonard Searle for useful information and discussions that benefited this work.  
This research made use of Astropy,\footnote{http://www.astropy.org} a community-developed core Python package for Astronomy \citep{ast13, ast18}.  Finally, we wish to extend special thanks to those of Hawaiian ancestry on whose sacred mountain we are privileged to be guests. 

\section*{Data Availability}
The data underlying this article will be shared on reasonable request to the corresponding author.

\vspace{5mm}
\noindent
\textit{Facilities}: \textit{Keck}/MOSFIRE, \textit{Keck}/LRIS, \textit{SDSS}

\noindent
\textit{Software}: Astropy \citep{ast13, ast18}, Corner \citep{for16}, IPython \citep{per07}, IRAF \citep{tod86, tod93}, Matplotlib \citep{hun07}, NumPy \citep{van11, har20}, Pandas \citep{pandas20}, SciPy \citep{oli07, mil11, vir20}

\bibliographystyle{mnras}
\bibliography{references}

\begin{thebibliography}{}
\makeatletter
\relax
\def\mn@urlcharsother{\let\do\@makeother \do\$\do\&\do\#\do\^\do\_\do\%\do\~}
\def\mn@doi{\begingroup\mn@urlcharsother \@ifnextchar [ {\mn@doi@}
  {\mn@doi@[]}}
\def\mn@doi@[#1]#2{\def\@tempa{#1}\ifx\@tempa\@empty \href
  {http://dx.doi.org/#2} {doi:#2}\else \href {http://dx.doi.org/#2} {#1}\fi
  \endgroup}
\def\mn@eprint#1#2{\mn@eprint@#1:#2::\@nil}
\def\mn@eprint@arXiv#1{\href {http://arxiv.org/abs/#1} {{\tt arXiv:#1}}}
\def\mn@eprint@dblp#1{\href {http://dblp.uni-trier.de/rec/bibtex/#1.xml}
  {dblp:#1}}
\def\mn@eprint@#1:#2:#3:#4\@nil{\def\@tempa {#1}\def\@tempb {#2}\def\@tempc
  {#3}\ifx \@tempc \@empty \let \@tempc \@tempb \let \@tempb \@tempa \fi \ifx
  \@tempb \@empty \def\@tempb {arXiv}\fi \@ifundefined
  {mn@eprint@\@tempb}{\@tempb:\@tempc}{\expandafter \expandafter \csname
  mn@eprint@\@tempb\endcsname \expandafter{\@tempc}}}

\bibitem[\protect\citeauthoryear{{Abazajian} et~al.,}{{Abazajian}
  et~al.}{2009}]{aba09}
{Abazajian} K.~N.,  et~al., 2009, \mn@doi [\apjs]
  {10.1088/0067-0049/182/2/543}, \href
  {https://ui.adsabs.harvard.edu/abs/2009ApJS..182..543A} {182, 543}

\bibitem[\protect\citeauthoryear{{Adelberger}, {Steidel}, {Shapley}, {Hunt},
  {Erb}, {Reddy}  \& {Pettini}}{{Adelberger} et~al.}{2004}]{ade04}
{Adelberger} K.~L.,  {Steidel} C.~C.,  {Shapley} A.~E.,  {Hunt} M.~P.,  {Erb}
  D.~K.,  {Reddy} N.~A.,   {Pettini} M.,  2004, \mn@doi [\apj]
  {10.1086/383221}, \href
  {https://ui.adsabs.harvard.edu/abs/2004ApJ...607..226A} {607, 226}

\bibitem[\protect\citeauthoryear{{Andrews} \& {Martini}}{{Andrews} \&
  {Martini}}{2013}]{and13}
{Andrews} B.~H.,  {Martini} P.,  2013, \mn@doi [\apj]
  {10.1088/0004-637X/765/2/140}, \href
  {https://ui.adsabs.harvard.edu/abs/2013ApJ...765..140A} {765, 140}

\bibitem[\protect\citeauthoryear{{Asplund}, {Grevesse}, {Sauval}  \&
  {Scott}}{{Asplund} et~al.}{2009}]{asp09}
{Asplund} M.,  {Grevesse} N.,  {Sauval} A.~J.,   {Scott} P.,  2009, \mn@doi
  [\araa] {10.1146/annurev.astro.46.060407.145222}, \href
  {https://ui.adsabs.harvard.edu/abs/2009ARA&A..47..481A} {47, 481}

\bibitem[\protect\citeauthoryear{{Astropy Collaboration} et~al.,}{{Astropy
  Collaboration} et~al.}{2013}]{ast13}
{Astropy Collaboration} et~al., 2013, \mn@doi [\aap]
  {10.1051/0004-6361/201322068}, \href
  {https://ui.adsabs.harvard.edu/abs/2013A&A...558A..33A} {558, A33}

\bibitem[\protect\citeauthoryear{{Astropy Collaboration} et~al.,}{{Astropy
  Collaboration} et~al.}{2018}]{ast18}
{Astropy Collaboration} et~al., 2018, \mn@doi [\aj] {10.3847/1538-3881/aabc4f},
  \href {https://ui.adsabs.harvard.edu/abs/2018AJ....156..123A} {156, 123}

\bibitem[\protect\citeauthoryear{{Azadi} et~al.,}{{Azadi} et~al.}{2017}]{aza17}
{Azadi} M.,  et~al., 2017, \mn@doi [\apj] {10.3847/1538-4357/835/1/27}, \href
  {https://ui.adsabs.harvard.edu/abs/2017ApJ...835...27A} {835, 27}

\bibitem[\protect\citeauthoryear{{Baldwin}, {Phillips}  \&
  {Terlevich}}{{Baldwin} et~al.}{1981}]{bal81}
{Baldwin} J.~A.,  {Phillips} M.~M.,   {Terlevich} R.,  1981, \mn@doi [\pasp]
  {10.1086/130766}, \href
  {https://ui.adsabs.harvard.edu/abs/1981PASP...93....5B} {93, 5}

\bibitem[\protect\citeauthoryear{{Bian}, {Kewley}  \& {Dopita}}{{Bian}
  et~al.}{2018}]{bia18}
{Bian} F.,  {Kewley} L.~J.,   {Dopita} M.~A.,  2018, \mn@doi [\apj]
  {10.3847/1538-4357/aabd74}, \href
  {https://ui.adsabs.harvard.edu/abs/2018ApJ...859..175B} {859, 175}

\bibitem[\protect\citeauthoryear{{Brinchmann}, {Pettini}  \&
  {Charlot}}{{Brinchmann} et~al.}{2008}]{bri08}
{Brinchmann} J.,  {Pettini} M.,   {Charlot} S.,  2008, \mn@doi [\mnras]
  {10.1111/j.1365-2966.2008.12914.x}, \href
  {https://ui.adsabs.harvard.edu/abs/2008MNRAS.385..769B} {385, 769}

\bibitem[\protect\citeauthoryear{{Bruzual} \& {Charlot}}{{Bruzual} \&
  {Charlot}}{2003}]{bru03}
{Bruzual} G.,  {Charlot} S.,  2003, \mn@doi [\mnras]
  {10.1046/j.1365-8711.2003.06897.x}, \href
  {https://ui.adsabs.harvard.edu/abs/2003MNRAS.344.1000B} {344, 1000}

\bibitem[\protect\citeauthoryear{{Calzetti}, {Armus}, {Bohlin}, {Kinney},
  {Koornneef}  \& {Storchi-Bergmann}}{{Calzetti} et~al.}{2000}]{cal00}
{Calzetti} D.,  {Armus} L.,  {Bohlin} R.~C.,  {Kinney} A.~L.,  {Koornneef} J.,
   {Storchi-Bergmann} T.,  2000, \mn@doi [\apj] {10.1086/308692}, \href
  {https://ui.adsabs.harvard.edu/abs/2000ApJ...533..682C} {533, 682}

\bibitem[\protect\citeauthoryear{{Chabrier}}{{Chabrier}}{2003}]{cha03}
{Chabrier} G.,  2003, \mn@doi [\pasp] {10.1086/376392}, \href
  {https://ui.adsabs.harvard.edu/abs/2003PASP..115..763C} {115, 763}

\bibitem[\protect\citeauthoryear{{Coil} et~al.,}{{Coil} et~al.}{2015}]{coi15}
{Coil} A.~L.,  et~al., 2015, \mn@doi [\apj] {10.1088/0004-637X/801/1/35}, \href
  {https://ui.adsabs.harvard.edu/abs/2015ApJ...801...35C} {801, 35}

\bibitem[\protect\citeauthoryear{{Conroy} \& {Gunn}}{{Conroy} \&
  {Gunn}}{2010}]{con10}
{Conroy} C.,  {Gunn} J.~E.,  2010, \mn@doi [\apj]
  {10.1088/0004-637X/712/2/833}, \href
  {https://ui.adsabs.harvard.edu/abs/2010ApJ...712..833C} {712, 833}

\bibitem[\protect\citeauthoryear{{Du} et~al.,}{{Du} et~al.}{2018}]{du18}
{Du} X.,  et~al., 2018, \mn@doi [\apj] {10.3847/1538-4357/aabfcf}, \href
  {https://ui.adsabs.harvard.edu/abs/2018ApJ...860...75D} {860, 75}

\bibitem[\protect\citeauthoryear{{Erb}, {Shapley}, {Pettini}, {Steidel},
  {Reddy}  \& {Adelberger}}{{Erb} et~al.}{2006a}]{erb06a}
{Erb} D.~K.,  {Shapley} A.~E.,  {Pettini} M.,  {Steidel} C.~C.,  {Reddy} N.~A.,
    {Adelberger} K.~L.,  2006a, \mn@doi [\apj] {10.1086/503623}, \href
  {https://ui.adsabs.harvard.edu/abs/2006ApJ...644..813E} {644, 813}

\bibitem[\protect\citeauthoryear{{Erb}, {Steidel}, {Shapley}, {Pettini},
  {Reddy}  \& {Adelberger}}{{Erb} et~al.}{2006b}]{erb06b}
{Erb} D.~K.,  {Steidel} C.~C.,  {Shapley} A.~E.,  {Pettini} M.,  {Reddy} N.~A.,
    {Adelberger} K.~L.,  2006b, \mn@doi [\apj] {10.1086/505341}, \href
  {https://ui.adsabs.harvard.edu/abs/2006ApJ...647..128E} {647, 128}

\bibitem[\protect\citeauthoryear{{Foreman-Mackey}}{{Foreman-Mackey}}{2016}]{for16}
{Foreman-Mackey} D.,  2016, \mn@doi [The Journal of Open Source Software]
  {10.21105/joss.00024}, \href
  {https://ui.adsabs.harvard.edu/abs/2016JOSS....1...24F} {1, 24}

\bibitem[\protect\citeauthoryear{{Freeman} et~al.,}{{Freeman}
  et~al.}{2019}]{fre19}
{Freeman} W.~R.,  et~al., 2019, \mn@doi [\apj] {10.3847/1538-4357/ab0655},
  \href {https://ui.adsabs.harvard.edu/abs/2019ApJ...873..102F} {873, 102}

\bibitem[\protect\citeauthoryear{{Grogin} et~al.,}{{Grogin}
  et~al.}{2011}]{gro11}
{Grogin} N.~A.,  et~al., 2011, \mn@doi [\apjs] {10.1088/0067-0049/197/2/35},
  \href {https://ui.adsabs.harvard.edu/abs/2011ApJS..197...35G} {197, 35}

\bibitem[\protect\citeauthoryear{{Hainline}, {Shapley}, {Greene}  \&
  {Steidel}}{{Hainline} et~al.}{2011}]{hai11}
{Hainline} K.~N.,  {Shapley} A.~E.,  {Greene} J.~E.,   {Steidel} C.~C.,  2011,
  \mn@doi [\apj] {10.1088/0004-637X/733/1/31}, \href
  {https://ui.adsabs.harvard.edu/abs/2011ApJ...733...31H} {733, 31}

\bibitem[\protect\citeauthoryear{Harris et~al.,}{Harris et~al.}{2020}]{har20}
Harris C.~R.,  et~al., 2020, \mn@doi [Nature] {10.1038/s41586-020-2649-2}, 585,
  357–362

\bibitem[\protect\citeauthoryear{{Hunter}}{{Hunter}}{2007}]{hun07}
{Hunter} J.~D.,  2007, Computing in Science \& Engineering, 9, 90

\bibitem[\protect\citeauthoryear{{Juneau} et~al.,}{{Juneau}
  et~al.}{2014}]{jun14}
{Juneau} S.,  et~al., 2014, \mn@doi [\apj] {10.1088/0004-637X/788/1/88}, \href
  {https://ui.adsabs.harvard.edu/abs/2014ApJ...788...88J} {788, 88}

\bibitem[\protect\citeauthoryear{{Kashino} et~al.,}{{Kashino}
  et~al.}{2019}]{kas19}
{Kashino} D.,  et~al., 2019, \mn@doi [\apjs] {10.3847/1538-4365/ab06c4}, \href
  {https://ui.adsabs.harvard.edu/abs/2019ApJS..241...10K} {241, 10}

\bibitem[\protect\citeauthoryear{{Kauffmann} et~al.,}{{Kauffmann}
  et~al.}{2003}]{kau03}
{Kauffmann} G.,  et~al., 2003, \mn@doi [\mnras]
  {10.1111/j.1365-2966.2003.07154.x}, \href
  {https://ui.adsabs.harvard.edu/abs/2003MNRAS.346.1055K} {346, 1055}

\bibitem[\protect\citeauthoryear{{Kewley}, {Dopita}, {Leitherer}, {Dav{\'e}},
  {Yuan}, {Allen}, {Groves}  \& {Sutherland}}{{Kewley} et~al.}{2013}]{kew13}
{Kewley} L.~J.,  {Dopita} M.~A.,  {Leitherer} C.,  {Dav{\'e}} R.,  {Yuan} T.,
  {Allen} M.,  {Groves} B.,   {Sutherland} R.,  2013, \mn@doi [\apj]
  {10.1088/0004-637X/774/2/100}, \href
  {https://ui.adsabs.harvard.edu/abs/2013ApJ...774..100K} {774, 100}

\bibitem[\protect\citeauthoryear{{Koekemoer} et~al.,}{{Koekemoer}
  et~al.}{2011}]{koe11}
{Koekemoer} A.~M.,  et~al., 2011, \mn@doi [\apjs] {10.1088/0067-0049/197/2/36},
  \href {https://ui.adsabs.harvard.edu/abs/2011ApJS..197...36K} {197, 36}

\bibitem[\protect\citeauthoryear{{Kriek}, {van Dokkum}, {Labb{\'e}}, {Franx},
  {Illingworth}, {Marchesini}  \& {Quadri}}{{Kriek} et~al.}{2009}]{kri09}
{Kriek} M.,  {van Dokkum} P.~G.,  {Labb{\'e}} I.,  {Franx} M.,  {Illingworth}
  G.~D.,  {Marchesini} D.,   {Quadri} R.~F.,  2009, \mn@doi [\apj]
  {10.1088/0004-637X/700/1/221}, \href
  {https://ui.adsabs.harvard.edu/abs/2009ApJ...700..221K} {700, 221}

\bibitem[\protect\citeauthoryear{{Kriek} et~al.,}{{Kriek} et~al.}{2015}]{kri15}
{Kriek} M.,  et~al., 2015, \mn@doi [\apjs] {10.1088/0067-0049/218/2/15}, \href
  {https://ui.adsabs.harvard.edu/abs/2015ApJS..218...15K} {218, 15}

\bibitem[\protect\citeauthoryear{{Law}, {Steidel}, {Shapley}, {Nagy}, {Reddy}
  \& {Erb}}{{Law} et~al.}{2012}]{law12}
{Law} D.~R.,  {Steidel} C.~C.,  {Shapley} A.~E.,  {Nagy} S.~R.,  {Reddy} N.~A.,
    {Erb} D.~K.,  2012, \mn@doi [\apj] {10.1088/0004-637X/745/1/85}, \href
  {https://ui.adsabs.harvard.edu/abs/2012ApJ...745...85L} {745, 85}

\bibitem[\protect\citeauthoryear{{Leung} et~al.,}{{Leung} et~al.}{2017}]{leu17}
{Leung} G. C.~K.,  et~al., 2017, \mn@doi [\apj] {10.3847/1538-4357/aa9024},
  \href {https://ui.adsabs.harvard.edu/abs/2017ApJ...849...48L} {849, 48}

\bibitem[\protect\citeauthoryear{{Liu}, {Shapley}, {Coil}, {Brinchmann}  \&
  {Ma}}{{Liu} et~al.}{2008}]{liu08}
{Liu} X.,  {Shapley} A.~E.,  {Coil} A.~L.,  {Brinchmann} J.,   {Ma} C.-P.,
  2008, \mn@doi [\apj] {10.1086/529030}, \href
  {https://ui.adsabs.harvard.edu/abs/2008ApJ...678..758L} {678, 758}

\bibitem[\protect\citeauthoryear{{Madau} \& {Dickinson}}{{Madau} \&
  {Dickinson}}{2014}]{mad14}
{Madau} P.,  {Dickinson} M.,  2014, \mn@doi [\araa]
  {10.1146/annurev-astro-081811-125615}, \href
  {https://ui.adsabs.harvard.edu/abs/2014ARA&A..52..415M} {52, 415}

\bibitem[\protect\citeauthoryear{{Mannucci}, {Cresci}, {Maiolino}, {Marconi}
  \& {Gnerucci}}{{Mannucci} et~al.}{2010}]{man10}
{Mannucci} F.,  {Cresci} G.,  {Maiolino} R.,  {Marconi} A.,   {Gnerucci} A.,
  2010, \mn@doi [\mnras] {10.1111/j.1365-2966.2010.17291.x}, \href
  {https://ui.adsabs.harvard.edu/abs/2010MNRAS.408.2115M} {408, 2115}

\bibitem[\protect\citeauthoryear{{Masters} et~al.,}{{Masters}
  et~al.}{2014}]{mas14}
{Masters} D.,  et~al., 2014, \mn@doi [\apj] {10.1088/0004-637X/785/2/153},
  \href {https://ui.adsabs.harvard.edu/abs/2014ApJ...785..153M} {785, 153}

\bibitem[\protect\citeauthoryear{{McLean} et~al.,}{{McLean}
  et~al.}{2012}]{mcl12}
{McLean} I.~S.,  et~al., 2012, in Ground-based and Airborne Instrumentation for
  Astronomy IV. p. 84460J, \mn@doi{10.1117/12.924794}

\bibitem[\protect\citeauthoryear{{Millman} \& {Aivazis}}{{Millman} \&
  {Aivazis}}{2011}]{mil11}
{Millman} K.~J.,  {Aivazis} M.,  2011, Computing in Science \& Engineering, 13,
  9

\bibitem[\protect\citeauthoryear{{Momcheva} et~al.,}{{Momcheva}
  et~al.}{2016}]{mom16}
{Momcheva} I.~G.,  et~al., 2016, \mn@doi [\apjs] {10.3847/0067-0049/225/2/27},
  \href {https://ui.adsabs.harvard.edu/abs/2016ApJS..225...27M} {225, 27}

\bibitem[\protect\citeauthoryear{{Mostardi}, {Shapley}, {Steidel}, {Trainor},
  {Reddy}  \& {Siana}}{{Mostardi} et~al.}{2015}]{mos15}
{Mostardi} R.~E.,  {Shapley} A.~E.,  {Steidel} C.~C.,  {Trainor} R.~F.,
  {Reddy} N.~A.,   {Siana} B.,  2015, \mn@doi [\apj]
  {10.1088/0004-637X/810/2/107}, \href
  {https://ui.adsabs.harvard.edu/abs/2015ApJ...810..107M} {810, 107}

\bibitem[\protect\citeauthoryear{{Oliphant}}{{Oliphant}}{2007}]{oli07}
{Oliphant} T.~E.,  2007, Computing in Science \& Engineering, 9, 10

\bibitem[\protect\citeauthoryear{{Perez} \& {Granger}}{{Perez} \&
  {Granger}}{2007}]{per07}
{Perez} F.,  {Granger} B.~E.,  2007, Computing in Science \& Engineering, 9, 21

\bibitem[\protect\citeauthoryear{{Persson} et~al.,}{{Persson}
  et~al.}{2013}]{per13}
{Persson} S.~E.,  et~al., 2013, \mn@doi [\pasp] {10.1086/671164}, \href
  {https://ui.adsabs.harvard.edu/abs/2013PASP..125..654P} {125, 654}

\bibitem[\protect\citeauthoryear{{Pettini} \& {Pagel}}{{Pettini} \&
  {Pagel}}{2004}]{pet04}
{Pettini} M.,  {Pagel} B. E.~J.,  2004, \mn@doi [\mnras]
  {10.1111/j.1365-2966.2004.07591.x}, \href
  {https://ui.adsabs.harvard.edu/abs/2004MNRAS.348L..59P} {348, L59}

\bibitem[\protect\citeauthoryear{{Reddy}, {Erb}, {Steidel}, {Shapley},
  {Adelberger}  \& {Pettini}}{{Reddy} et~al.}{2005}]{red05}
{Reddy} N.~A.,  {Erb} D.~K.,  {Steidel} C.~C.,  {Shapley} A.~E.,  {Adelberger}
  K.~L.,   {Pettini} M.,  2005, \mn@doi [\apj] {10.1086/444588}, \href
  {https://ui.adsabs.harvard.edu/abs/2005ApJ...633..748R} {633, 748}

\bibitem[\protect\citeauthoryear{{Reddy}, {Steidel}, {Pettini}, {Adelberger},
  {Shapley}, {Erb}  \& {Dickinson}}{{Reddy} et~al.}{2008}]{red08}
{Reddy} N.~A.,  {Steidel} C.~C.,  {Pettini} M.,  {Adelberger} K.~L.,  {Shapley}
  A.~E.,  {Erb} D.~K.,   {Dickinson} M.,  2008, \mn@doi [\apjs]
  {10.1086/521105}, \href
  {https://ui.adsabs.harvard.edu/abs/2008ApJS..175...48R} {175, 48}

\bibitem[\protect\citeauthoryear{{Reddy}, {Pettini}, {Steidel}, {Shapley},
  {Erb}  \& {Law}}{{Reddy} et~al.}{2012}]{red12}
{Reddy} N.~A.,  {Pettini} M.,  {Steidel} C.~C.,  {Shapley} A.~E.,  {Erb} D.~K.,
    {Law} D.~R.,  2012, \mn@doi [\apj] {10.1088/0004-637X/754/1/25}, \href
  {https://ui.adsabs.harvard.edu/abs/2012ApJ...754...25R} {754, 25}

\bibitem[\protect\citeauthoryear{{Reddy} et~al.,}{{Reddy} et~al.}{2015}]{red15}
{Reddy} N.~A.,  et~al., 2015, \mn@doi [\apj] {10.1088/0004-637X/806/2/259},
  \href {https://ui.adsabs.harvard.edu/abs/2015ApJ...806..259R} {806, 259}

\bibitem[\protect\citeauthoryear{{Reddy} et~al.,}{{Reddy} et~al.}{2018}]{red18}
{Reddy} N.~A.,  et~al., 2018, \mn@doi [\apj] {10.3847/1538-4357/aaa3e7}, \href
  {https://ui.adsabs.harvard.edu/abs/2018ApJ...853...56R} {853, 56}

\bibitem[\protect\citeauthoryear{{Reddy} et~al.,}{{Reddy} et~al.}{2021}]{red21}
{Reddy} N.~A.,  et~al., 2021, arXiv e-prints, \href
  {https://ui.adsabs.harvard.edu/abs/2021arXiv210805363R} {p. arXiv:2108.05363}

\bibitem[\protect\citeauthoryear{{Rudie} et~al.,}{{Rudie} et~al.}{2012}]{rud12}
{Rudie} G.~C.,  et~al., 2012, \mn@doi [\apj] {10.1088/0004-637X/750/1/67},
  \href {https://ui.adsabs.harvard.edu/abs/2012ApJ...750...67R} {750, 67}

\bibitem[\protect\citeauthoryear{{Rudie}, {Steidel}, {Pettini}, {Trainor},
  {Strom}, {Hummels}, {Reddy}  \& {Shapley}}{{Rudie} et~al.}{2019}]{rud19}
{Rudie} G.~C.,  {Steidel} C.~C.,  {Pettini} M.,  {Trainor} R.~F.,  {Strom}
  A.~L.,  {Hummels} C.~B.,  {Reddy} N.~A.,   {Shapley} A.~E.,  2019, \mn@doi
  [\apj] {10.3847/1538-4357/ab4255}, \href
  {https://ui.adsabs.harvard.edu/abs/2019ApJ...885...61R} {885, 61}

\bibitem[\protect\citeauthoryear{{Runco} et~al.,}{{Runco} et~al.}{2021}]{run21}
{Runco} J.~N.,  et~al., 2021, \mn@doi [\mnras] {10.1093/mnras/stab119}, \href
  {https://ui.adsabs.harvard.edu/abs/2021MNRAS.502.2600R} {502, 2600}

\bibitem[\protect\citeauthoryear{{Sanders} et~al.,}{{Sanders}
  et~al.}{2015}]{san15}
{Sanders} R.~L.,  et~al., 2015, \mn@doi [\apj] {10.1088/0004-637X/799/2/138},
  \href {https://ui.adsabs.harvard.edu/abs/2015ApJ...799..138S} {799, 138}

\bibitem[\protect\citeauthoryear{{Sanders} et~al.,}{{Sanders}
  et~al.}{2016}]{san16}
{Sanders} R.~L.,  et~al., 2016, \mn@doi [\apj] {10.3847/0004-637X/816/1/23},
  \href {https://ui.adsabs.harvard.edu/abs/2016ApJ...816...23S} {816, 23}

\bibitem[\protect\citeauthoryear{{Sanders} et~al.,}{{Sanders}
  et~al.}{2018}]{san18}
{Sanders} R.~L.,  et~al., 2018, \mn@doi [\apj] {10.3847/1538-4357/aabcbd},
  \href {https://ui.adsabs.harvard.edu/abs/2018ApJ...858...99S} {858, 99}

\bibitem[\protect\citeauthoryear{{Sanders} et~al.,}{{Sanders}
  et~al.}{2020}]{san20b}
{Sanders} R.~L.,  et~al., 2020, \mn@doi [\mnras] {10.1093/mnras/stz3032}, \href
  {https://ui.adsabs.harvard.edu/abs/2020MNRAS.491.1427S} {491, 1427}

\bibitem[\protect\citeauthoryear{{Sanders} et~al.,}{{Sanders}
  et~al.}{2021}]{san21}
{Sanders} R.~L.,  et~al., 2021, \mn@doi [\apj] {10.3847/1538-4357/abf4c1},
  \href {https://ui.adsabs.harvard.edu/abs/2021ApJ...914...19S} {914, 19}

\bibitem[\protect\citeauthoryear{{Shapley}, {Coil}, {Ma}  \& {Bundy}}{{Shapley}
  et~al.}{2005}]{sha05}
{Shapley} A.~E.,  {Coil} A.~L.,  {Ma} C.-P.,   {Bundy} K.,  2005, \mn@doi
  [\apj] {10.1086/497630}, \href
  {https://ui.adsabs.harvard.edu/abs/2005ApJ...635.1006S} {635, 1006}

\bibitem[\protect\citeauthoryear{{Shapley} et~al.,}{{Shapley}
  et~al.}{2015}]{sha15}
{Shapley} A.~E.,  et~al., 2015, \mn@doi [\apj] {10.1088/0004-637X/801/2/88},
  \href {https://ui.adsabs.harvard.edu/abs/2015ApJ...801...88S} {801, 88}

\bibitem[\protect\citeauthoryear{{Shapley} et~al.,}{{Shapley}
  et~al.}{2019}]{sha19}
{Shapley} A.~E.,  et~al., 2019, \mn@doi [\apjl] {10.3847/2041-8213/ab385a},
  \href {https://ui.adsabs.harvard.edu/abs/2019ApJ...881L..35S} {881, L35}

\bibitem[\protect\citeauthoryear{{Shivaei} et~al.,}{{Shivaei}
  et~al.}{2015}]{shi15}
{Shivaei} I.,  et~al., 2015, \mn@doi [\apj] {10.1088/0004-637X/815/2/98}, \href
  {https://ui.adsabs.harvard.edu/abs/2015ApJ...815...98S} {815, 98}

\bibitem[\protect\citeauthoryear{{Skelton} et~al.,}{{Skelton}
  et~al.}{2014}]{ske14}
{Skelton} R.~E.,  et~al., 2014, \mn@doi [\apjs] {10.1088/0067-0049/214/2/24},
  \href {https://ui.adsabs.harvard.edu/abs/2014ApJS..214...24S} {214, 24}

\bibitem[\protect\citeauthoryear{{Steidel}, {Adelberger}, {Shapley}, {Pettini},
  {Dickinson}  \& {Giavalisco}}{{Steidel} et~al.}{2003}]{ste03}
{Steidel} C.~C.,  {Adelberger} K.~L.,  {Shapley} A.~E.,  {Pettini} M.,
  {Dickinson} M.,   {Giavalisco} M.,  2003, \mn@doi [\apj] {10.1086/375772},
  \href {https://ui.adsabs.harvard.edu/abs/2003ApJ...592..728S} {592, 728}

\bibitem[\protect\citeauthoryear{{Steidel}, {Shapley}, {Pettini}, {Adelberger},
  {Erb}, {Reddy}  \& {Hunt}}{{Steidel} et~al.}{2004}]{ste04}
{Steidel} C.~C.,  {Shapley} A.~E.,  {Pettini} M.,  {Adelberger} K.~L.,  {Erb}
  D.~K.,  {Reddy} N.~A.,   {Hunt} M.~P.,  2004, \mn@doi [\apj]
  {10.1086/381960}, \href
  {https://ui.adsabs.harvard.edu/abs/2004ApJ...604..534S} {604, 534}

\bibitem[\protect\citeauthoryear{{Steidel} et~al.,}{{Steidel}
  et~al.}{2014}]{ste14}
{Steidel} C.~C.,  et~al., 2014, \mn@doi [\apj] {10.1088/0004-637X/795/2/165},
  \href {https://ui.adsabs.harvard.edu/abs/2014ApJ...795..165S} {795, 165}

\bibitem[\protect\citeauthoryear{{Steidel}, {Strom}, {Pettini}, {Rudie},
  {Reddy}  \& {Trainor}}{{Steidel} et~al.}{2016}]{ste16}
{Steidel} C.~C.,  {Strom} A.~L.,  {Pettini} M.,  {Rudie} G.~C.,  {Reddy} N.~A.,
    {Trainor} R.~F.,  2016, \mn@doi [\apj] {10.3847/0004-637X/826/2/159}, \href
  {https://ui.adsabs.harvard.edu/abs/2016ApJ...826..159S} {826, 159}

\bibitem[\protect\citeauthoryear{{Strom}, {Steidel}, {Rudie}, {Trainor},
  {Pettini}  \& {Reddy}}{{Strom} et~al.}{2017}]{str17}
{Strom} A.~L.,  {Steidel} C.~C.,  {Rudie} G.~C.,  {Trainor} R.~F.,  {Pettini}
  M.,   {Reddy} N.~A.,  2017, \mn@doi [\apj] {10.3847/1538-4357/836/2/164},
  \href {https://ui.adsabs.harvard.edu/abs/2017ApJ...836..164S} {836, 164}

\bibitem[\protect\citeauthoryear{{Strom}, {Steidel}, {Rudie}, {Trainor}  \&
  {Pettini}}{{Strom} et~al.}{2018}]{str18}
{Strom} A.~L.,  {Steidel} C.~C.,  {Rudie} G.~C.,  {Trainor} R.~F.,   {Pettini}
  M.,  2018, \mn@doi [\apj] {10.3847/1538-4357/aae1a5}, \href
  {https://ui.adsabs.harvard.edu/abs/2018ApJ...868..117S} {868, 117}

\bibitem[\protect\citeauthoryear{{Strom}, {Rudie}, {Steidel}  \&
  {Trainor}}{{Strom} et~al.}{2021}]{str21}
{Strom} A.~L.,  {Rudie} G.~C.,  {Steidel} C.~C.,   {Trainor} R.~F.,  2021,
  arXiv e-prints, \href {https://ui.adsabs.harvard.edu/abs/2021arXiv211106410S}
  {p. arXiv:2111.06410}

\bibitem[\protect\citeauthoryear{{Tody}}{{Tody}}{1986}]{tod86}
{Tody} D.,  1986, in {Crawford} D.~L.,  ed.,  Society of Photo-Optical
  Instrumentation Engineers (SPIE) Conference Series Vol. 627, Instrumentation
  in astronomy VI. p.~733, \mn@doi{10.1117/12.968154}

\bibitem[\protect\citeauthoryear{{Tody}}{{Tody}}{1993}]{tod93}
{Tody} D.,  1993, in {Hanisch} R.~J.,  {Brissenden} R.~J.~V.,   {Barnes} J.,
  eds,  Astronomical Society of the Pacific Conference Series Vol. 52,
  Astronomical Data Analysis Software and Systems II. p.~173

\bibitem[\protect\citeauthoryear{{Topping}, {Shapley}, {Reddy}, {Sanders},
  {Coil}, {Kriek}, {Mobasher}  \& {Siana}}{{Topping} et~al.}{2020}]{top20a}
{Topping} M.~W.,  {Shapley} A.~E.,  {Reddy} N.~A.,  {Sanders} R.~L.,  {Coil}
  A.~L.,  {Kriek} M.,  {Mobasher} B.,   {Siana} B.,  2020, \mn@doi [\mnras]
  {10.1093/mnras/staa1410}, \href
  {https://ui.adsabs.harvard.edu/abs/2020MNRAS.495.4430T} {495, 4430}

\bibitem[\protect\citeauthoryear{{Tremonti} et~al.,}{{Tremonti}
  et~al.}{2004}]{tre04}
{Tremonti} C.~A.,  et~al., 2004, \mn@doi [\apj] {10.1086/423264}, \href
  {https://ui.adsabs.harvard.edu/abs/2004ApJ...613..898T} {613, 898}

\bibitem[\protect\citeauthoryear{{Turner}, {Schaye}, {Steidel}, {Rudie}  \&
  {Strom}}{{Turner} et~al.}{2014}]{tur14}
{Turner} M.~L.,  {Schaye} J.,  {Steidel} C.~C.,  {Rudie} G.~C.,   {Strom}
  A.~L.,  2014, \mn@doi [\mnras] {10.1093/mnras/stu1801}, \href
  {https://ui.adsabs.harvard.edu/abs/2014MNRAS.445..794T} {445, 794}

\bibitem[\protect\citeauthoryear{{Veilleux} \& {Osterbrock}}{{Veilleux} \&
  {Osterbrock}}{1987}]{vei87}
{Veilleux} S.,  {Osterbrock} D.~E.,  1987, \mn@doi [\apjs] {10.1086/191166},
  \href {https://ui.adsabs.harvard.edu/abs/1987ApJS...63..295V} {63, 295}

\bibitem[\protect\citeauthoryear{{Virtanen} et~al.,}{{Virtanen}
  et~al.}{2020}]{vir20}
{Virtanen} P.,  et~al., 2020, \mn@doi [Nature Methods]
  {10.1038/s41592-019-0686-2}, \href
  {https://ui.adsabs.harvard.edu/abs/2020NatMe..17..261V} {17, 261}

\bibitem[\protect\citeauthoryear{{Williams}, {Quadri}, {Franx}, {van Dokkum}
  \& {Labb{\'e}}}{{Williams} et~al.}{2009}]{wil09}
{Williams} R.~J.,  {Quadri} R.~F.,  {Franx} M.,  {van Dokkum} P.,   {Labb{\'e}}
  I.,  2009, \mn@doi [\apj] {10.1088/0004-637X/691/2/1879}, \href
  {https://ui.adsabs.harvard.edu/abs/2009ApJ...691.1879W} {691, 1879}

\bibitem[\protect\citeauthoryear{{Wright}, {Larkin}, {Graham}  \&
  {Ma}}{{Wright} et~al.}{2010}]{wri10}
{Wright} S.~A.,  {Larkin} J.~E.,  {Graham} J.~R.,   {Ma} C.-P.,  2010, \mn@doi
  [\apj] {10.1088/0004-637X/711/2/1291}, \href
  {https://ui.adsabs.harvard.edu/abs/2010ApJ...711.1291W} {711, 1291}

\bibitem[\protect\citeauthoryear{{Yeh}, {Verdolini}, {Krumholz}, {Matzner}  \&
  {Tielens}}{{Yeh} et~al.}{2013}]{yeh13}
{Yeh} S. C.~C.,  {Verdolini} S.,  {Krumholz} M.~R.,  {Matzner} C.~D.,
  {Tielens} A. G.~G.~M.,  2013, \mn@doi [\apj] {10.1088/0004-637X/769/1/11},
  \href {https://ui.adsabs.harvard.edu/abs/2013ApJ...769...11Y} {769, 11}

\bibitem[\protect\citeauthoryear{{York} et~al.,}{{York} et~al.}{2000}]{yor00}
{York} D.~G.,  et~al., 2000, \mn@doi [\aj] {10.1086/301513}, \href
  {https://ui.adsabs.harvard.edu/abs/2000AJ....120.1579Y} {120, 1579}

\bibitem[\protect\citeauthoryear{pandas~development team}{pandas~development
  team}{2020}]{pandas20}
pandas~development team T.,  2020, pandas-dev/pandas: Pandas,
  \mn@doi{10.5281/zenodo.3509134}, \url
  {https://doi.org/10.5281/zenodo.3509134}

\bibitem[\protect\citeauthoryear{{van der Walt}, {Colbert}  \&
  {Varoquaux}}{{van der Walt} et~al.}{2011}]{van11}
{van der Walt} S.,  {Colbert} S.~C.,   {Varoquaux} G.,  2011, Computing in
  Science \& Engineering, 13, 22

\makeatother
\end{thebibliography}

\appendix

\section{CSF Modeling} \label{sec:csf_models}

In this Appendix, we investigate the galaxy properties of the MOSDEF and KBSS $z\sim2$ spectroscopic and $z\sim2$ targeted samples assuming a CSF history instead of a delayed-$\tau$ star formation history. 
The SEDs are fit using the stellar population templates from \citet{bru03} assuming a \citet{cha03} IMF. The \citet{bru03} templates do not account for emission-lines or nebular continuum emission. 
Following previous studies \citep{red18, du18}, we adopt two combinations of extinction curves and metallicity: a 1.4$Z_{\odot}$ metallicity with the \citet{cal00} attenuation curve (hereafter referred to as ``1.4 $Z_{\odot}$+Calzetti'', and a 0.28$Z_{\odot}$ metallicity with the Small Magellanic Cloud (SMC) attenuation curve ``0.28 $Z_{\odot}$+SMC''. 
Additionally, the code allows stellar population ages between 10 Myr and 5 Gyr and E(B-V) values between 0.0 and 0.6. 

We employ several constraints on the models based on past studies of $z\sim2$ galaxies. 
Following \citet{du18}, we use the 1.4 $Z_{\odot}$+Calzetti grid for galaxies with log$_{10}(M_{\ast}/M_{\odot}) \geq 10.45$ and the 0.28 $Z_{\odot}$+SMC grid for galaxies with log$_{10}(M_{\ast}/M_{\odot}) \leq 10.45$. This choice in assumed metallicity based on stellar mass is consistent with the mass-metallicity relationship, MZR, (e.g., \citealt{tre04, erb06a, and13, ste14, san15}).
We set a minimum stellar population age of 50 Myr, which is based on typical dynamical timescales at $z\sim2$ \citep{red12}. 

We probe the same set of galaxy properties as with the FAST delayed-$\tau$ models: $M_{\ast}$, SFR(SED), sSFR(SED), age of the stellar population (now parameterized by $t$ instead of $t/\tau$), $A_{\rm{V}}$, and UVJ colors. 
Once again, we used IRAF/{\it sbands} to estimate the UVJ colors from the best-fit SEDs. 
Using the $M_{\ast}$ from this fitting method, 5/259 galaxies from the MOSDEF $z\sim2$ spectroscopic sample have $\log_{10}$($M_{\ast}/M_{\odot}$) $<$ 9.0 and are removed from this sample. This is close to the number of galaxies removed from the MOSDEF $z\sim2$ spectroscopic sample with the FAST fitting methodology (9 galaxies). Therefore, the MOSDEF $z\sim2$ spectroscopic sample consists of 254 galaxies using the CSF models. 
The KBSS $z\sim2$ spectroscopic sample and both the MOSDEF and KBSS $z\sim2$ targeted samples have the same number of galaxies as they did in the main analysis.

\begin{figure*}
    \centering
     \subfloat[]{
       \includegraphics[width=0.32\linewidth]{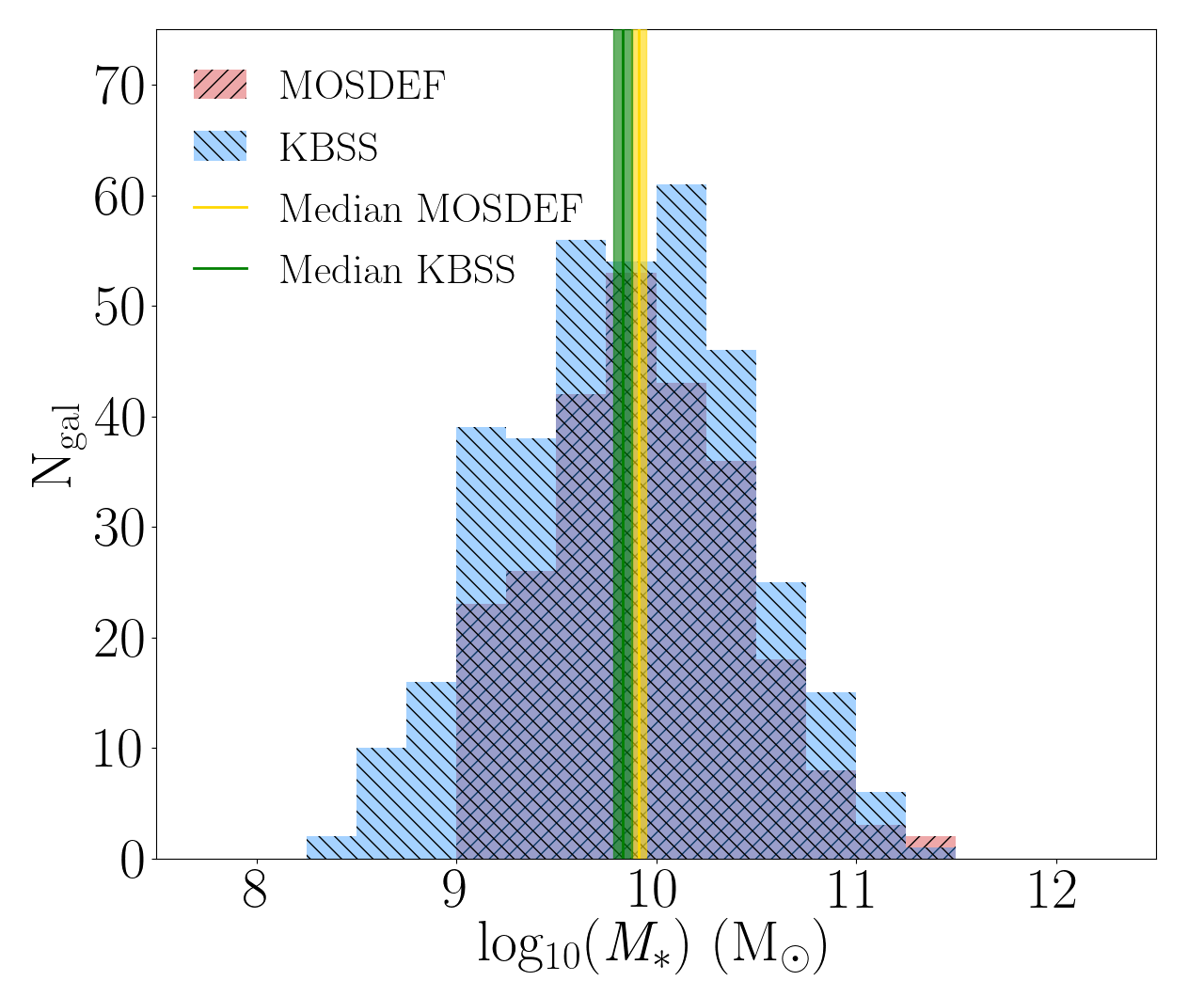}
     }
     \hfill
     \subfloat[]{
       \includegraphics[width=0.32\linewidth]{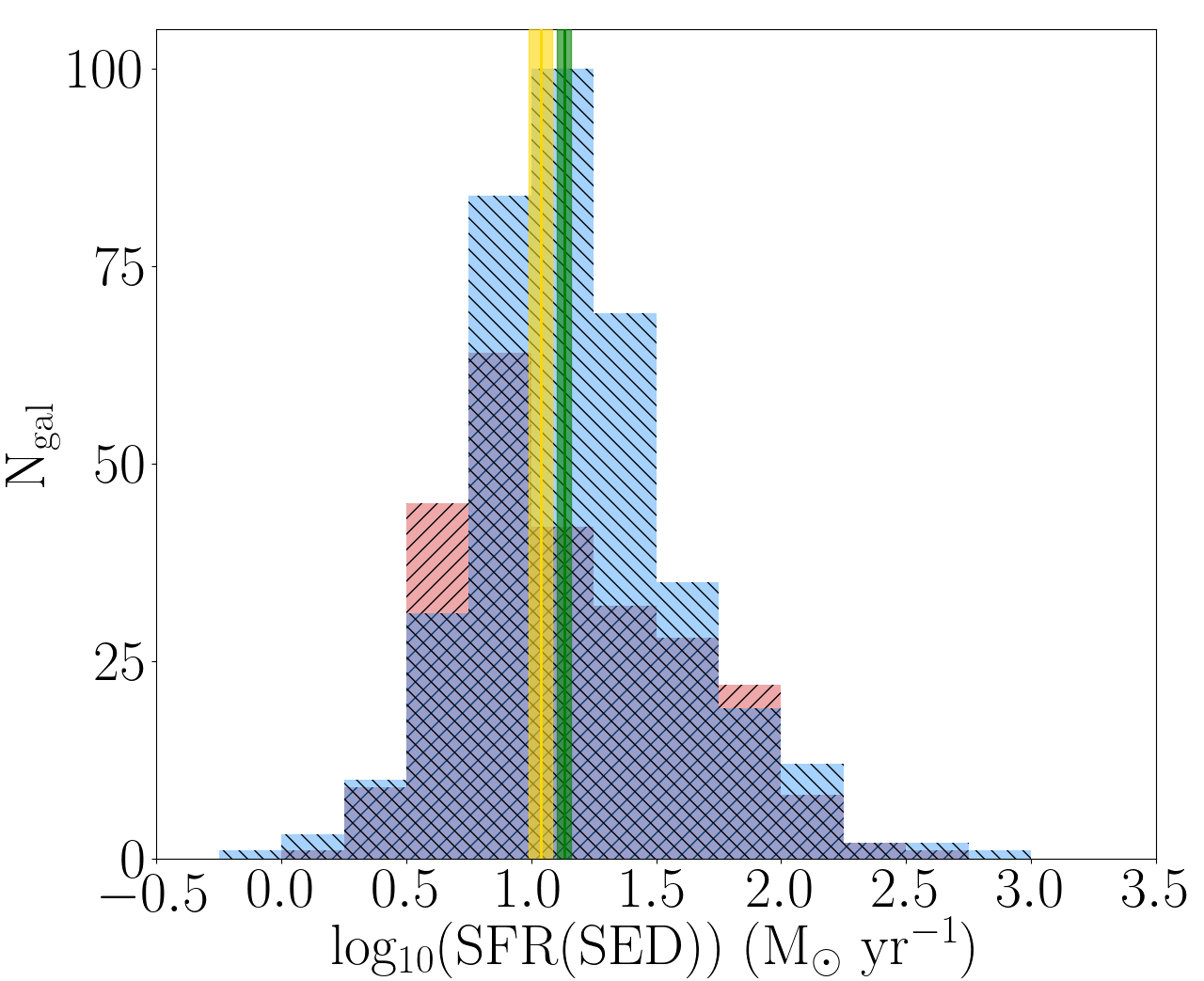}
     }
     \hfill
     \subfloat[]{
       \includegraphics[width=0.32\linewidth]{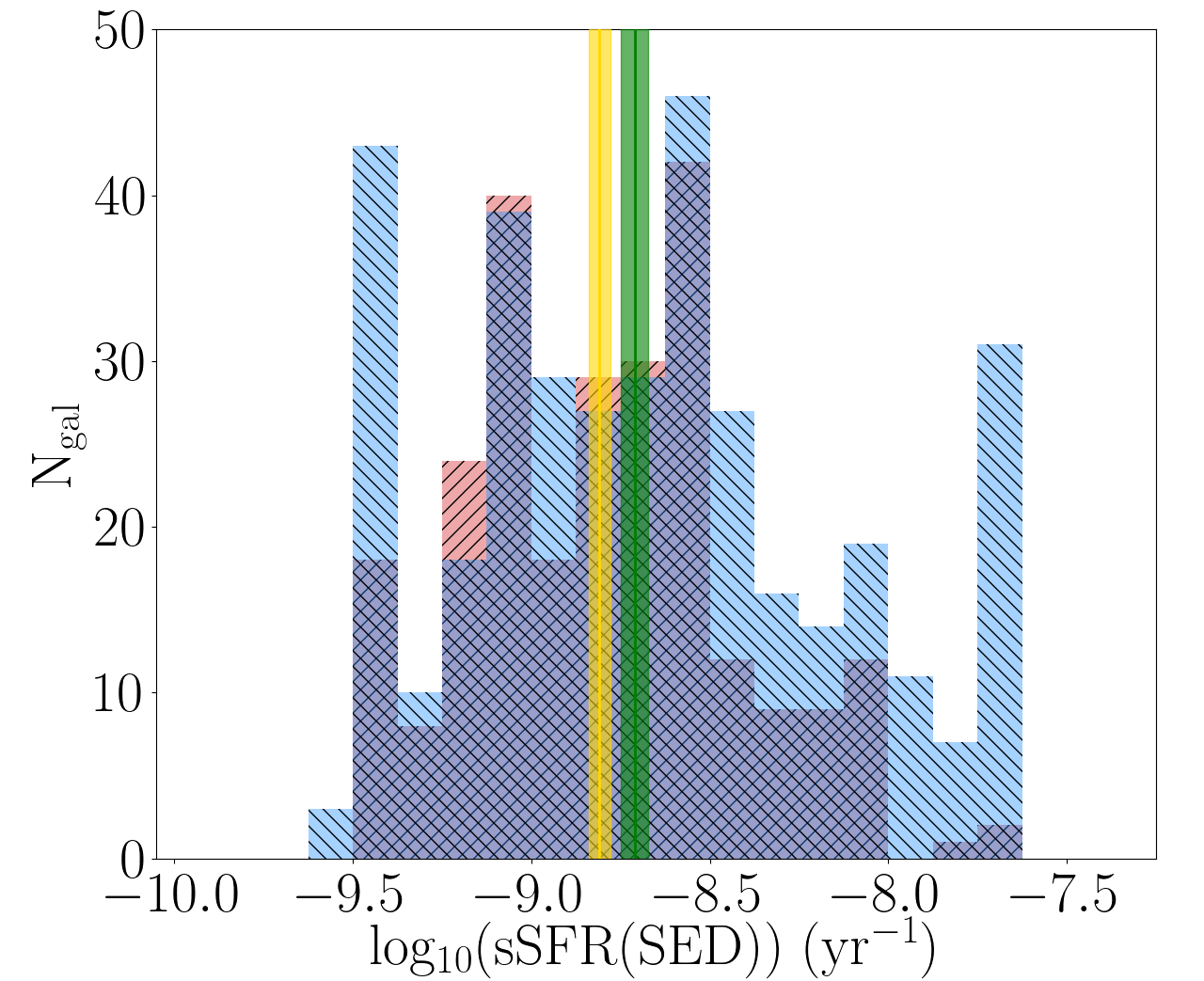}
     }
     \hfill
     \subfloat[]{
       \includegraphics[width=0.32\linewidth]{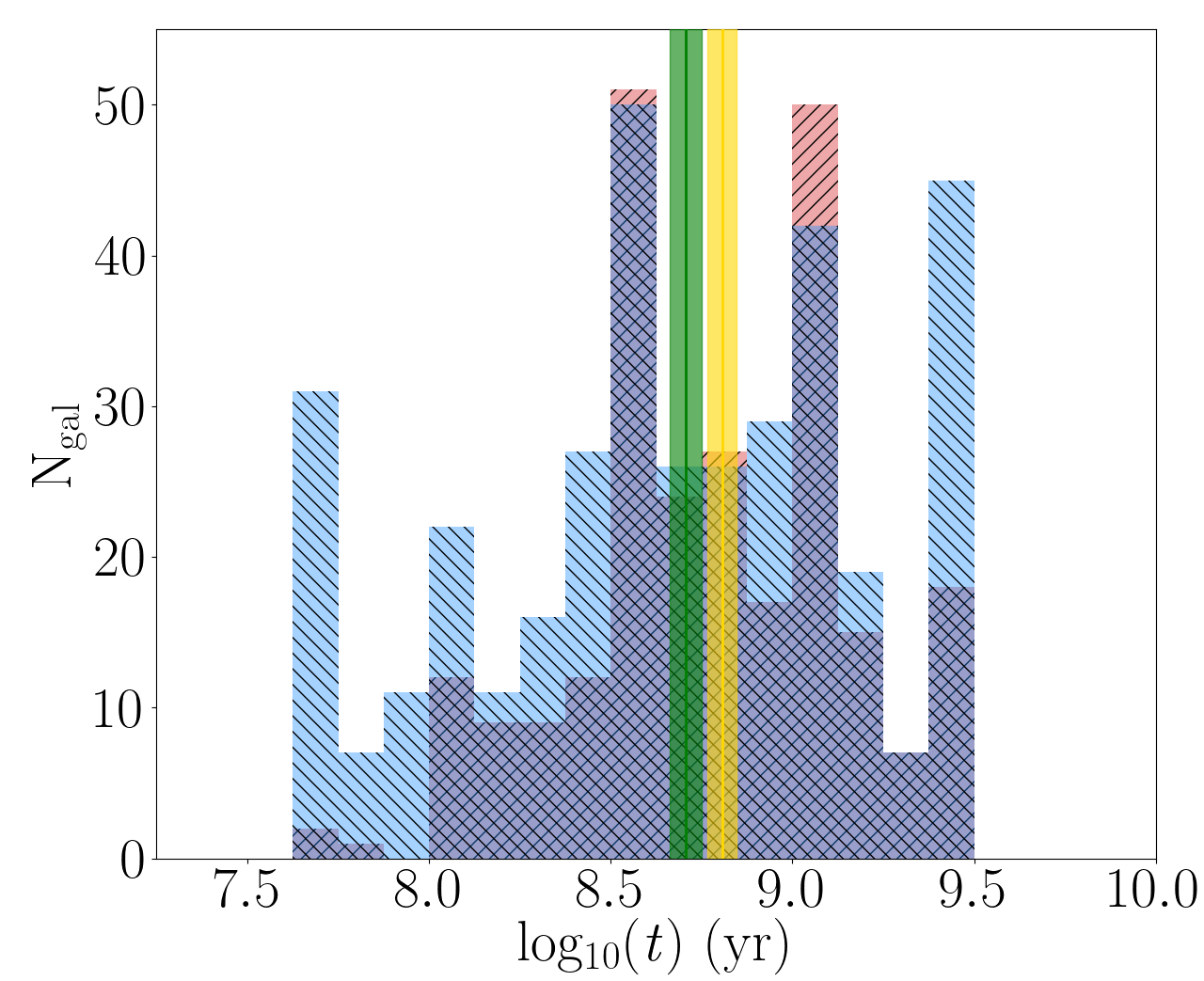}
     }
     \hfill
     \subfloat[]{
       \includegraphics[width=0.32\linewidth]{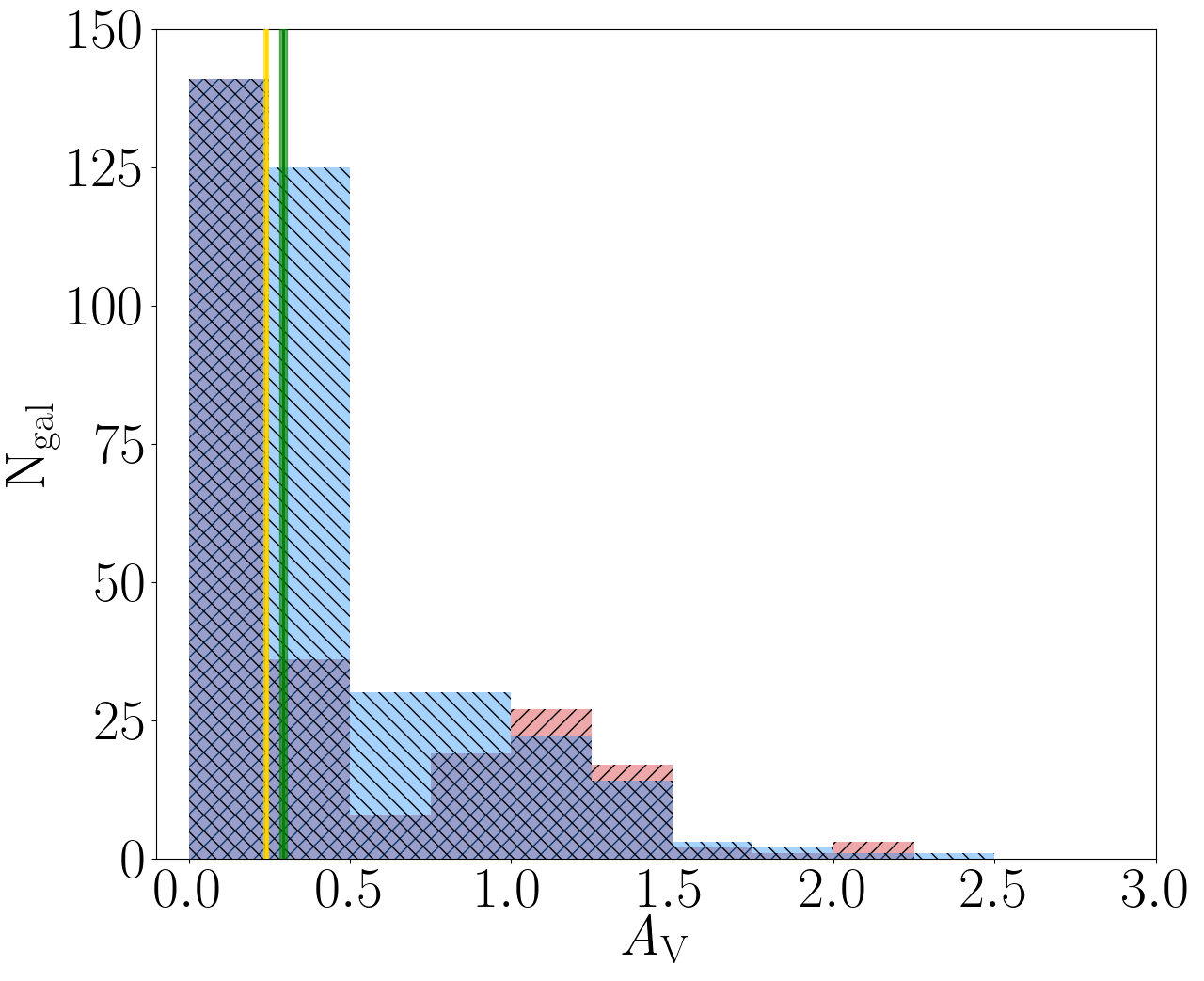}
     }
     \hfill
     \subfloat[]{
       \includegraphics[width=0.32\linewidth]{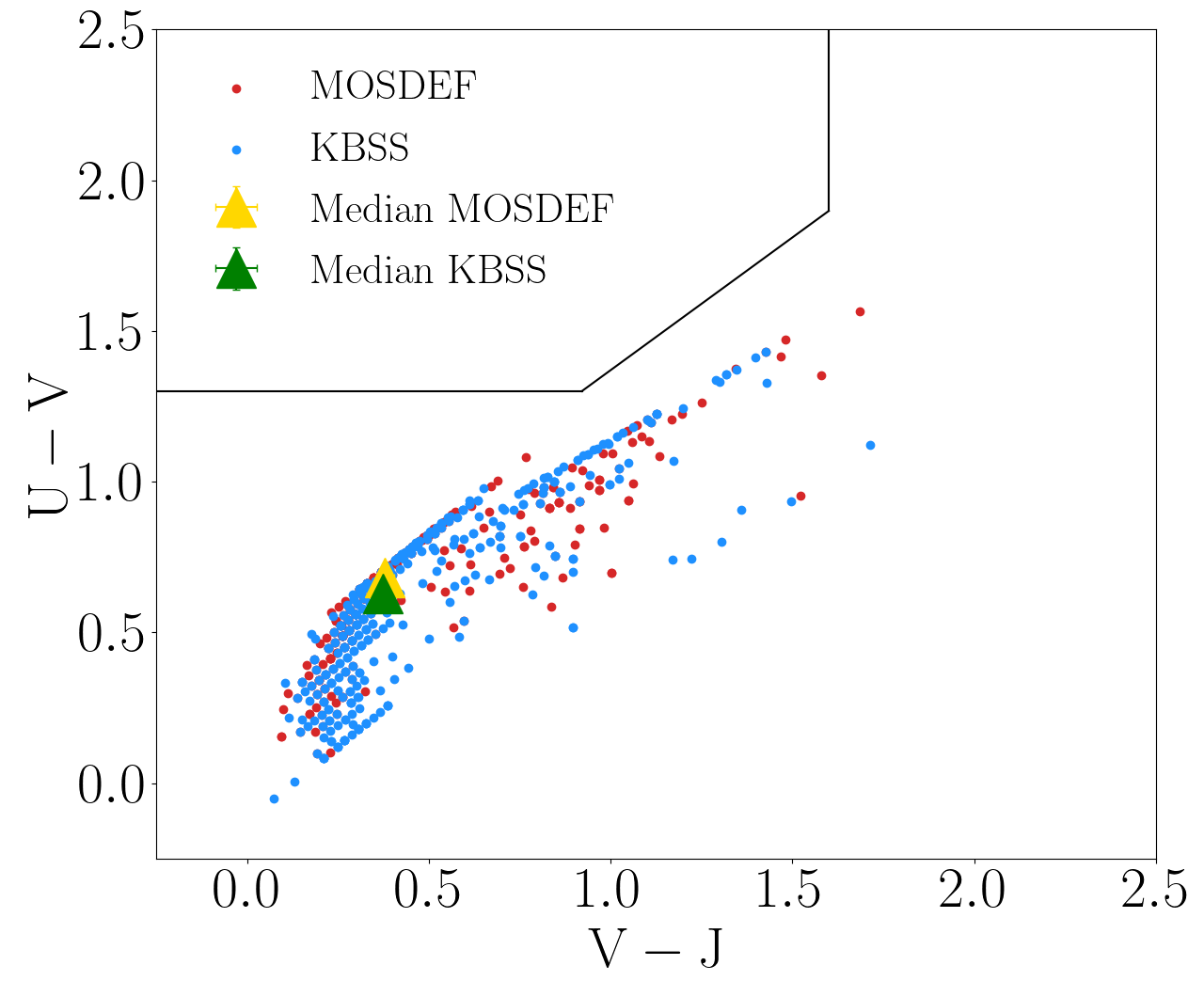}
     }
    \caption{Distribution of physical properties for the MOSDEF (red) and KBSS (blue) $z\sim2$ spectroscopic samples. The sample medians, with uncertainties are shown in yellow (green) for the MOSDEF (KBSS) samples, and are given in Table \ref{tab:z2_spec_sample_properties_csf}. The following galaxy properties are shown: (a) $M_{\ast}$, (b) SFR(SED), (c) sSFR(SED), (d) log$_{10}$($t$) of the stellar population using a constant star formation model, (e) $A_{\rm{V}}$, and (f) the UVJ diagram.  We find that the MOSDEF spectroscopic sample has a higher age, lower SFR(SED), lower sSFR(SED), and $A_{\rm{V}}$, and redder U$-$V color compared to the KBSS spectroscopic sample.}
    \label{fig:z2_spec_sample_properties_csf}
\end{figure*}

\begin{figure*}
    \centering
     \subfloat[]{
       \includegraphics[width=0.32\linewidth]{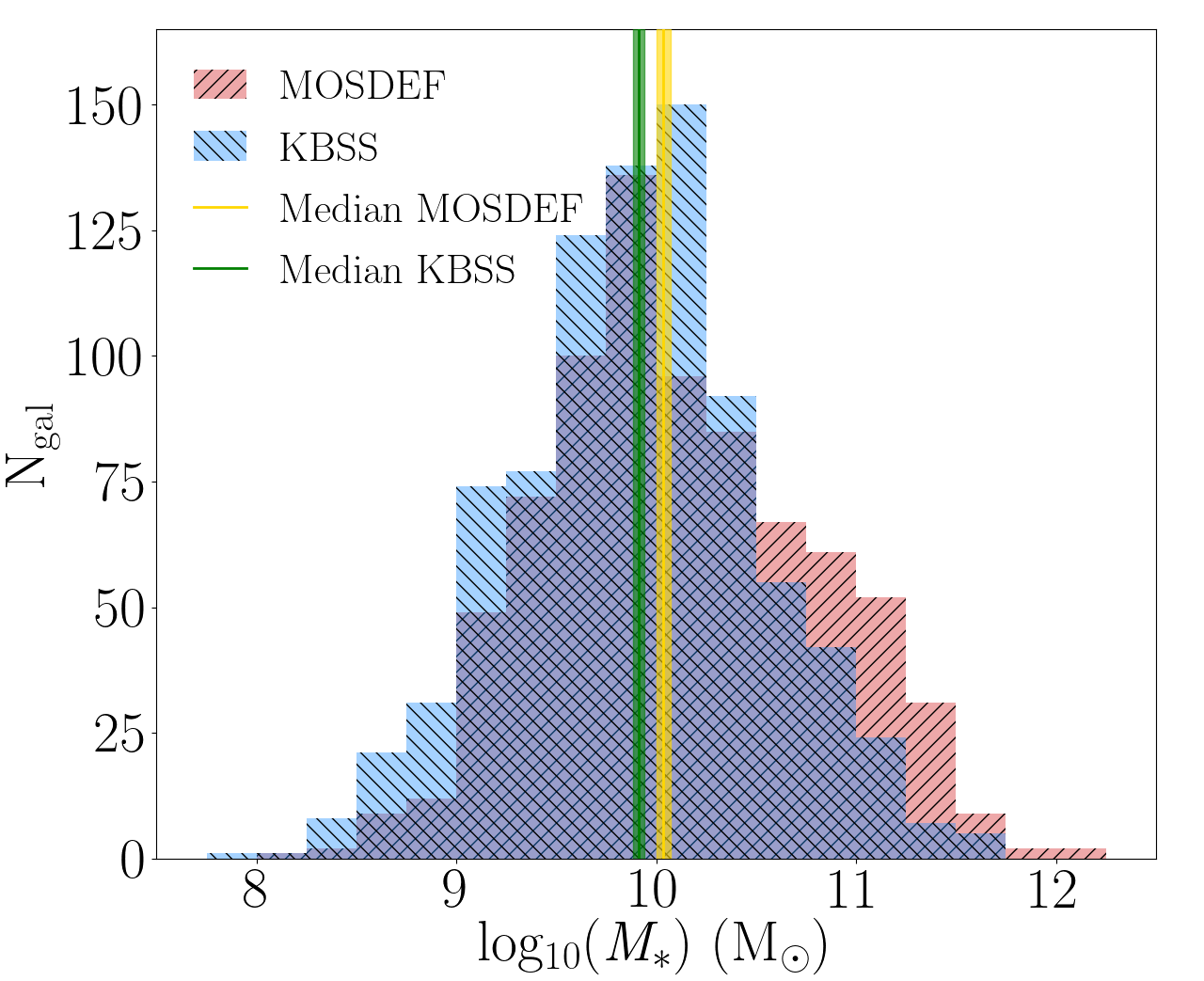}
     }
     \hfill
     \subfloat[]{
       \includegraphics[width=0.32\linewidth]{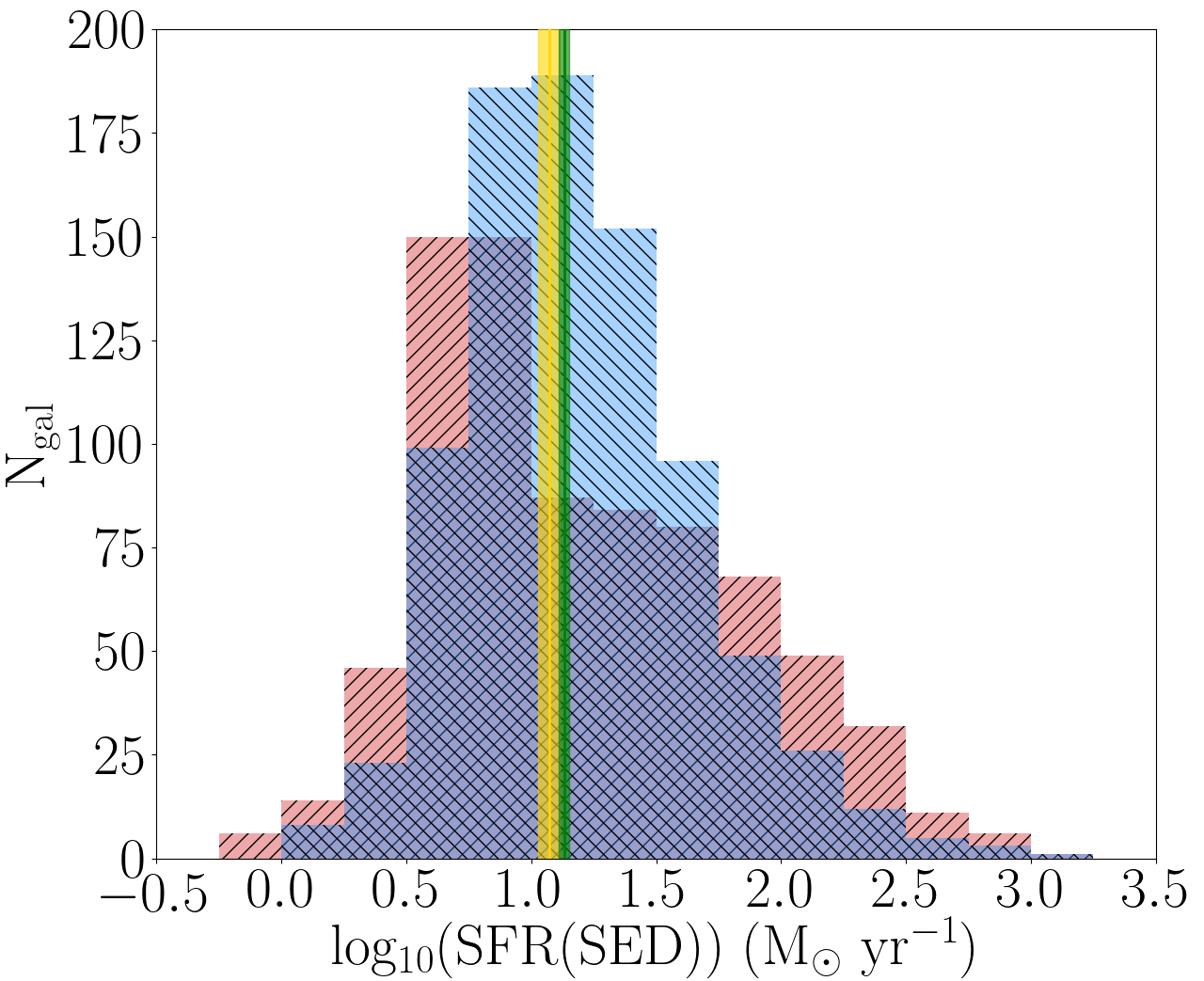}
     }
     \hfill
     \subfloat[]{
       \includegraphics[width=0.32\linewidth]{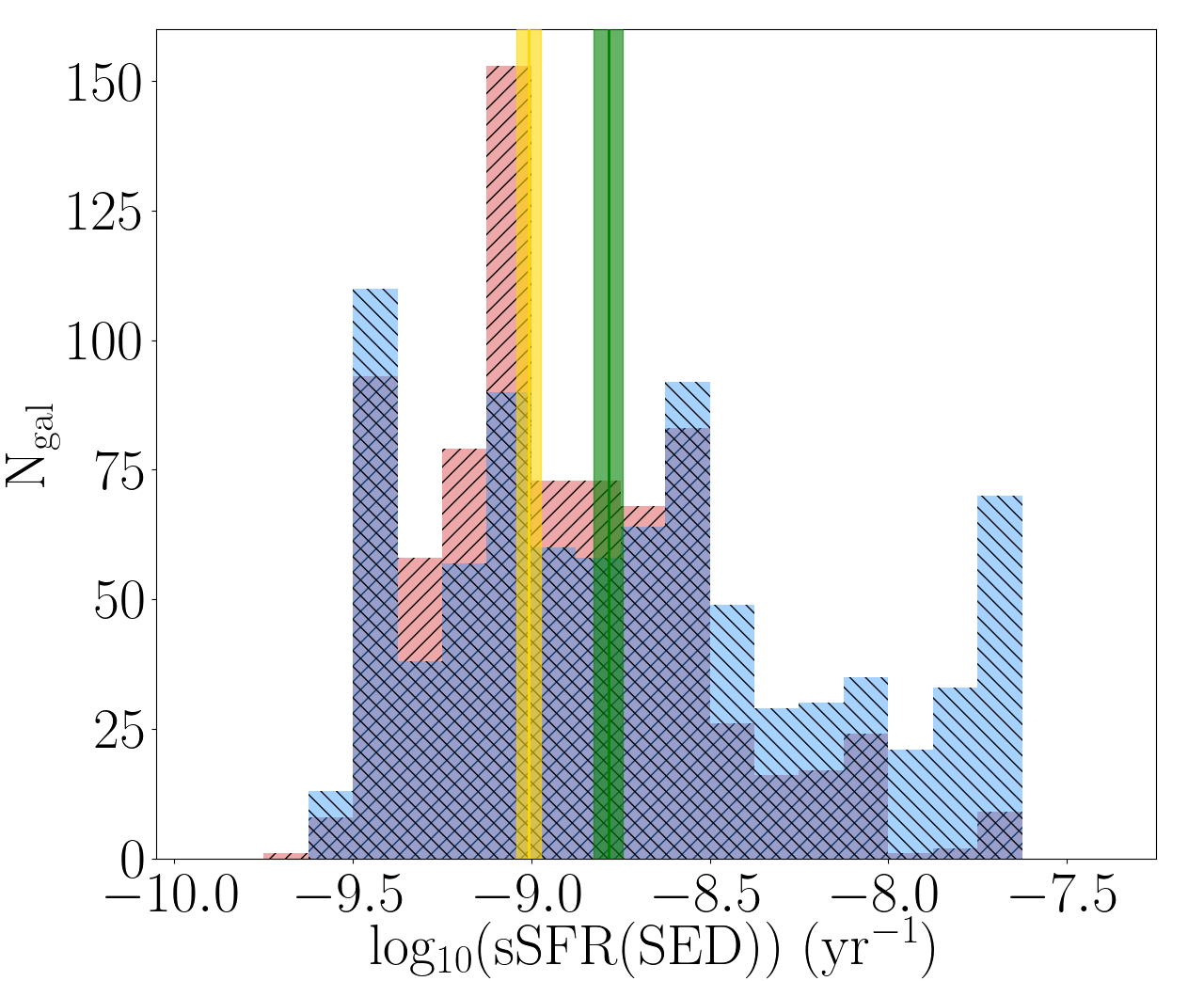}
     }
     \hfill
     \subfloat[]{
       \includegraphics[width=0.32\linewidth]{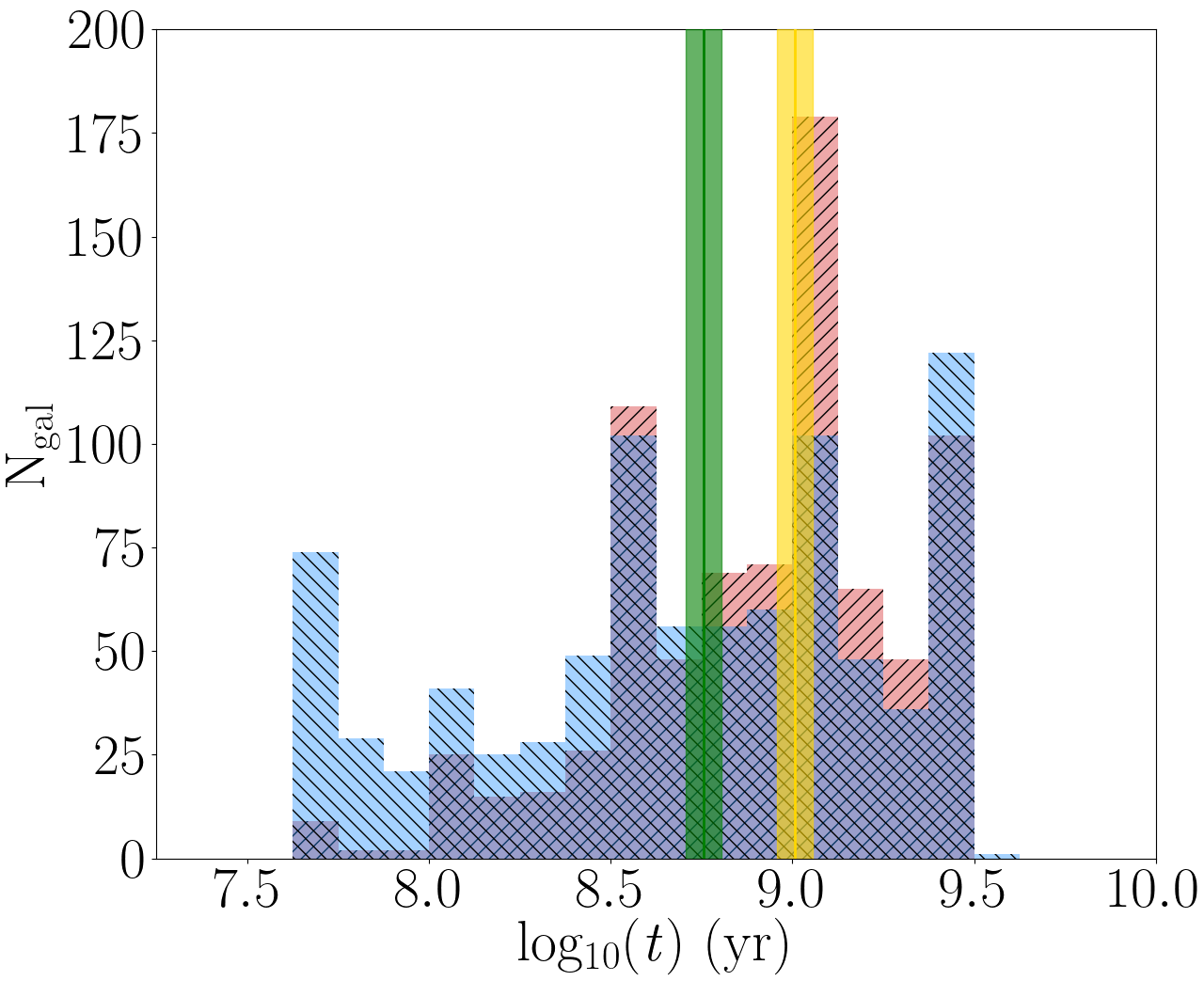}
     }
     \hfill
     \subfloat[]{
       \includegraphics[width=0.32\linewidth]{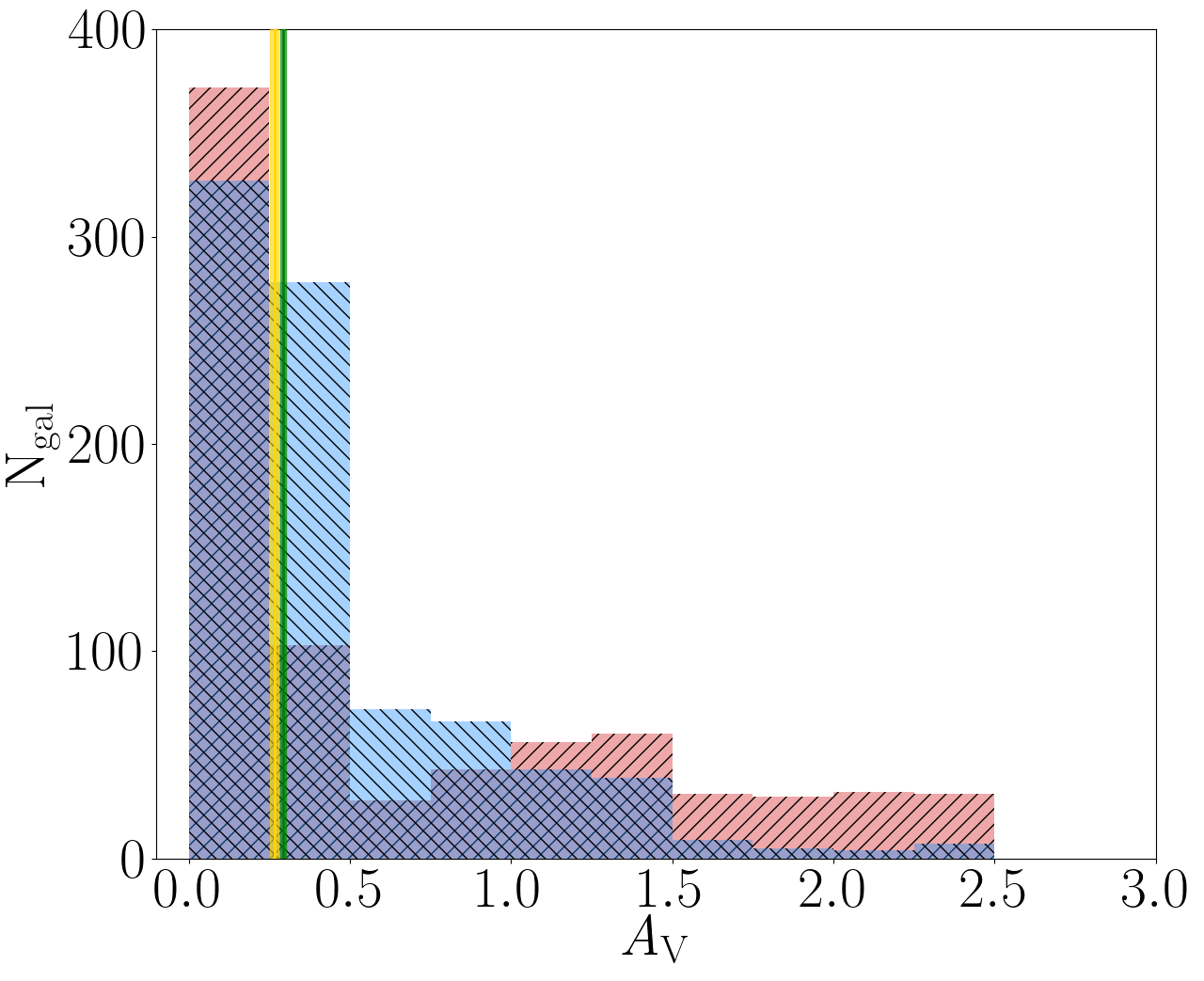}
     }
     \hfill
     \subfloat[]{
       \includegraphics[width=0.32\linewidth]{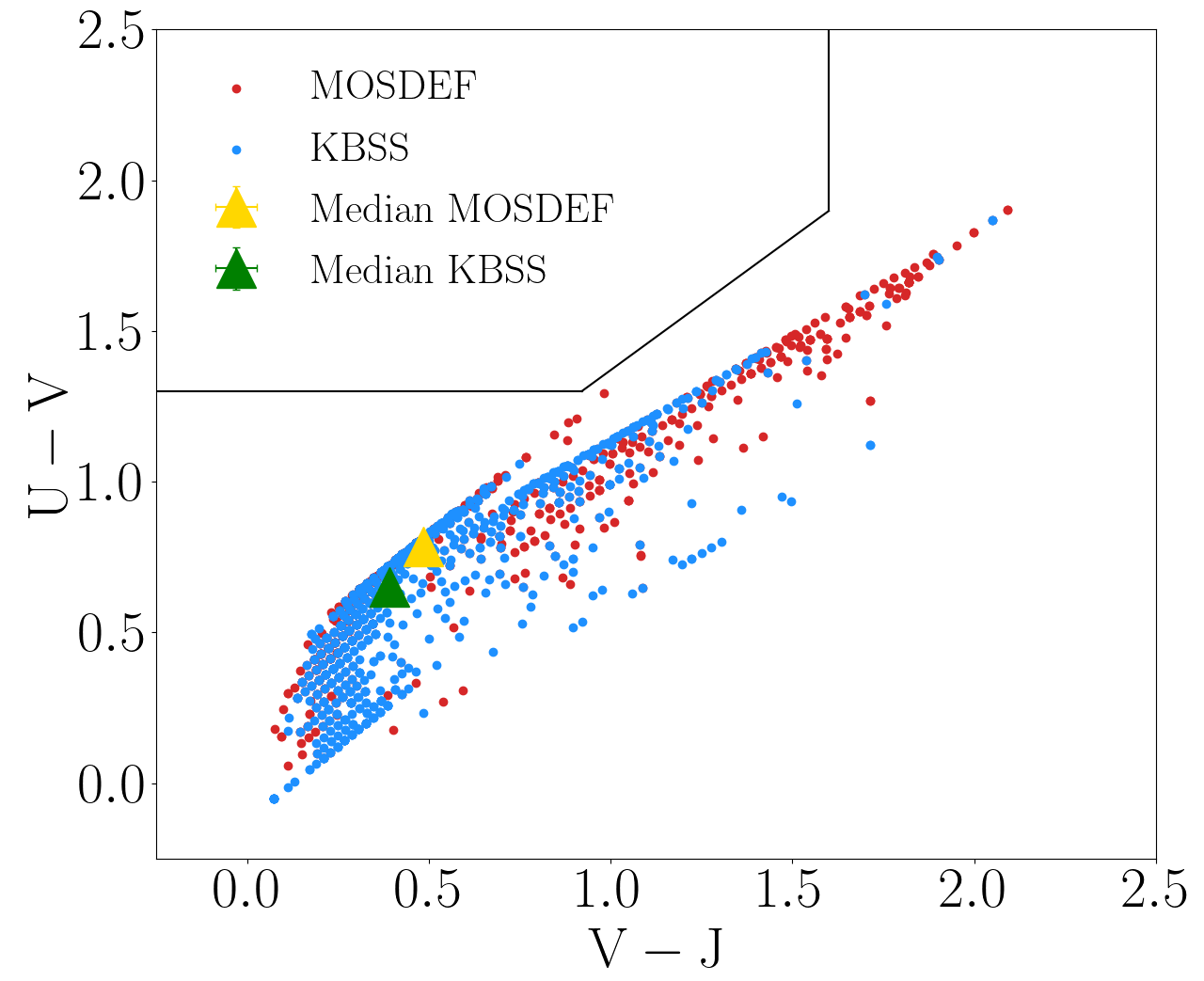}
     }
    \caption{Distribution of physical properties for the MOSDEF (red) and KBSS (blue) $z\sim2$ targeted samples. The sample medians, with uncertainties are shown in yellow (green) for the MOSDEF (KBSS) samples, and are given in Table \ref{tab:z2_targeted_sample_properties_csf}. The following galaxy properties are shown: (a) $M_{\ast}$, (b) SFR(SED), (c) sSFR(SED), (d) log$_{10}$($t$) of the stellar population using a constant star formation model, (e) $A_{\rm{V}}$, and (f) the UVJ diagram.  We find that the MOSDEF $z\sim2$ targeted sample has a higher $M_{\ast}$ and age, lower SFR(SED) and lower sSFR(SED), and redder UVJ colors compared to the KBSS $z\sim2$ targeted sample.}
    \label{fig:z2_targeted_sample_properties_csf}
\end{figure*}

\begin{figure*}
    \includegraphics[width=0.49\linewidth]{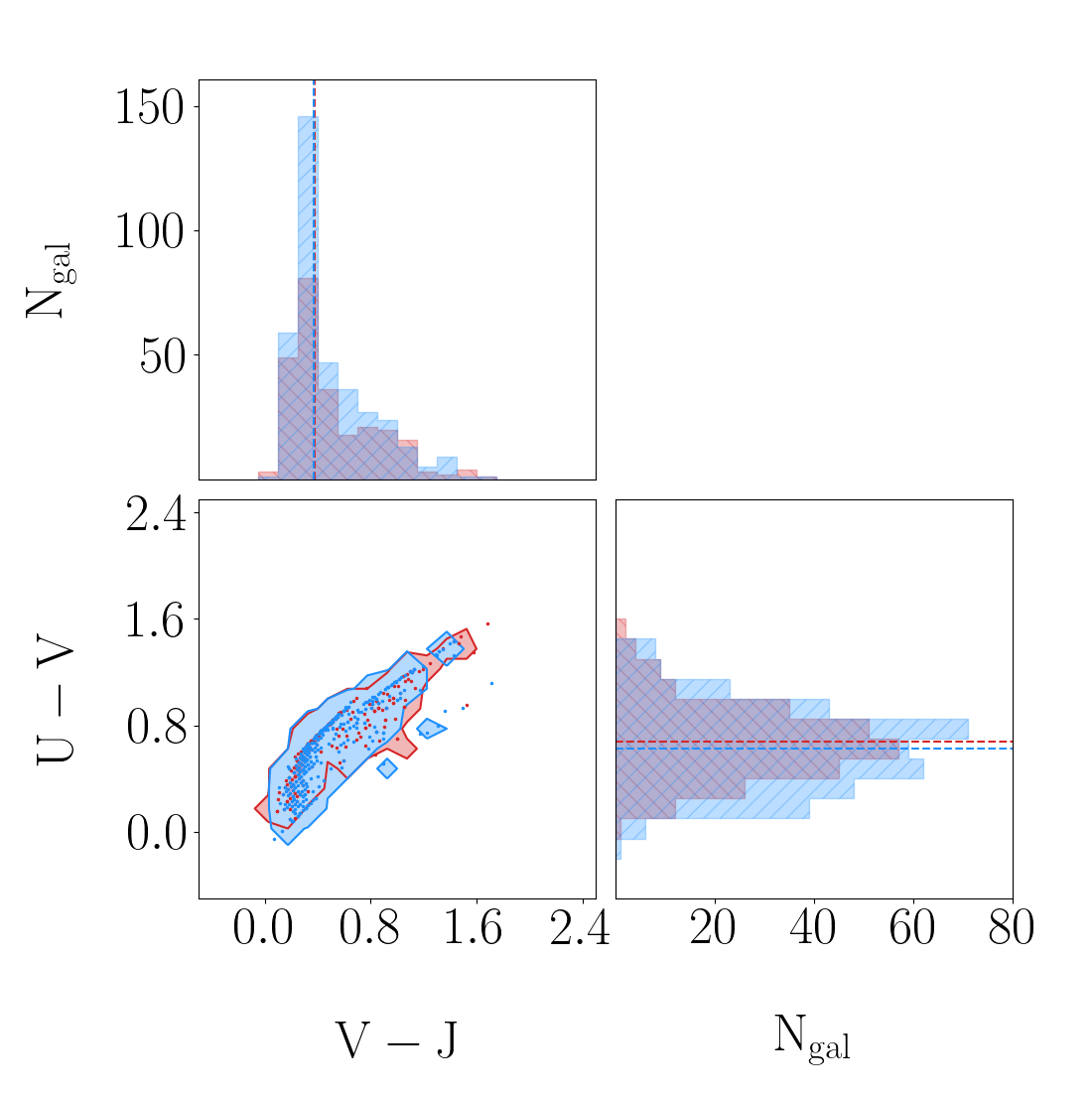}
    \includegraphics[width=0.49\linewidth]{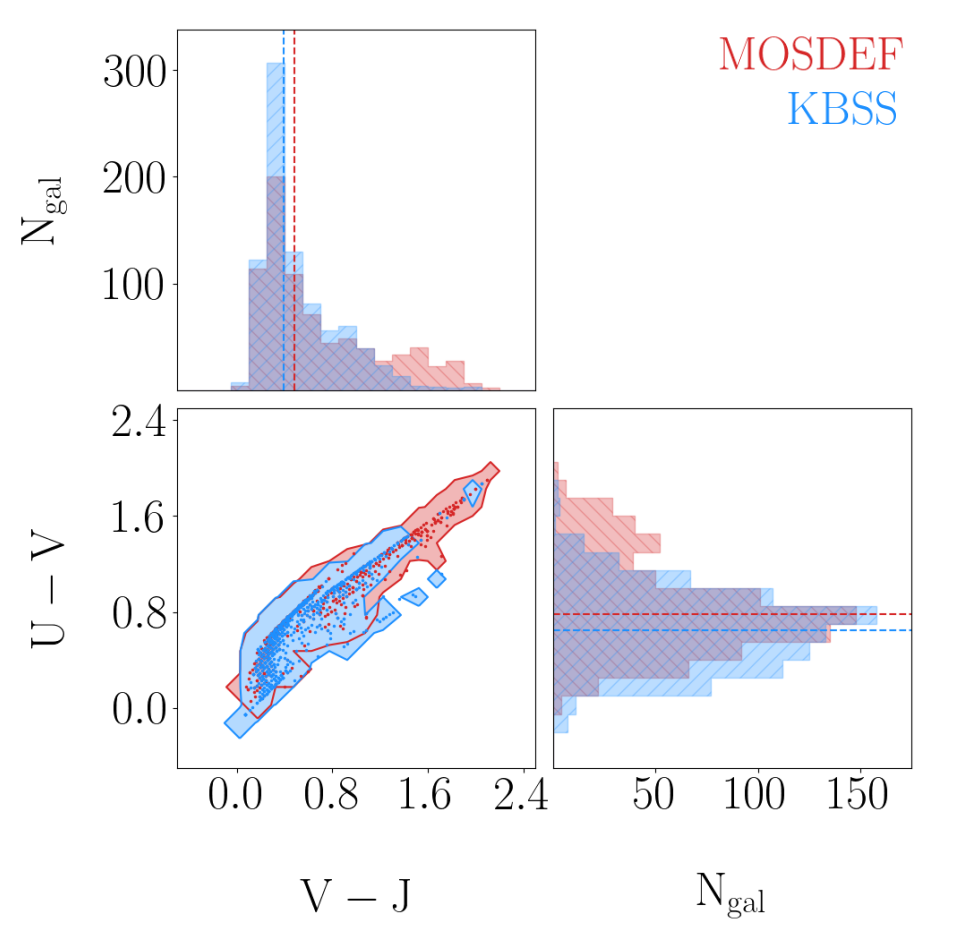}
    \caption{Corner plots comparing the distributions of U$-$V and V$-$J colors for the $z\sim2$ spectroscopic (left) and targeted (right) samples. MOSDEF and KBSS are shown in red and blue, respectively. The dashed lines on the 1D histograms mark the median values for the distributions of U$-$V and V$-$J colors given in Tables \ref{tab:z2_spec_sample_properties_csf} and \ref{tab:z2_targeted_sample_properties_csf}, and the contours in the 2D panel trace the 3$\sigma$ regions in UVJ color-color space. The individual data points are also shown in the 2D distribution. Similar to Figure \ref{fig:z2_zspec_sample_uvj_corner} and \ref{fig:z2_targeted_sample_uvj_corner}, it is shown here that while the $z\sim2$ spectroscopic samples span a similar range in UVJ space, the MOSDEF $z\sim2$ targeted sample extends to redder UVJ colors than the KBSS $z\sim2$ targeted sample.}
    \label{fig:mosdef_kbss_csf_uvj_corner}
\end{figure*}

Figures \ref{fig:z2_spec_sample_properties_csf} and \ref{fig:z2_targeted_sample_properties_csf} show the histograms and UVJ diagram for the galaxy properties of the $z\sim2$ spectroscopic and $z\sim2$ targeted samples, respectively. Tables \ref{tab:z2_spec_sample_properties_csf} and \ref{tab:z2_targeted_sample_properties_csf} contain the median values of the distributions for the $z\sim2$ spectroscopic and $z\sim2$ targeted samples, respectively. Additionally, Figure \ref{fig:mosdef_kbss_csf_uvj_corner} gives corner plots showing the 1D histograms and 2D distribution with contour maps of the UVJ colors for the $z\sim2$ spectroscopic and $z\sim2$ targeted samples.

We find that the MOSDEF $z\sim2$ spectroscopic sample has a slightly larger median $t$ (i.e., older) and U$-$V color (i.e., redder) and slightly smaller median SFR(SED), sSFR(SED), and $A_{\rm{V}}$ compared to the KBSS $z\sim2$ spectroscopic sample. The remaining galaxy properties, $M_{\ast}$ and V$-$J color, agree within the median uncertainties. The p-values estimated from the KS test are greater than $3\sigma$ for $t$, sSFR(SED), and $A_{\rm{V}}$. 
The result that the KBSS $z\sim2$ spectroscopic sample is younger than its MOSDEF counterpart is similar to the results using the FAST SED fitting from Section \ref{subsec:spec_sample_galaxy_properties}. Also similar is that the 3$\sigma$ contours occupy similar regions for the UVJ colors in Figure \ref{fig:mosdef_kbss_csf_uvj_corner}. 
The differences in the median SFR(SED), sSFR(SED), $A_{\rm{V}}$, and U$-$V color were not seen in Section \ref{subsec:spec_sample_galaxy_properties}. 
Similar to the delayed-$\tau$ model results, there is a population (49/369 or 13.3\%) of young KBSS galaxies with intense star formation (i.e., log$_{10}(t/$yr$)\leq8.0$ and log$_{10}$(sSFR(SED)/yr$^{-1}$)$\geq-8.0$). There are only 3/254 (1.2\%) MOSDEF galaxies in this regime. 

When looking at the $z\sim2$ targeted samples, we find that MOSDEF has a larger median $M_{\ast}$ and $t$, a smaller median sSFR(SED), and redder UVJ colors compared to KBSS. The medians for SFR(SED) and $A_{\rm{V}}$ agree within the uncertainties. These results are all highly significant, with p-values greater than $4\sigma$ for all galaxy properties. 
The results from the CSF models find differences in the median $t$ and U$-$V color of MOSDEF and KBSS $z\sim2$ targeted samples, which was not seen in the FAST fitting. 
On the other hand, the results from the CSF models find no differences in the median SFR(SED), as found in the results from FAST. 
Similar to the delayed-$\tau$ models, 124/850 (14.6\%) of KBSS galaxies are young (log$_{10}$($t$/yr)$\leq8.0$) with intense star formation (log$_{10}$(sSFR(SED)/yr$^{-1}$)$\geq-8.0$) compared to only 12/786 galaxies (1.5\%) in the MOSDEF sample. Meanwhile, 134/786 (17.0\%) of MOSDEF galaxies are extremely reddened (U$-$V and V$-$J $\geq$ 1.25) compared to 22/850 galaxies (2.6\%) in the KBSS sample. Additionally, the 3$\sigma$ contour map for MOSDEF extends to redder UVJ colors compared to KBSS. 
We note here that the distribution of inferred U-V and V-J colors for the MOSDEF $z\sim2$ targeted sample changes when assuming CSF models, relative to the results for the delayed-$\tau$ modeling presented in Section~\ref{subsec:targeted_sample_sed_properties}. Specifically, the CSF models do not accurately capture the SED shapes of the most massive, low sSFR(SED) galaxies in the MOSDEF $z\sim2$ targeted sample, and therefore place them (erroneously) outside of the quiescent region of the UVJ diagram. 

\begin{figure*}
    \includegraphics[width=0.49\linewidth]{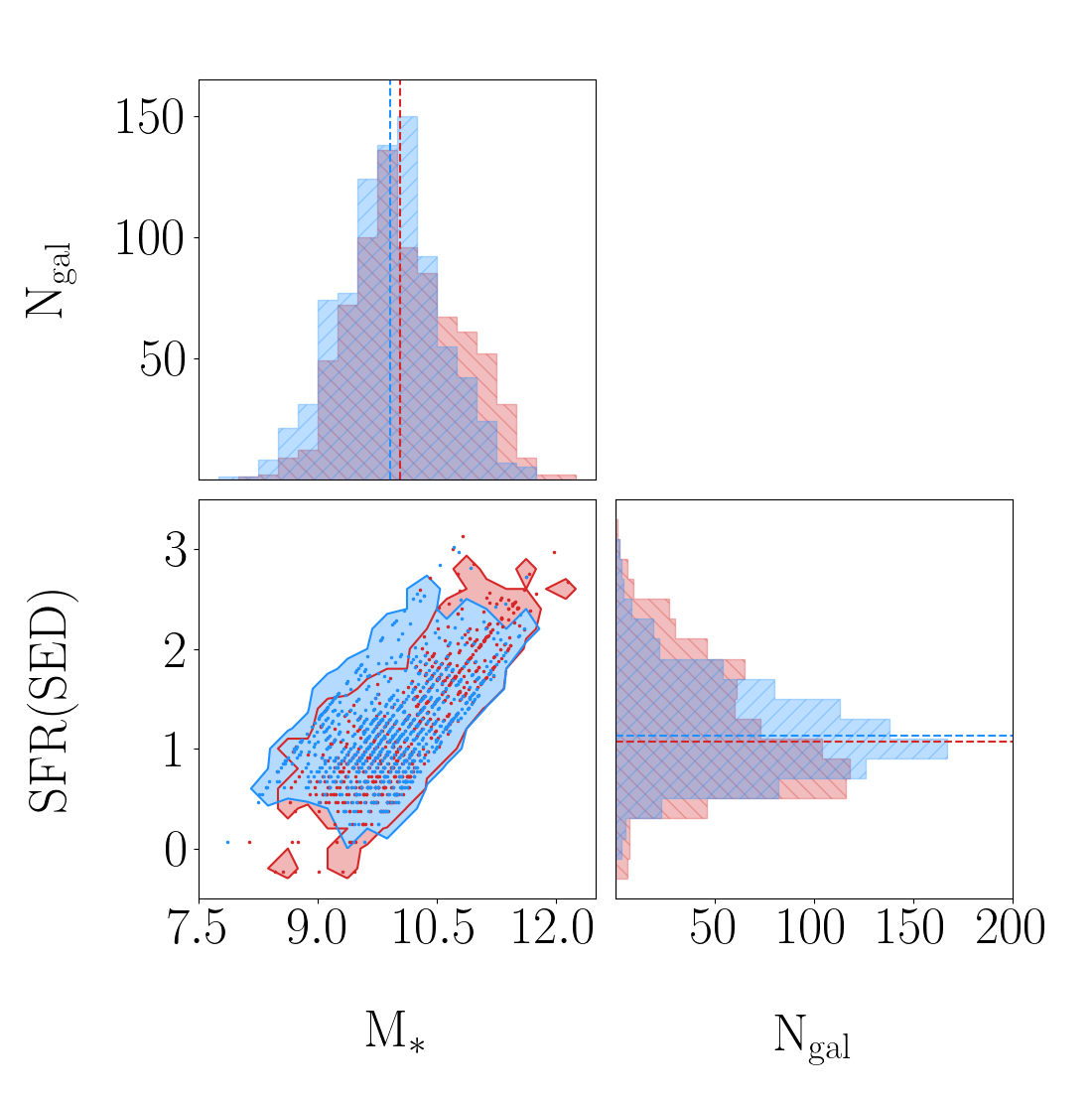}
    \includegraphics[width=0.49\linewidth]{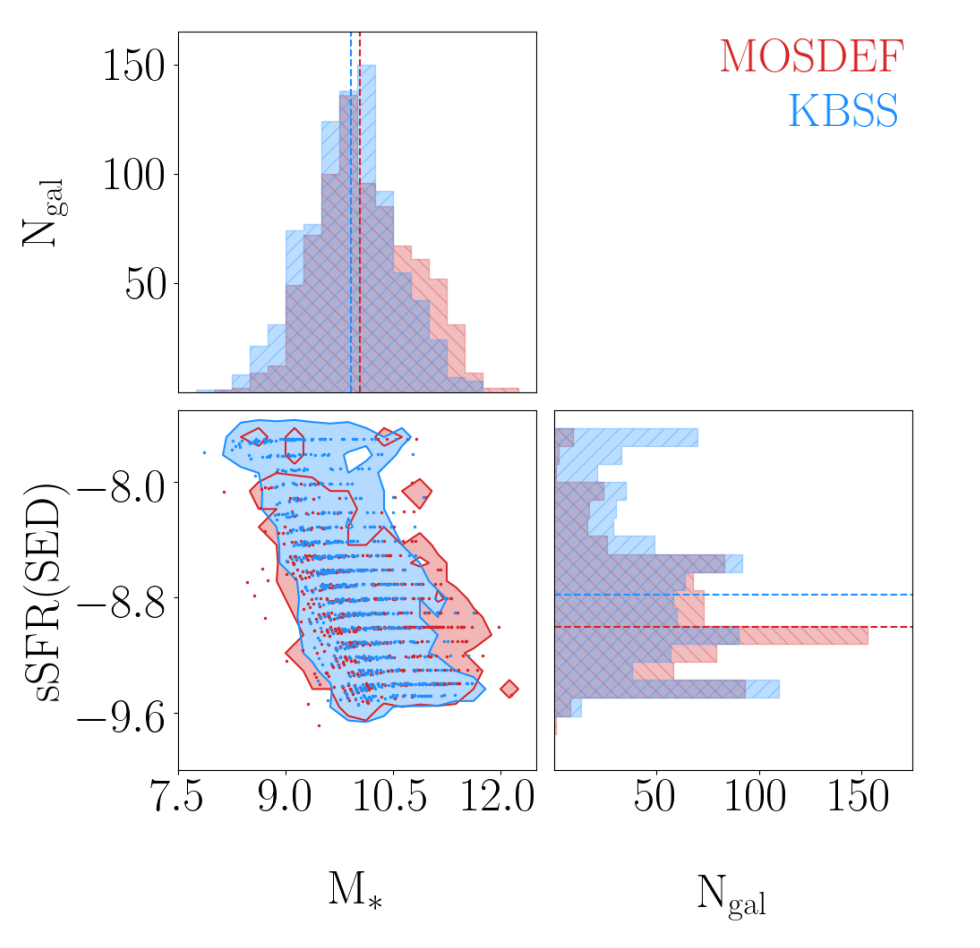}
    \includegraphics[width=0.49\linewidth]{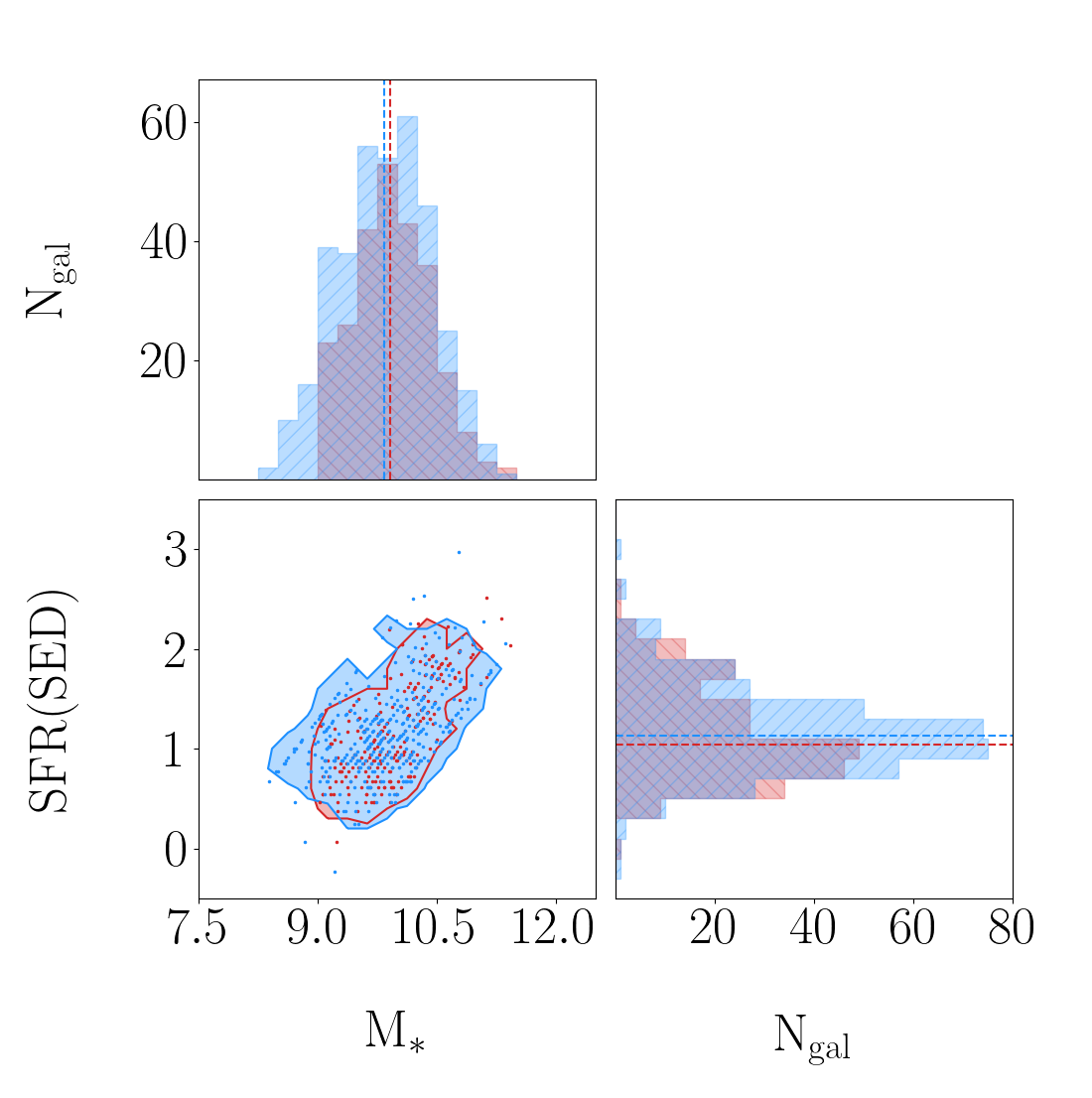}
    \includegraphics[width=0.49\linewidth]{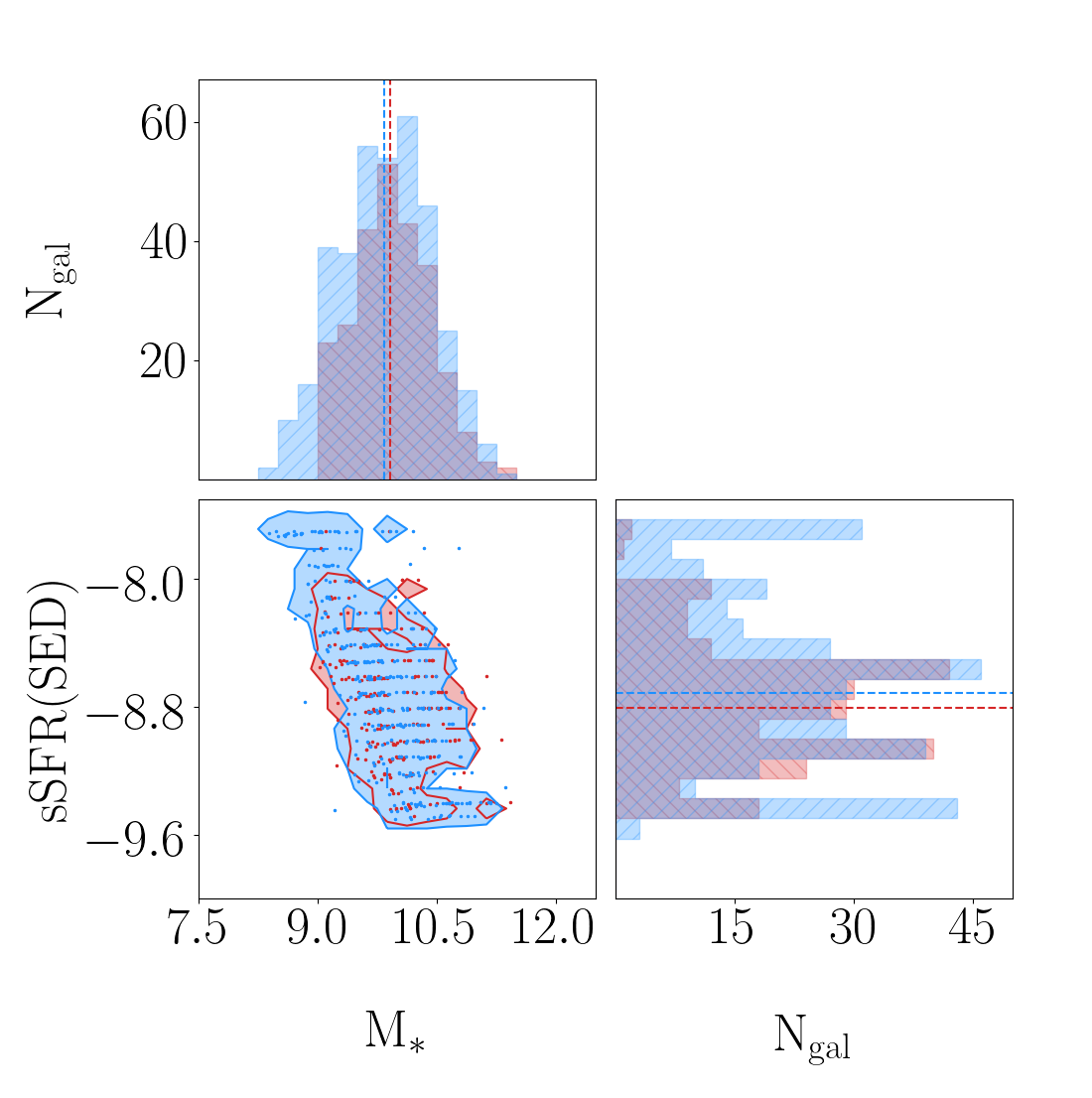}
    \caption{Top: $z\sim2$ targeted samples. Bottom: $z\sim2$ spectroscopic samples. The left (right) column contains corner plots comparing SFR(SED) vs. $M_{\ast}$ (sSFR(SED) vs. $M_{\ast}$). MOSDEF and KBSS are shown in red and blue, respectively. The dashed lines on the 1D histograms mark the median values for the distributions, and the contours on the diagonal trace the 3$\sigma$ regions in 2D space. The individual data points are also shown in the 2D distribution. These plots utilize CSF fitting, and are analogous to Figures \ref{fig:mosdef_kbss_survey_properties} ($z\sim2$ targeted sample) and \ref{fig:mosdef_kbss_spectroscopic_properties} ($z\sim2$ spectroscopic sample).}
    \label{fig:mosdef_kbss_csf_binned_plots}
\end{figure*}

Figure \ref{fig:mosdef_kbss_csf_binned_plots} contains corner plots showing both the 1D histograms and 2D contour mapping for SFR(SED) vs. $M_{\ast}$ and sSFR(SED) vs. $M_{\ast}$ for the MOSDEF and KBSS spectroscopic (top panels) and targeted (lower panels) samples. We find mostly similar results to the delayed-$\tau$ FAST results in the main body of the text (see Figures \ref{fig:mosdef_kbss_survey_properties} and \ref{fig:mosdef_kbss_csf_binned_plots}). KBSS extends to lower $M_{\ast}$ and higher sSFR(SED) than MOSDEF in both the $z\sim2$ spectroscopic and $z\sim2$ targeted samples. The distributions for the KBSS $z\sim2$ spectroscopic and $z\sim2$ targeted samples are very similar in both SFR(SED) vs. $M_{\ast}$ and sSFR(SED) vs $M_{\ast}$ space. Meanwhile, we find the same differences between the MOSDEF $z\sim2$ spectroscopic and $z\sim2$ targeted samples. The 3$\sigma$ contour for the MOSDEF spectroscopic sample reaches a similar upper mass, lower sSFR(SED) limit as the KBSS sample, while the MOSDEF $z\sim2$ targeted sample extends to higher $M_{\ast}$. This once again shows that MOSDEF is incomplete in this high mass regime. As we discuss previously in the paper, KBSS does not target the high mass galaxies for which MOSDEF is not successful in obtaining high quality S/N spectra.

While the results for the delayed-$\tau$ and CSF models do not perfectly align (i.e., the relative median values between the MOSDEF and KBSS $z\sim2$ spectroscopic and $z\sim2$ targeted samples sometimes vary based on which fitting method is used, and the rest-frame UVJ colors of the most massive galaxies in the MOSDEF $z\sim2$ targeted sample are not captured by the CSF models), the key takeaways of the study concerning the galaxy properties are the same for both models.
\begin{enumerate}
    \item The MOSDEF and KBSS $z\sim2$ spectroscopic samples are more similar to each other than the $z\sim2$ targeted galaxy samples.
    \item The KBSS spectroscopic and $z\sim2$ targeted samples have a subset of young galaxies with intense star formation that are mostly absent from the MOSDEF samples.
    \item The MOSDEF $z\sim2$ targeted sample contains extremely red (i.e., high UVJ colors) galaxies that are mostly absent from the KBSS $z\sim2$ targeted sample.
    \item The subset of extremely red MOSDEF galaxies in the $z\sim2$ targeted sample are not represented in the spectroscopic sample, highlighting that MOSDEF is incomplete in this regime. 
\end{enumerate}

As stated in Section \ref{subsec:sed_fitting}, these key takeaways remain unchanged if we use $\tau$-rising models instead of CSF or delayed-$\tau$ models.

\begin{table*}
    \centering
    \begin{tabular}{rrrrr}
        \multicolumn{5}{c}{Median Values for Physical Properties of the MOSDEF and KBSS $z\sim2$ Spectroscopic Samples (CSF Models)} \\
        \hline\hline
        Physical Property & MOSDEF Median & KBSS Median & p-value & Statistical significance \\
        (1) & (2) & (3) & (4) & (5) \\
        \hline
 log$_{10}(M_{\ast}/M_{\odot}$) & 9.91 $\pm$ 0.04 & 9.83 $\pm$ 0.05 & 0.023 & 2.27$\sigma$ \\
 log$_{10}$($t$/yr) & 8.81 $\pm$ 0.04 & 8.71 $\pm$ 0.04 & 0.00066 & 3.41$\sigma$ \\
 log$_{10}$(SFR(SED)/M$_{\odot}$/yr$^{-1}$) & 1.04 $\pm$ 0.05 & 1.13 $\pm$ 0.03 & 0.026 & 2.23$\sigma$ \\
 log$_{10}$(sSFR(SED)/yr$^{-1}$) & $-$8.81 $\pm$ 0.03 & $-$8.71 $\pm$ 0.04 & 0.00056 & 3.45$\sigma$ \\
 $A_{\rm{V}}$ & 0.24 $\pm$ 0.01 & 0.29 $\pm$ 0.01 & 0.00019 & 3.73$\sigma$ \\
 U$-$V & 0.68 $\pm$ 0.02 & 0.63 $\pm$ 0.02 & 0.057 & 1.90$\sigma$ \\
 V$-$J & 0.37 $\pm$ 0.02 & 0.37 $\pm$ 0.02 & 0.46 & 0.74$\sigma$ \\
        \hline
    \end{tabular}
    \caption{
  Col. (1): Physical property of the galaxies in the sample shown in Figure \ref{fig:z2_spec_sample_properties_csf}.
  Col. (2): Median value with uncertainty of the MOSDEF $z\sim2$ spectroscopic sample.
  Col. (3): Median value with uncertainty of the KBSS $z\sim2$ spectroscopic sample.
  Col. (4): Two-tailed p-value, based on the K-S test, estimating the probability that the null hypothesis can be rejected.
  Col. (5): Statistical significance (i.e. the $\sigma$ value) that the p-value corresponds to. }
    \label{tab:z2_spec_sample_properties_csf}
\end{table*}

\begin{table*}
    \centering
    \begin{tabular}{rrrrr}
        \multicolumn{5}{c}{Median Values for Physical Properties of the MOSDEF and KBSS $z\sim2$ Targeted Samples (CSF Models)} \\
        \hline\hline
        Physical Property & MOSDEF Median & KBSS Median & p-value & Statistical significance \\
        (1) & (2) & (3) & (4) & (5) \\
        \hline
 log$_{10}(M_{\ast}/M_{\odot}$) & 10.04 $\pm$ 0.04 & 9.91 $\pm$ 0.03 & 2.5e$-$7 & 5.16$\sigma$ \\
 log$_{10}$($t$/yr) & 9.01 $\pm$ 0.05 & 8.76 $\pm$ 0.05 & 7.8e$-$14 & 7.47$\sigma$ \\
 log$_{10}$(SFR(SED)/M$_{\odot}$/yr$^{-1}$) & 1.07 $\pm$ 0.04 & 1.13 $\pm$ 0.02 & 7.2e$-$6 & 4.49$\sigma$ \\
 log$_{10}$(sSFR(SED)/yr$^{-1}$) & $-$9.01 $\pm$ 0.03 & $-$8.78 $\pm$ 0.04 & 1.6e$-$14 & 7.68$\sigma$ \\
 $A_{\rm{V}}$ & 0.27 $\pm$ 0.01 & 0.29 $\pm$ 0.01 & 3.5e$-$13 & 7.27$\sigma$ \\
 U$-$V & 0.78 $\pm$ 0.01 & 0.65 $\pm$ 0.02 & 1.2e$-$10 & 6.44$\sigma$ \\
 V$-$J & 0.48 $\pm$ 0.02 & 0.39 $\pm$ 0.01 & 7.3e$-$9 & 5.78$\sigma$ \\
        \hline
    \end{tabular}
    \caption{
  Col. (1): Physical property of the galaxies in the sample shown in Figure \ref{fig:z2_targeted_sample_properties_csf}.
  Col. (2): Median value with uncertainty of the MOSDEF $z\sim2$ targeted sample.
  Col. (3): Median value with uncertainty of the KBSS $z\sim2$ targeted sample.
  Col. (4): Two-tailed p-value, based on the K-S test, estimating the probability that the null hypothesis can be rejected.
  Col. (5): Statistical significance (i.e. the $\sigma$ value) that the p-value corresponds to. }
    \label{tab:z2_targeted_sample_properties_csf}
\end{table*}

\section{Emission-line Fitting Examples} \label{sec:emline_fitting_example}

\begin{figure*}
    \centering
     \subfloat[]{
       \includegraphics[width=0.49\linewidth]{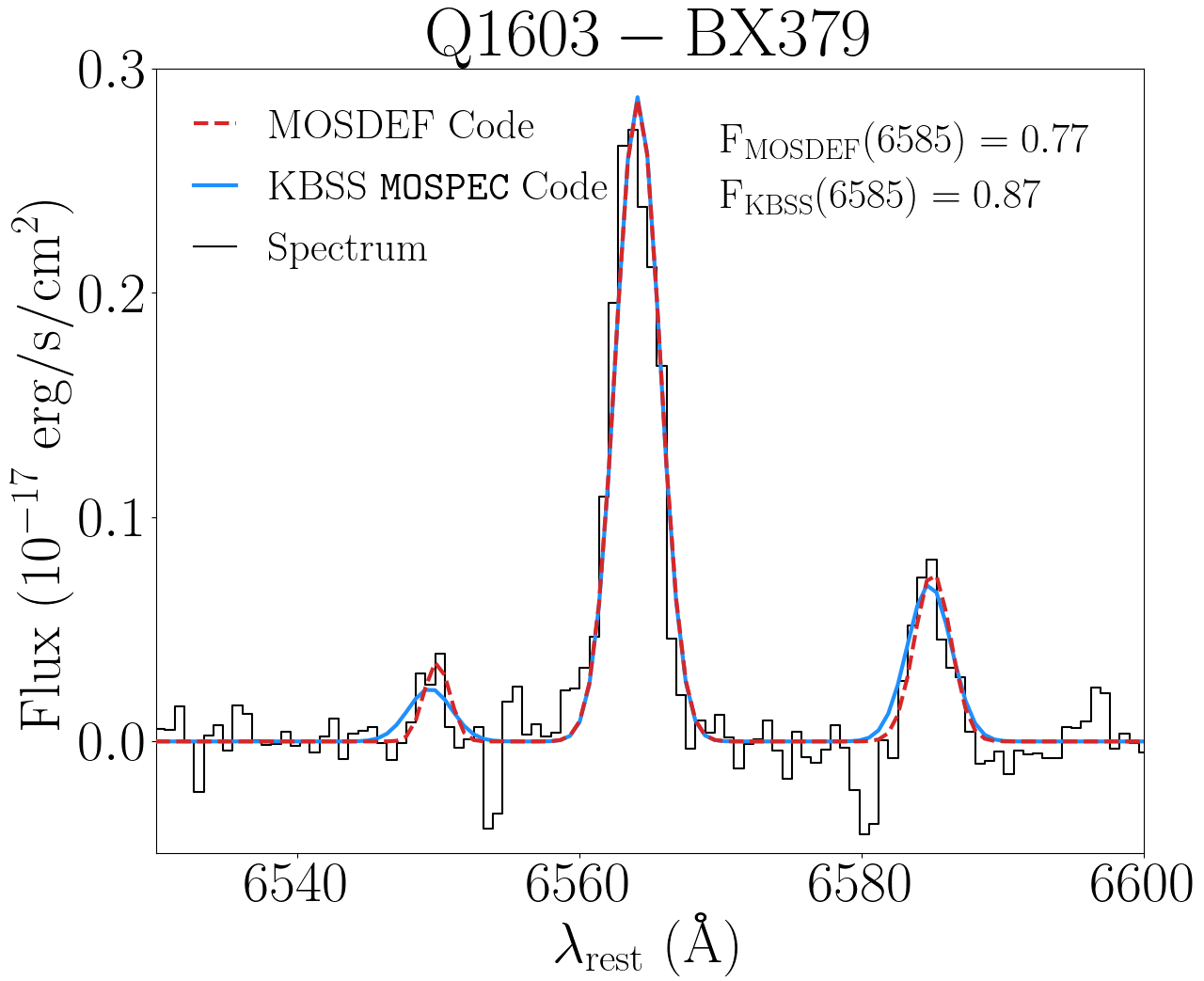}
     }
     \hfill
     \subfloat[]{
       \includegraphics[width=0.49\linewidth]{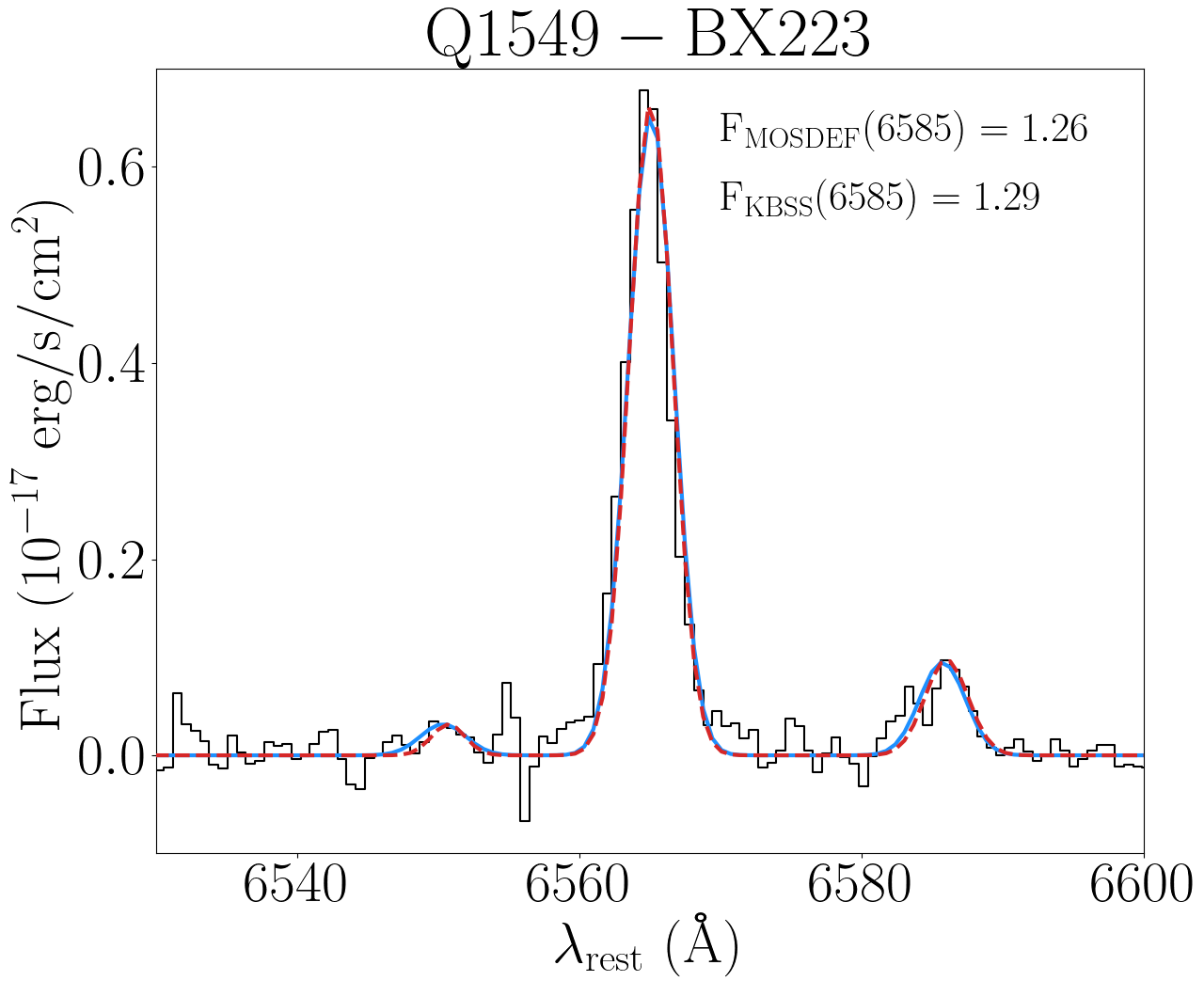}
     }
     \hfill
     \subfloat[]{
       \includegraphics[width=0.49\linewidth]{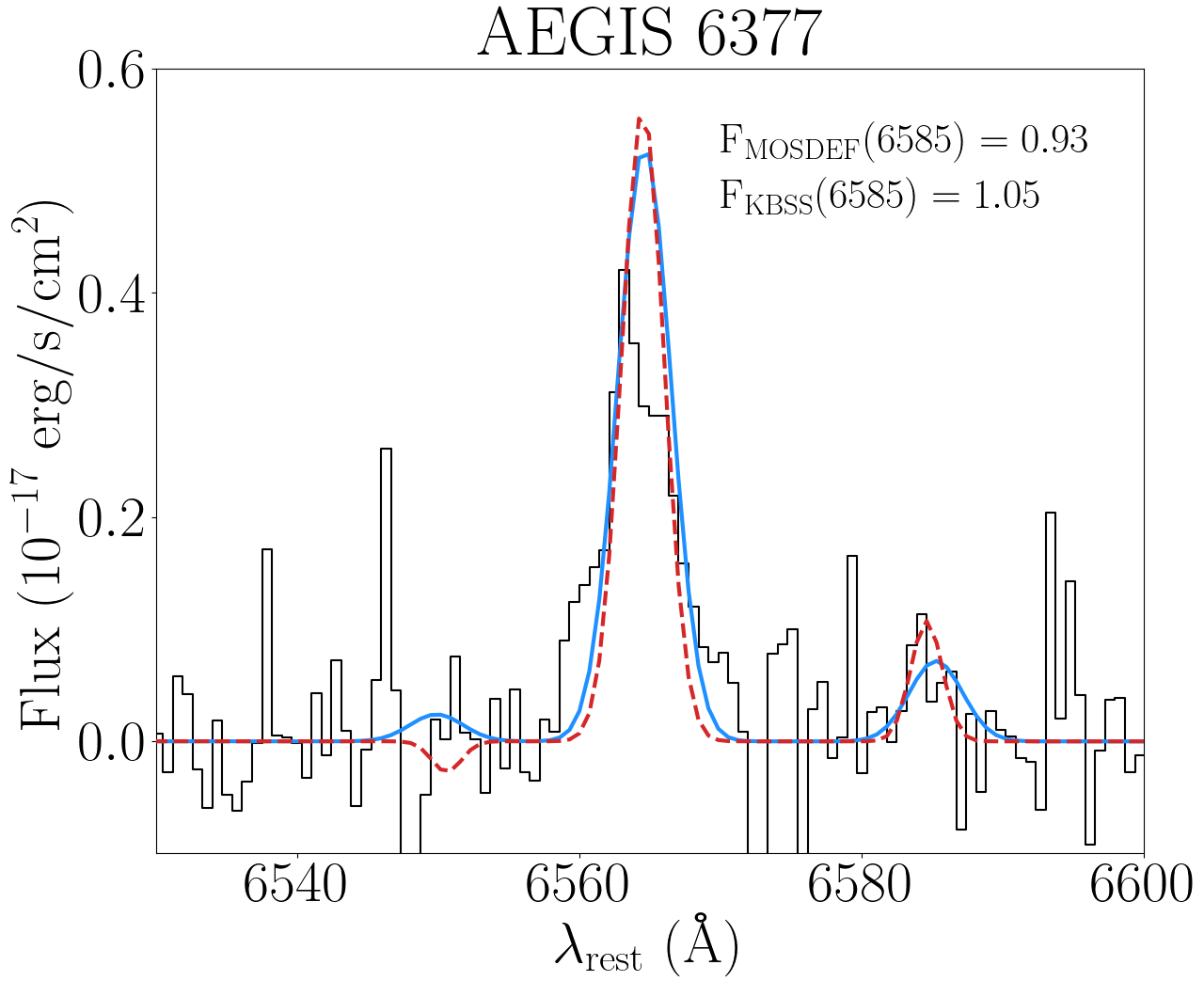}
     }
     \hfill
     \subfloat[]{
       \includegraphics[width=0.49\linewidth]{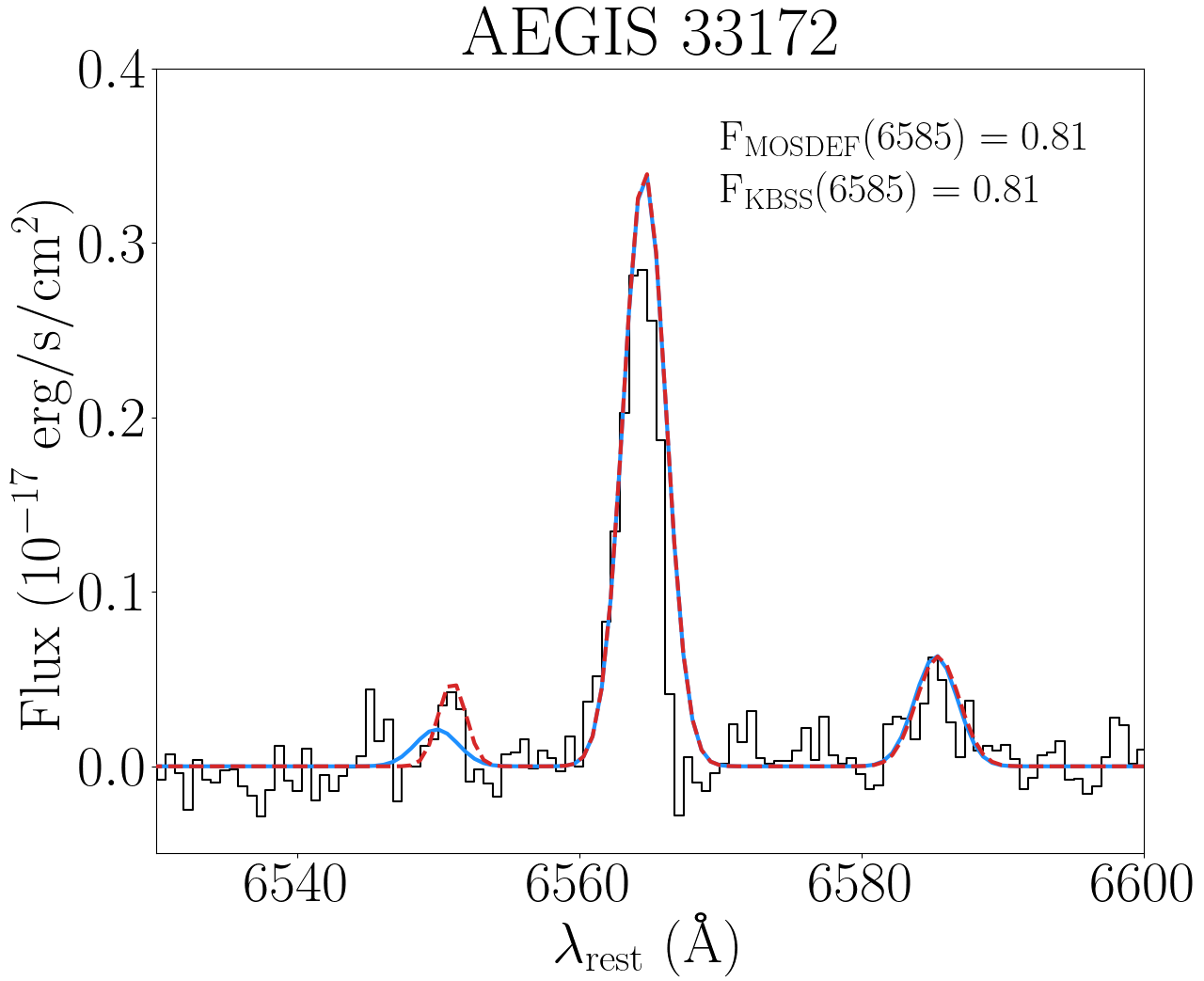}
     }
    \caption{Examples showing how the MOSDEF (red) and KBSS {\tt MOSPEC} (blue) codes fit the H$\alpha$ and [N~\textsc{II}]$\lambda$6585 emission-lines. The codes also simultaneously fit the  [N~\textsc{II}]$\lambda$6550; however, we did not use it for this study. We pick the following galaxies to highlight the following details about the code. (a) Q1603-BX379, for which we display the typical [N~\textsc{II}]$\lambda$6585 offset between the two codes (i.e., the KBSS {\tt MOSPEC} code is $\sim$13\% larger). The MOSDEF code uses the Gaussian fit for the preferred flux. (b) Q1549-BX223, for which we display similar [N~\textsc{II}]$\lambda$6585 fluxes, where the MOSDEF code uses the Gaussian fit for the preferred flux. (c) AEGIS 6377, for which we display the typical [N~\textsc{II}]$\lambda$6585 offset between the two codes where the MOSDEF code uses the Gaussian fit for the preferred flux. (d) AEGIS 33172, for which we display similar [N~\textsc{II}]$\lambda$6585 fluxes, where the MOSDEF code uses the Gaussian flux for the preferred flux. 
    Panel (c) shows that the Gaussian peaks for [N~\textsc{II}]$\lambda$6585 do not always perfectly agree in the two codes.} 
    \label{fig:example_emline_fits}
\end{figure*}

Here we provide examples of the MOSDEF and KBSS emission-line fitting. 
In these examples, we focus on the fit of [N~\textsc{II}]$\lambda$6585 because weaker emission-lines exhibited the most flux variance when one compares the MOSDEF and KBSS {\tt MOSPEC} codes. 
The fits to H$\alpha$ and [N~\textsc{II}]$\lambda$6550 (not relevant to this study) are included in the plots due to their proximity to [N~\textsc{II}]$\lambda$6585. 
Figure \ref{fig:example_emline_fits} shows the fits to four example spectra, two from the MOSDEF $z\sim2$ spectroscopic sample and two from the KBSS $z\sim2$ spectroscopic sample. Each panel includes the spectra as well as the fits from both the MOSDEF and KBSS {\tt MOSPEC} codes. 

As discussed in Section \ref{subsec:discussion_niibpt_offset}, the MOSDEF Gaussian fit to the [N~\textsc{II}]$\lambda$6585 line estimates a lower flux on average compared to the integrated bandpass flux. Meanwhile, the KBSS Gaussian fit estimates a higher flux on average compared to the integrated bandpass flux. 
The assumptions used by the two methods are both reasonable, given the differences in typical S/N of the MOSDEF and KBSS surveys. The goal of this paper has not been to determine which fitting method is more correct, but instead to show that different assumptions in the emission-line fitting can lead to small systematic variations in measured flux.

\bsp
\label{lastpage}
\end{document}